\documentclass[twocolumn]{aastex62}
\NewPageAfterKeywords
\usepackage{hyperref}
\usepackage[hyphenbreaks]{breakurl}
\usepackage{comment}
\usepackage{threeparttablex}
\usepackage{xstring}
\providecommand{\sci}[1]{\protect\ensuremath{\times 10^{\StrSubstitute[0]{#1}{e}{}}}}

\newcommand{\TOIs}{58 }

\graphicspath{{./}{figures/}}

\received{2021 November 25} 
\revised{2022 March 3} 
\accepted{2022 March 23}
\submitjournal{AJ}

\begin{document}

\title{A Dearth of Close-In Stellar Companions to M-dwarf TESS Objects of Interest}

\correspondingauthor{Catherine A. Clark}
\email{catclark@nau.edu}

\author[0000-0002-2361-5812]{Catherine A. Clark}
\affil{Northern Arizona University, 527 South Beaver Street, Flagstaff, AZ 86011, USA}
\affil{Lowell Observatory, 1400 West Mars Hill Road, Flagstaff, AZ 86001, USA}

\author[0000-0002-8552-158X]{Gerard T. van Belle}
\affil{Lowell Observatory, 1400 West Mars Hill Road, Flagstaff, AZ 86001, USA}

\author[0000-0002-5741-3047]{David R. Ciardi}
\affil{NASA Exoplanet Science Institute Caltech/IPAC, Pasadena, CA 91125, USA}

\author[0000-0003-2527-1598]{Michael B. Lund}
\affil{NASA Exoplanet Science Institute Caltech/IPAC, Pasadena, CA 91125, USA}

\author[0000-0002-2532-2853]{Steve~B.~Howell}
\affil{NASA Ames Research Center, Moffett Field, CA 94035, USA}

\author[0000-0002-0885-7215]{Mark E. Everett}
\affil{National Optical Infrared Research Laboratory, 950 N. Cherry Ave., Tucson, AZ 85719, USA}

\author[0000-0002-5627-5471]{Charles A. Beichman}
\affil{NASA Exoplanet Science Institute Caltech/IPAC, Pasadena, CA 91125, USA}

\author[0000-0001-6031-9513]{Jennifer G. Winters}
\affil{Center for Astrophysics $\vert$ Harvard \& Smithsonian, 60 Garden Street, Cambridge, MA 02138, USA}



\begin{abstract}

TESS has proven to be a powerful resource for finding planets, including those that orbit the most prevalent stars in our galaxy: the M dwarfs. Identification of stellar companions (both bound and unbound) has become a standard component of the transiting planet confirmation process in order to assess the level of light curve dilution and the possibility of the target being a false positive. Studies of stellar companions have also enabled investigations into stellar multiplicity in planet-hosting systems, which has wide-ranging implications for both exoplanet detection and characterization, as well as for the formation and evolution of planetary systems. Speckle and AO imaging are some of the most efficient and effective tools for revealing close-in stellar companions; we therefore present observations of \TOIs M-dwarf TOIs obtained using a suite of speckle imagers at the 3.5-m WIYN telescope, the 4.3-m Lowell Discovery Telescope, and the 8.1-m Gemini North and South telescopes. These observations, as well as near-infrared adaptive optics images obtained for a subset (14) of these TOIs, revealed only two close-in stellar companions. Upon surveying the literature, and cross-matching our sample with Gaia, SUPERWIDE, and the catalog from \citet{El-Badry2021MNRAS.506.2269E}, we reveal an additional 15 widely-separated common proper motion companions. We also evaluate the potential for undetected close-in companions. Taking into consideration the sensitivity of the observations, our findings suggest that the orbital period distribution of stellar companions to planet-hosting M dwarfs is shifted to longer periods compared to the expected distribution for field M dwarfs.

\end{abstract}



\section{Introduction} \label{sec:intro}

Ever since the first exoplanets were discovered 30 years ago \citep{Wolszczan1992Natur.355..145W}, astronomers have been on the hunt for worlds like our own. One way we are searching for exoplanets is with the Transiting Exoplanet Survey Satellite \citep[TESS;][]{Ricker2015JATIS...1a4003R}. TESS monitors the brightness of tens of thousands of stars in the sky at once, obtaining highly precise photometric time series, in the form of light curves, for each star. Exoplanets appear in these light curves as dips in the brightness, indicating that the planet is passing in front of, or transiting, its host star. However, one factor that can inhibit or change the detection of a planetary transit is the presence of an unseen stellar companion.

It has been shown that there is an observational bias against detecting Earth-sized, transiting planets due to ``third light'' contamination of the light curves, which is caused by stellar companions \citep{Lester2021AJ....162...75L}. This third light contamination can lead to additional obstacles in planet characterization including underestimated planet radii \citep{Ciardi2015ApJ...805...16C}, skewed planet radius distributions and occurrence rates \citep{Hirsch2017AJ....153..117H,Teske2018AJ....156..292T, Bouma2018AJ....155..244B}, incorrect characterization of both stars' properties \citep{FurlanHowell2020ApJ...898...47F}, and improper mean density and atmospheric values \citep{Howell2020FrASS...7...10H}. Additionally, close-in stellar companions (typically $<50-100$ au) have dynamical influences on planetary formation processes and the planets in their systems, including the perturbation and truncation of protoplanetary disks \citep{Jang-Condell2015ApJ...799..147J}, the gravitational excitement of planetesimals causing collisional destruction \citep{RafikovSilsbee2015aApJ...798...69R, RafikovSilsbee2015bApJ...798...70R}, and the scattering or ejection of planets that have formed \citep{HaghighipourRaymond2007ApJ...666..436H}. Though wider binaries have been thought to bear little impact on planetary architectures, recent studies have suggested that stellar companions within a few hundred au might affect the formation or orbital properties of giant planets \citep[e.g.][]{Fontanive2019MNRAS.485.4967F, FontaniveBardalezGagliuffi2021FrASS...8...16F, Hirsch2021AJ....161..134H, Mustill2021arXiv210315823M, Su2021AJ....162..272S}. Stellar companions can thus hinder planet formation and/or long-term stability of any inner planets that they host.

And stellar companions to solar-type exoplanet hosts (those of spectral types F, G, or K) are not rare: 40-50\% of solar-type Kepler \citep{Borucki2011ApJ...728..117B}, K2 \citep{Howell2014PASP..126..398H}, and TESS exoplanet host stars host stellar companions as well \citep[e.g.,][]{Horch2014ApJ...795...60H, Deacon2016MNRAS.455.4212D, Hirsch2017AJ....153..117H, Matson2018AJ....156...31M, Ziegler2018AJ....156..259Z, Howell2021AJ....161..164H}, which is consistent with the multiplicity rate for solar-type field stars \citep{Raghavan2010ApJS..190....1R}.

High-resolution imaging has become a standard process to eliminate the various ``astrophysical false positives'' that plague transit surveys \citep{Howell2011AJ....142...19H}, including the effects of close-in stellar companions, and has been used frequently for the validation and confirmation of transiting planets as well. In particular, speckle interferometry, or speckle imaging, has proven to be a powerful method for exploring the region around exoplanet hosts (within $\sim1\arcsec$) to search for close-in stellar companions that may be contaminating light curves from transit surveys.

Because of its ability to resolve close-in stellar companions, high-resolution imaging also offers the opportunity to study the orbital period distribution for planet-hosting binary systems. Work on Kepler, K2, and now TESS suggests that the stellar companions to solar-type exoplanet hosts have longer orbital periods than the companions to solar-type field stars \citep[e.g.,][]{Kraus2012ApJ...745...19K, Bergfors2013MNRAS.428..182B, Wang2014ApJ...791..111W, Kraus2016AJ....152....8K, Ziegler2020AJ....159...19Z, Howell2021AJ....161..164H, Lester2021AJ....162...75L, MoeKratter2021MNRAS.507.3593M}. Independent of these studies, an unbiased survey that combined radial velocity and imaging data of nearby solar-type stars found that the occurrence rate of planets in wide binaries is the same as (or perhaps higher than) that of single stars, but that close binaries (with separations $< 100$ au) have a much lower rate of planet occurrence \citep{Hirsch2021AJ....161..134H}. It has therefore been shown that the orbital period distribution may be shifted between planet-hosting and non-planet-hosting stars, at least for solar-type exoplanet hosts.

However, a systematic study with the specific goal of understanding the frequency and orbital period distribution of stellar companions to planet-hosting M dwarfs has not been carried out, although there have been specific searches for companions around low-mass stars in general \citep[e.g.,][]{Winters2019AJ....157..216W, Salama2021AJ....162..102S}. The M dwarfs have proven to be favorable targets for detecting Earth-sized planets in the habitable zone; however, if stellar companions are present, they could be hiding the Earth-sized planets. If the stellar multiplicity rate of planet-hosting M dwarfs is similar to that of the field M dwarfs \citep[$\sim27\%$;][]{Winters2019AJ....157..216W}, then many Earth-sized planets could be missed by transit surveys such as Kepler, K2, and TESS.

In order to understand the stellar multiplicity rate of planet-hosting M dwarfs, we used speckle imaging to search for stellar companions to \TOIs M-dwarf TESS Objects of Interest (TOIs). We employed four speckle imagers in both the northern and southern hemispheres to carry out this task. Optical speckle imaging is a powerful resource for the detection of stellar companions; however, the M dwarfs emit most strongly in the near-infrared. We have therefore supplemented our survey with near-infrared adaptive optics (AO) data. While these data were not collected specifically for our survey, they were obtained as part of the Exoplanet Follow-Up Observing Program (ExoFOP)\footnote{\url{https://exofop.ipac.caltech.edu/tess}}, and enable us to detect later-type and substellar companions that our speckle observations may have missed. In order to survey the orbital period distribution for planet-hosting M-dwarf binaries beyond the fields of view of the speckle and infrared cameras, we also cross-matched our sample with Gaia EDR3 \citep{Gaia2021A&A...649A...1G}, SUPERWIDE \citep{HartmanLepine2020ApJS..247...66H}, and \citet{El-Badry2021MNRAS.506.2269E} to search for common proper motion (CPM) companions (those that are moving in the same direction and at the same velocity as their M-dwarf TOI primaries). The specific objective of this work is therefore not only to provide high-resolution imaging data to the exoplanet scientists  searching for planets in the TESS data, but also to investigate the multiplicity rate and orbital period distribution for M-dwarf exoplanet hosts.

In Section \ref{sec:observations}, we describe our target selection, observations, and data reduction process. In Section \ref{sec:stellarcompanions}, we provide observed properties for the stellar companions revealed by our high-resolution imaging, the stellar companions known to the literature, and the potential CPM companions found with Gaia EDR3, SUPERWIDE, and \citet{El-Badry2021MNRAS.506.2269E}. We also discuss potential close-in companions. We estimate astrophysical parameters, including projected separation, mass ratio, and orbital period, for the stellar companions that have observational data. In Section \ref{sec:missed_companions}, we evaluate the completeness of the high-resolution imaging, and discuss possible properties of the potential close-in companions. In Section \ref{sec:implications}, we discuss the implications of our findings for the orbital period distribution for planet-hosting M dwarfs. We summarize our conclusions and discuss future work in Section \ref{sec:summary}.

\section{Observations} \label{sec:observations}

We used four speckle imagers located around the world to observe \TOIs M-dwarf TOIs. We also supplemented our observations with AO data obtained from the ExoFOP.

\subsection{Target Selection}

We used the TESS mission’s list of TOIs that is made public on the ExoFOP to create our sample. As of UT February 16 2022, there were 5243 TOIs in this list. Using the ExoFOP TOI table, we chose TOIs with stellar $\rm T_{\rm eff} < 4000$ K in order to select M-dwarf host stars \citep{Ciardi2011AJ....141..108C}. We chose TOIs below this temperature threshold without regard for previously known multiplicity. The size of the TOI sample was primarily set by the accessibility to targets from various sites, the availability of telescope time, and the limitations of weather; while not all TOIs in the temperature range of interest were observed, the set of M-dwarf TOIs presented here provide a reasonably-sized sample to test the frequency of stellar companions to planet-hosting M dwarfs.

\begin{figure*}
    \centering
    \includegraphics[width=\textwidth]{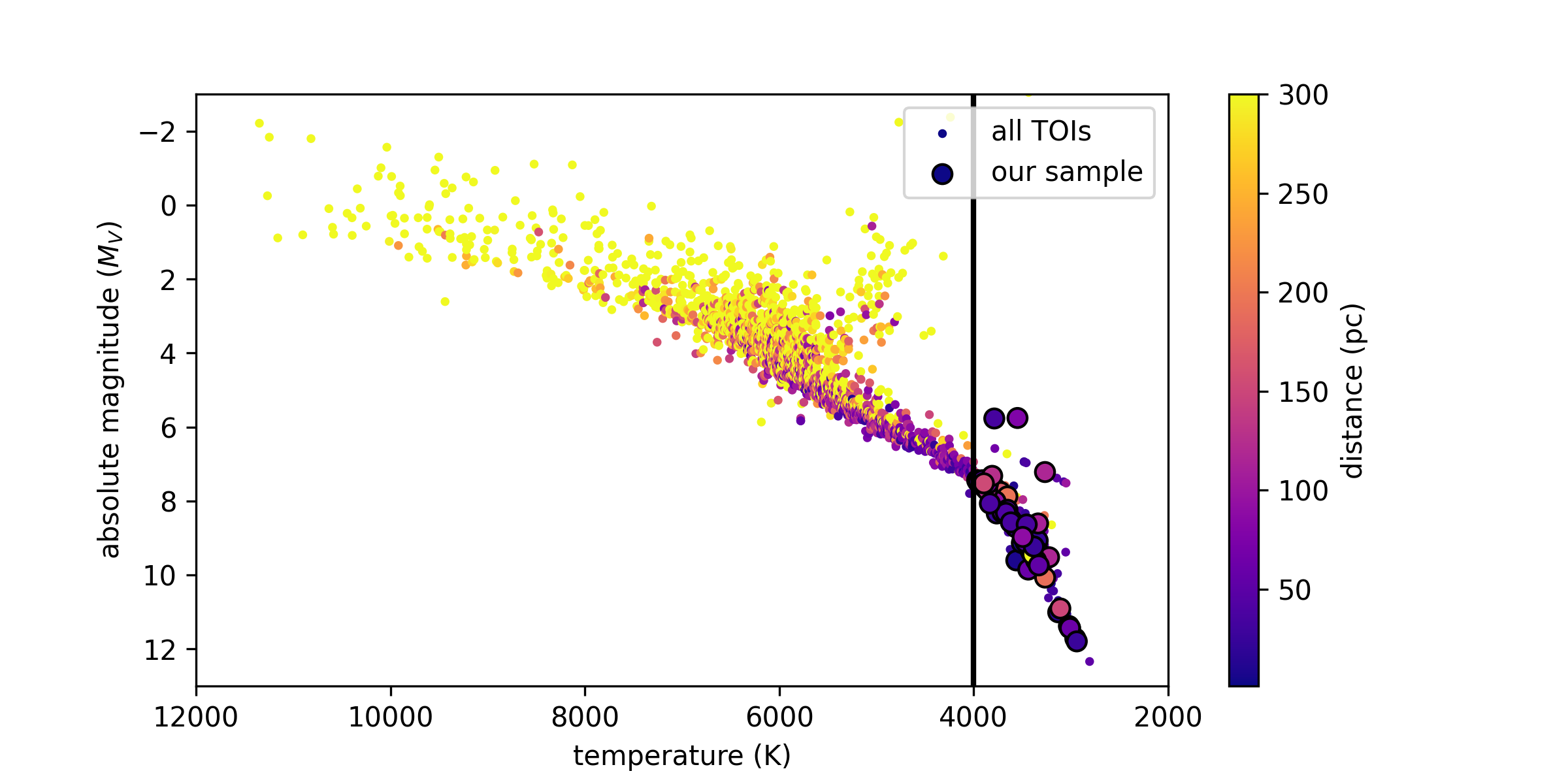}
    \caption{H-R diagram comparing the \TOIs M-dwarf TOIs in our sample (the larger, outlined circles) to the full sample of TOIs listed in the ExoFOP TOI table. This plot uses stellar temperature and distance values obtained from the ExoFOP TOI table, and we calculated absolute magnitude values using the stellar temperature and radius values listed in this table. The color of each point corresponds to the distance. The black line represents the temperature cutoff for our sample ($\rm T_{\rm eff} < 4000$ K). Discussion of the three M-dwarf TOIs that are elevated above the Main Sequence occurs in Section \ref{subsec:close-in_companions}.}
    \label{fig:h-r_diagram}
\end{figure*}

\begin{figure*}
    \centering
    \includegraphics[width=0.49\textwidth]{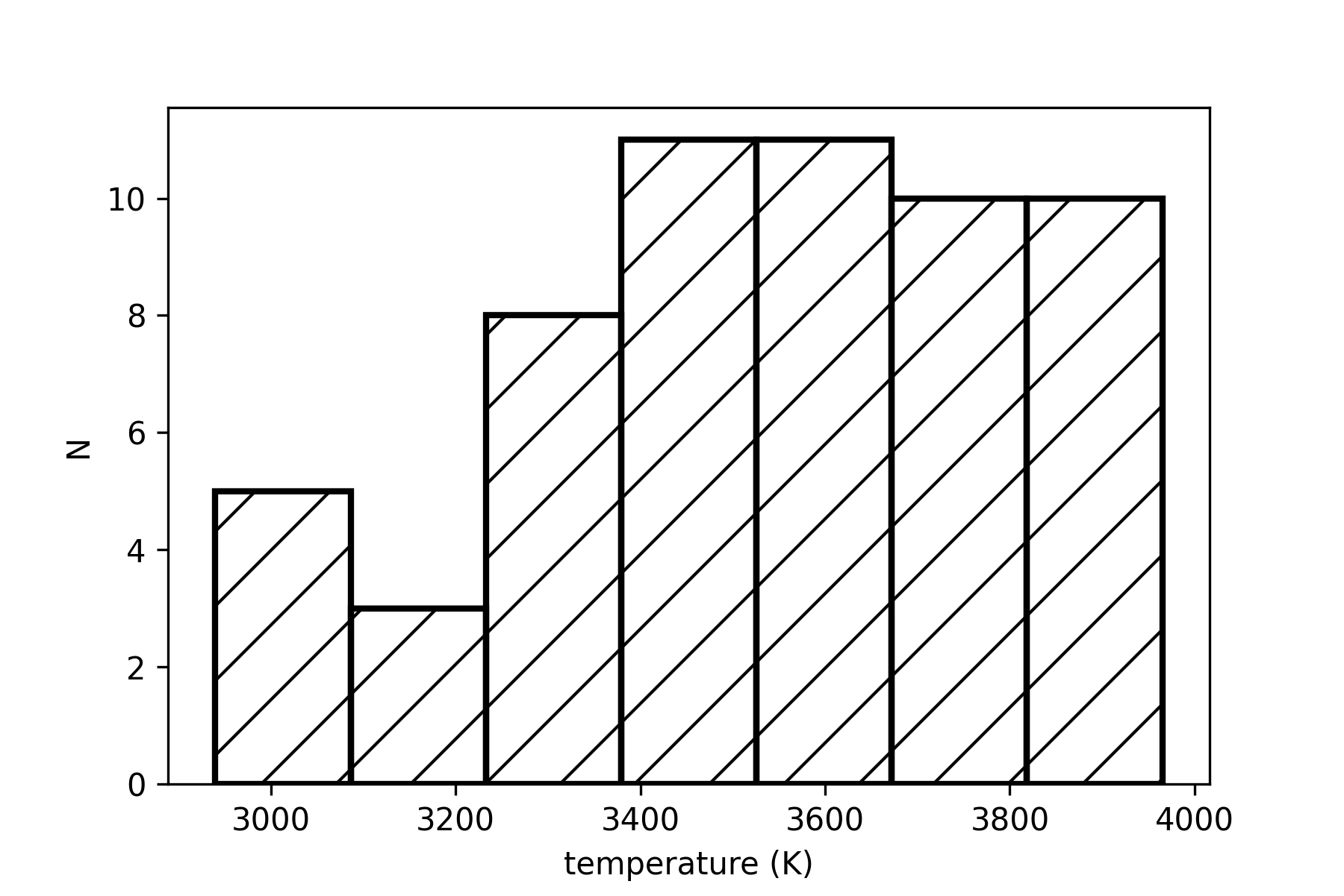}
    \includegraphics[width=0.49\textwidth]{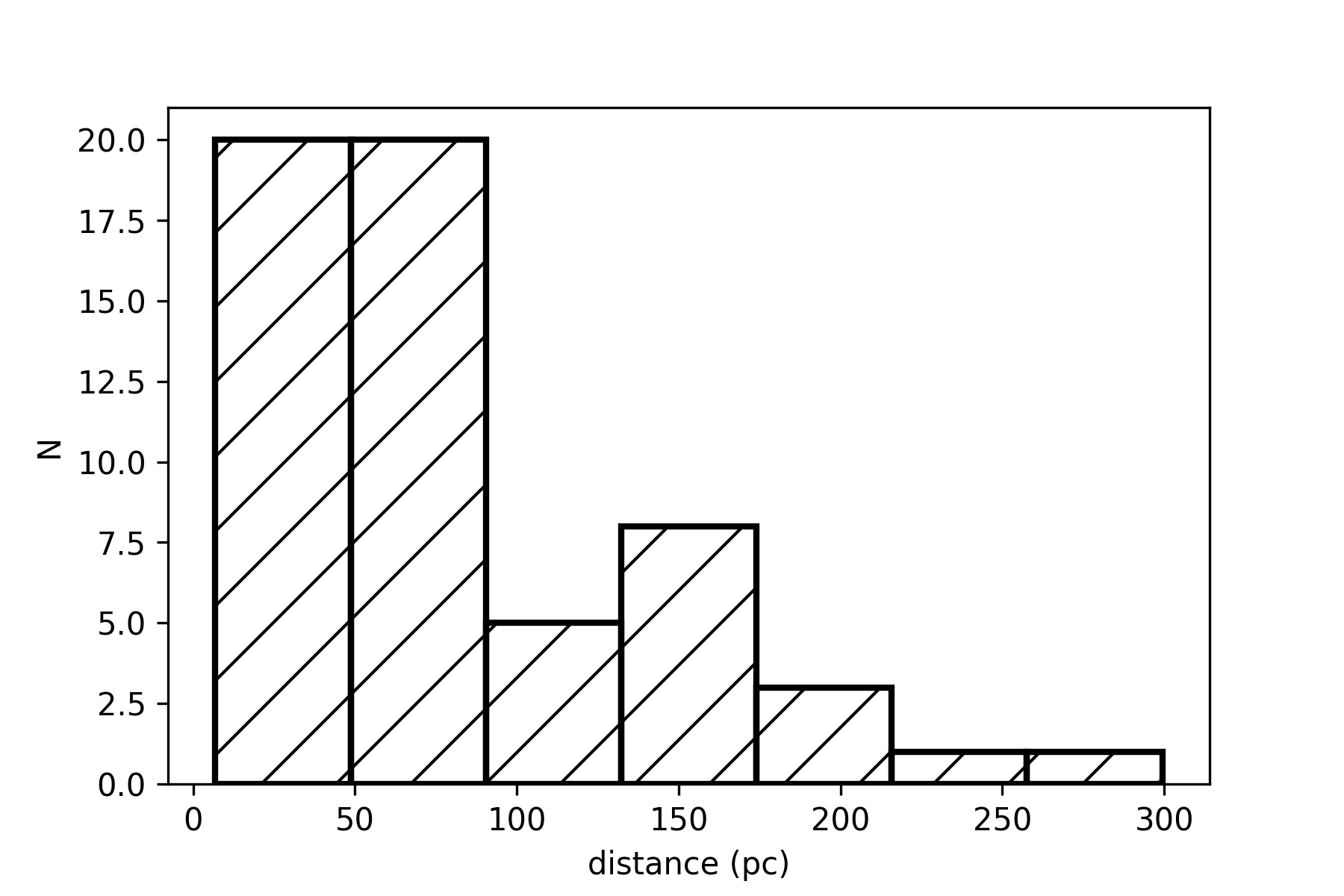}
    \caption{The temperature (left) and distance (right) distributions for the \TOIs M-dwarf TOIs in our sample. For a magnitude-limited sample, hotter stars are intrinsically brighter and therefore more physically distant. Since the stars in our sample are cool and faint, our spatial resolution into, and achieved inner working angle for, these systems is significantly better than it would be for solar-type stars.}
    \label{fig:distributions_sample}
\end{figure*}

An H-R diagram comparing these \TOIs TOIs to the the full sample of TOIs is shown in Figure \ref{fig:h-r_diagram}. The temperature and distance distributions for the M-dwarf TOIs in our sample are shown in Figure \ref{fig:distributions_sample}. In Table \ref{table:TOIs}, we list relevant stellar parameters for the targets in our sample, including the stellar effective temperature, TESS magnitude, and distance as obtained from the ExoFOP. The ExoFOP uses distances primarily from Gaia DR2 \citep{Gaia2018A&A...616A...1G} through the TESS Input Catalog Version 8 \citep{Stassun2019AJ....158..138S}.

\clearpage


\begin{ThreePartTable}
\begin{TableNotes}
\footnotesize
\item [*] A distance for TOI 734 was not listed in the ExoFOP, so one was calculated using its parallax from Gaia EDR3.
\end{TableNotes}
\begin{longtable}{c c c c} 
 \caption{M-dwarf TESS Objects of Interest in our sample} \\
 \hline
 Target & TESS Mag & $\rm T_{\rm eff}$ & Distance \\
 & & (K) & (pc) \\
 \hline
 \endfirsthead
 \multicolumn{4}{c}%
 {\tablename\ \thetable\ -- \textit{Continued from previous column}} \\
 \hline
 Target & TESS Mag & $\rm T_{\rm eff}$ & Distance \\
 & & (K) & (pc) \\
 \hline
 \endhead
 \hline
 \multicolumn{4}{c}{\textit{Continued on next column}} \\
 \endfoot
 \insertTableNotes
 \endlastfoot
 TOI 198 & 9.9 & 3763 & 23.7 \\
 TOI 244 & 10.3 & 3342 & 22.0 \\
 TOI 256 & 11.3 & 3131 & 15.0 \\
 TOI 277 & 11.7 & 3748 & 64.7 \\
 TOI 278 & 13.2 & 2955 & 44.4 \\
 TOI 436 & 13.1 & 3550 & 77.5 \\
 TOI 442 & 10.7 & 3779 & 52.6 \\
 TOI 455 & 8.8 & 3562 & 6.9 \\
 TOI 468 & 13.3 & 3724 & 169.3 \\
 TOI 482 & 13.1 & 3692 & 173.8 \\
 TOI 488 & 11.2 & 3332 & 27.4 \\
 TOI 497 & 13.2 & 3333 & 56.3 \\
 TOI 507 & 13.2 & 3338 & 110.9 \\
 TOI 513 & 14.7 & 3266 & 190.2 \\
 TOI 516 & 14.1 & 3109 & 150.5 \\
 TOI 519 & 14.4 & 3225 & 115.6 \\
 TOI 521 & 12.4 & 3439 & 60.9 \\
 TOI 526 & 12.3 & 3617 & 70.9 \\
 TOI 528 & 13.4 & 3965 & 192.4 \\
 TOI 529 & 14.1 & 3021 & 63.1 \\
 TOI 530 & 13.5 & 3688 & 148.8 \\
 TOI 531 & 13.8 & 3462 & 167.3 \\
 TOI 532 & 12.7 & 3946 & 135.0 \\
 TOI 538 & 14.1 & 3411 & 133.2 \\
 TOI 549 & 14.2 & 3009 & 80.7 \\
 TOI 552 & 13.8 & 3654 & 195.0 \\
 TOI 557 & 11.6 & 3883 & 76.0 \\
 TOI 562 & 8.7 & 3505 & 9.4 \\
 TOI 573 & 12.5 & 3404 & 88.4 \\
 TOI 620 & 10.2 & 3633 & 33.0 \\
 TOI 633 & 13.9 & 3267 & 116.5 \\
 TOI 643 & 13.7 & 3359 & 87.8 \\
 TOI 654 & 12.2 & 3433 & 57.8 \\
 TOI 674 & 11.9 & 3467 & 46.1 \\
 TOI 716 & 12.4 & 3776 & 88.8 \\
 TOI 717 & 11.4 & 3788 & 34.7 \\
 TOI 727 & 11.0 & 3653 & 43.0 \\
 TOI 734\tnote{*} & 15.2 & 3010 & 62.7 \\
 TOI 736 & 13.6 & 2940 & 26.5 \\
 TOI 737 & 14.8 & 3394 & 299.4 \\
 TOI 749 & 13.9 & 3602 & 232.1 \\
 TOI 756 & 12.6 & 3614 & 86.2 \\
 TOI 782 & 12.3 & 3331 & 52.5 \\
 TOI 797 & 11.7 & 3704 & 56.2 \\
 TOI 821 & 11.2 & 3669 & 39.6 \\
 TOI 876 & 11.5 & 3955 & 78.2 \\
 TOI 900 & 11.9 & 3861 & 121.0 \\
 TOI 912 & 10.5 & 3566 & 26.1 \\
 TOI 1201 & 10.9 & 3948 & 37.9 \\
 TOI 1235 & 9.9 & 3912 & 39.6 \\
 TOI 1238 & 11.3 & 3853 & 70.7 \\
 TOI 1266 & 11.0 & 3618 & 36.0 \\
 TOI 1467 & 10.6 & 3834 & 37.4 \\
 TOI 1468 & 10.9 & 3382 & 24.7 \\
 TOI 1634 & 11.0 & 3455 & 35.3 \\
 TOI 1635 & 13.4 & 3496 & 89.7 \\
 TOI 1638 & 12.1 & 3810 & 126.3 \\
 TOI 1639 & 13.0 & 3896 & 154.0 \\ [1ex]
 \hline
 \label{table:TOIs}
\end{longtable}
\end{ThreePartTable}

\subsection{Observational Routine}

In total, we observed \TOIs TOIs between UT October 11 2019 and UT December 6 2020, and obtained AO data from the ExoFOP observed between UT November 22 2018 and UT August 8 2021.

We employed the NN-EXPLORE Exoplanet and Stellar Speckle Imager \citep[NESSI;][]{Scott2018PASP..130e4502S} at the 3.5-m WIYN telescope, the Differential Speckle Survey Instrument \citep[DSSI;][]{Horch2009AJ....137.5057H} at the 4.3-m Lowell Discovery Telescope (LDT), and the ‘Alopeke \citep{ScottHowell2018SPIE10701E..0GS, Scott2021FrASS...8..138S} and Zorro \citep{Scott2019ESS.....433015S, Scott2021FrASS...8..138S} speckle imagers at the 8.1-m Gemini North and Gemini South telescopes to carry out our speckle observations of M-dwarf TOIs.

NESSI, DSSI, ‘Alopeke, and Zorro are similar dual-channel imagers that use a dichroic filter to split the collimated beam at $\sim$700 nm into two channels, which are then imaged on separate high-speed readout EMCCDs. 
On a standard observing night, the telescope was kept close to the meridian to minimize residual atmospheric dispersion. Objects in the target list were ordered in right ascension and split into small groups, and an unresolved, bright, single star from the Bright Star Catalog \citep{HoffleitJaschek1982bsc..book.....H} at a similar sky position was chosen as a point source calibration object for each group. Depending on target brightness, one to nine data cubes were obtained for each star at two wavelengths simultaneously. These data cubes consist of 1000 exposures. Each exposure is 40 ms in duration at the LDT and at WIYN, and 60 ms in duration at the Gemini telescopes. These short exposures are necessary in order to ``freeze'' out the atmosphere from our observations. With these short exposure lengths, the observing cadence was roughly 2-10 minutes per target. The data cubes were then stored as multi-extension FITS files.

Additionally, 14 of our M-dwarf targets were observed with near-infrared (NIR) adaptive optics (AO) imaging at Palomar Observatory and/or Keck Observatory. While the optical speckle data provide higher resolution, the NIR AO imaging provides complementary observations with greater sensitivity to redder objects. The AO data were acquired using the Palomar 200-inch Hale Telescope with the Palomar High Angular Resolution Observer \citep[PHARO;][]{Hayward2001PASP..113..105H} behind the natural guide star AO system P3K \citep{Dekany2013}, and/or the the 10-m Keck II Telescope with the NIRC2 instrument behind the natural guide star AO system \citep{Wizinowich2000}.

The PHARO observations were taken in a standard 5-point quincunx dither pattern with steps of 4\arcsec\ in the narrow-band $Br-\gamma$ filter $(\lambda_o = 2.1686; \Delta\lambda = 0.0326~\mu$m). Each dither position was observed three times, offset in position from each other by $0.5\arcsec$, for a total of 15 frames. The integration times varied with the brightness of the target but were typically $1-10$ sec per frame for a total on-source time of 1-150 seconds. PHARO has a pixel scale of 0.025\arcsec\ per pixel for a total field of view of $\sim25\arcsec$. 

The NIRC2 data were acquired in a standard 3-point dither pattern that is used with NIRC2 to avoid the left lower quadrant of the detector, which is typically noisier than the other three quadrants. The dither pattern step size was 3\arcsec\ and was repeated twice, with each dither offset from the previous dither by $0.5\arcsec$. NIRC2 was used in the narrow-angle mode that has a pixel scale of approximately 0.0099442\arcsec\ per pixel for a total field of view of $\sim10\arcsec$. The Keck observations were made in both the narrow-band $Br-\gamma$ filter $(\lambda_o = 2.1686; \Delta\lambda = 0.0326~\mu$m) and the narrow-band $J-cont$ filter $(\lambda_o = 1.2132; \Delta\lambda = 0.0198~\mu$m) with typical integration times in each filter of 1-2 second per frame for a total of 9-18 seconds on target.

\begin{figure*}
    \centering
    \includegraphics{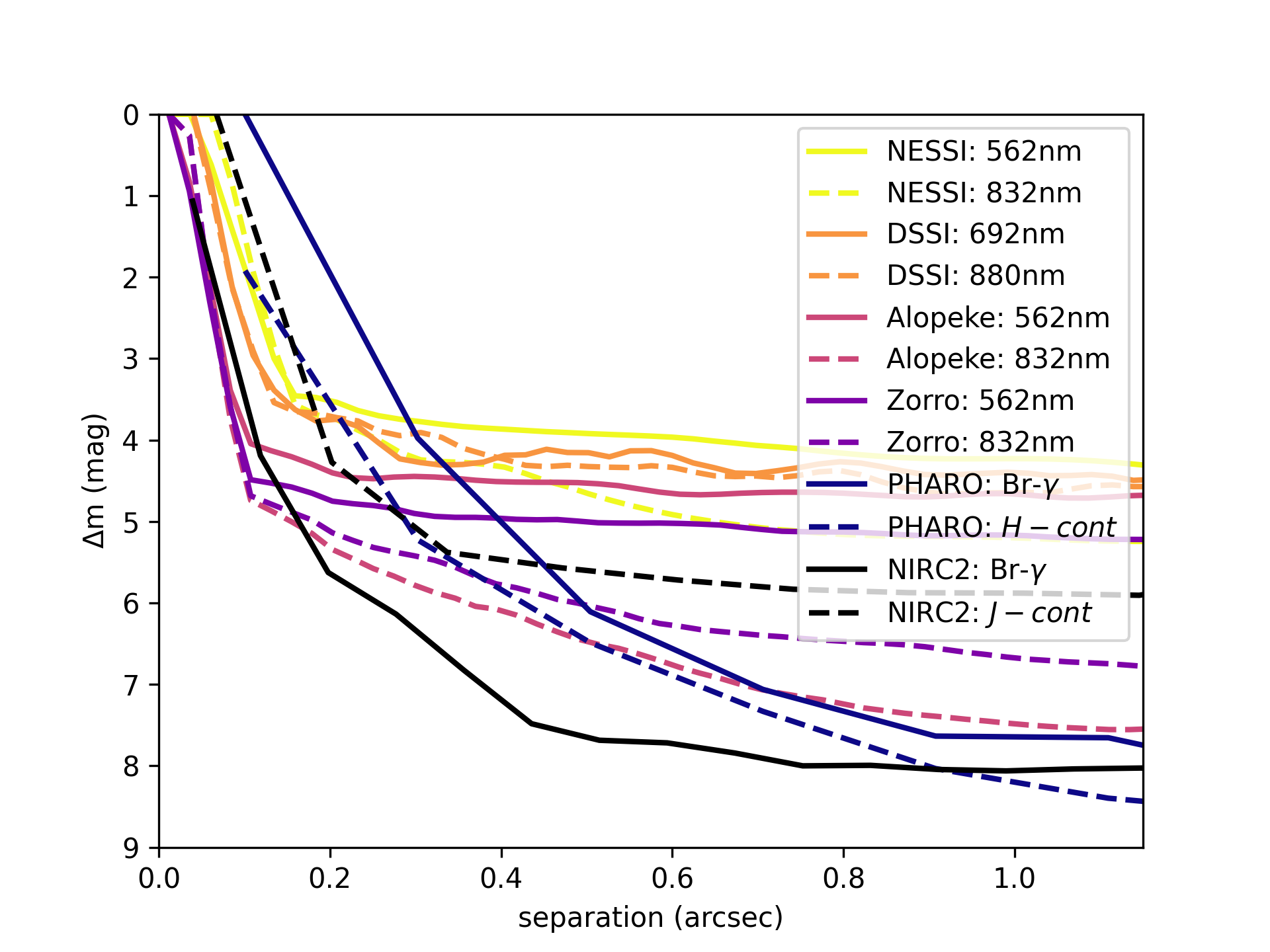}
    \caption{The median contrast curve for each filter for each of the six speckle and AO instruments used in this survey. It should be noted that the NIRC2 contrast curves extend to roughly $4\arcsec$, and that the PHARO contrast curves extend to roughly $10\arcsec$. The larger apertures of Gemini North, Gemini South, PHARO, and NIRC2 allow us to probe larger delta magnitudes as compared to NESSI and DSSI, as shown by their median contrast curves.}
    \label{fig:median_contrast_curves}
\end{figure*}

The date, telescope, image type, bandpass, and $5\sigma$ $\Delta$m contrast limits obtained in the observation at 0.2\arcsec\ and 1.0\arcsec\ are listed in Table \ref{table:observations}. The median contrast curve for each filter and instrument used in this survey is shown in Figure \ref{fig:median_contrast_curves}.

\onecolumngrid
\begin{longtable}{c c c c c c c} 
 \caption{Summary of observations}\\
 \hline
 Target & UT Date & Telescope & Image & Bandpass & $\Delta$m & $\Delta$m \\
 & YYYY-MM-DD & & Type & & 0.2\arcsec\ & 1.0\arcsec\ \\
 \hline
 \endfirsthead
 \multicolumn{7}{c}
 {\tablename\ \thetable\ -- \textit{Continued from previous page}} \\
 \hline
 Target & UT Date & Telescope & Image & Bandpass & $\Delta$m & $\Delta$m \\
 & YYYY-MM-DD & & Type & & 0.2\arcsec\ & 1.0\arcsec\ \\
 \hline
 \endhead
 \hline \multicolumn{7}{c}{\textit{Continued on next page}} \\
 \endfoot
 \endlastfoot
TOI 198 & 2018-11-22 & Keck & AO & $J-cont$ & 4.46 & 8.32 \\
 & 2018-11-22 & Keck & AO & Br-$\gamma$ & 5.49 & 8.33 \\
 & 2020-08-04 & Gemini North & Speckle & 562 nm & 4.06 & 4.40 \\
 & 2020-08-04 & Gemini North & Speckle & 832 nm & 5.31 & 7.60 \\
TOI 244 & 2020-08-04 & Gemini North & Speckle & 562 nm & 3.90 & 4.14 \\
 & 2020-08-04 & Gemini North & Speckle & 832 nm & 5.35 & 6.61 \\
TOI 256 & 2018-12-22 & Hale & AO & Br-$\gamma$ & 2.64 & 7.54 \\
 & 2020-08-10 & Gemini North & Speckle & 562 nm & 4.16 & 4.56 \\
 & 2020-08-10 & Gemini North & Speckle & 832 nm & 5.08 & 8.10 \\
TOI 277 & 2020-08-10 & Gemini North & Speckle & 562 nm & 4.11 & 4.54 \\
 & 2020-08-10 & Gemini North & Speckle & 832 nm & 5.47 & 7.48 \\
TOI 278 & 2018-12-24 & Hale & AO & Br-$\gamma$ & 3.00 & 6.59 \\
 & 2019-10-13 & WIYN & Speckle & 562 nm & 3.46 & 4.23 \\
 & 2019-10-13 & WIYN & Speckle & 832 nm & 3.47 & 4.77 \\
TOI 436 & 2020-01-14 & Gemini South & Speckle & 562 nm & 4.61 & 5.16 \\
 & 2020-01-14 & Gemini South & Speckle & 832 nm & 5.12 & 6.45 \\
TOI 442 & 2020-01-14 & Gemini South & Speckle & 562 nm & 4.44 & 4.89 \\
 & 2020-01-14 & Gemini South & Speckle & 832 nm & 4.82 & 7.23 \\
TOI 455 & 2020-12-06 & Gemini North & Speckle & 562 nm & 4.38 & 4.66 \\
 & 2020-12-06 & Gemini North & Speckle & 832 nm & 5.57 & 7.94 \\
TOI 468 & 2020-02-10 & LDT & Speckle & 692 nm & 3.86 & 4.12 \\
 & 2020-02-10 & LDT & Speckle & 880 nm & 4.16 & 4.74 \\
TOI 482 & 2019-10-11 & WIYN & Speckle & 562 nm & 3.34 & 4.11 \\
 & 2019-10-11 & WIYN & Speckle & 832 nm & 3.59 & 4.61 \\
TOI 488 & 2019-04-18 & Hale & AO & Br-$\gamma$ & 3.08 & 6.21 \\
 & 2020-02-09 & LDT & Speckle & 692 nm & 4.29 & 4.73 \\
 & 2020-02-09 & LDT & Speckle & 880 nm & 3.48 & 4.73 \\
TOI 497 & 2020-02-09 & LDT & Speckle & 692 nm & 3.05 & 3.96 \\
 & 2020-02-09 & LDT & Speckle & 880 nm & 3.77 & 3.75 \\
TOI 507 & 2020-02-09 & LDT & Speckle & 692 nm & 4.00 & 4.07 \\
 & 2020-02-09 & LDT & Speckle & 880 nm & 4.36 & 4.69 \\
TOI 513 & 2020-02-09 & LDT & Speckle & 692 nm & 3.06 & 3.62 \\
 & 2020-02-09 & LDT & Speckle & 880 nm & 2.90 & 3.90 \\
TOI 516 & 2020-02-09 & LDT & Speckle & 692 nm & 3.50 & 4.40 \\
 & 2020-02-09 & LDT & Speckle & 880 nm & 3.63 & 4.67 \\
TOI 519 & 2020-02-10 & LDT & Speckle & 692 nm & 2.23 & 3.41 \\
 & 2020-02-10 & LDT & Speckle & 880 nm & 1.42 & 3.63 \\
TOI 521 & 2020-02-10 & LDT & Speckle & 692 nm & 3.74 & 4.65 \\
 & 2020-02-10 & LDT & Speckle & 880 nm & 3.83 & 4.68 \\
 & 2020-12-05 & Hale & AO & Br-$\gamma$ & 3.46 & 7.78 \\
TOI 526 & 2019-10-10 & WIYN & Speckle & 562 nm & 4.16 & 4.79 \\
 & 2019-10-10 & WIYN & Speckle & 832 nm & 3.85 & 5.37 \\
 & 2021-12-05 & Hale & AO & Br-$\gamma$ & 2.70 & 7.66 \\
TOI 528 & 2020-02-09 & LDT & Speckle & 692 nm & 4.40 & 4.73 \\
 & 2020-02-09 & LDT & Speckle & 880 nm & 4.40 & 4.72 \\
TOI 529 & 2020-02-10 & LDT & Speckle & 692 nm & 4.44 & 4.76 \\
 & 2020-02-10 & LDT & Speckle & 880 nm & 4.34 & 4.72 \\
TOI 530 & 2019-10-13 & WIYN & Speckle & 562 nm & 3.71 & 5.20 \\
 & 2019-10-13 & WIYN & Speckle & 832 nm & 3.38 & 4.54 \\
TOI 531 & 2020-02-10 & LDT & Speckle & 692 nm & 4.56 & 4.69 \\
 & 2020-02-10 & LDT & Speckle & 880 nm & 4.79 & 4.79 \\
TOI 532 & 2019-10-10 & WIYN & Speckle & R & 4.07 & 4.86 \\
 & 2019-10-10 & WIYN & Speckle & z' & 4.23 & 5.93 \\
 & 2020-12-05 & Hale & AO & Br-$\gamma$ & 2.64 & 7.41 \\
TOI 538 & 2019-10-11 & WIYN & Speckle & 562 nm & 3.58 & 3.91 \\
 & 2019-10-11 & WIYN & Speckle & 832 nm & 2.72 & 3.53 \\
TOI 549 & 2020-02-10 & LDT & Speckle & 692 nm & 2.68 & 3.18 \\
 & 2020-02-10 & LDT & Speckle & 880 nm & 2.98 & 3.59 \\
TOI 552 & 2020-02-10 & LDT & Speckle & 692 nm & 1.82 & 3.08 \\
 & 2020-02-10 & LDT & Speckle & 880 nm & 2.64 & 4.02 \\
TOI 557 & 2019-10-12 & WIYN & Speckle & 562 nm & 3.51 & 4.48 \\
 & 2019-10-12 & WIYN & Speckle & 832 nm & 3.69 & 5.49 \\
 & 2019-10-14 & WIYN & Speckle & 562 nm & 3.42 & 3.64 \\
 & 2019-10-14 & WIYN & Speckle & 832 nm & 3.92 & 4.92 \\
TOI 562 & 2020-03-15 & Gemini South & Speckle & 562 nm & 5.70 & 6.97 \\
 & 2020-03-15 & Gemini South & Speckle & 832 nm & 5.05 & 7.65 \\
TOI 573 & 2020-03-15 & Gemini South & Speckle & 562 nm & 4.65 & 5.18 \\
 & 2020-03-15 & Gemini South & Speckle & 832 nm & 4.82 & 6.34 \\
TOI 620 & 2020-03-16 & Gemini South & Speckle & 562 nm & 4.83 & 5.79 \\
 & 2020-03-16 & Gemini South & Speckle & 832 nm & 4.68 & 8.01 \\
TOI 633 & 2020-03-16 & Gemini South & Speckle & 562 nm & 4.81 & 4.99 \\
 & 2020-03-16 & Gemini South & Speckle & 832 nm & 4.62 & 5.63 \\
 & 2021-02-24 & Hale & AO & Br-$\gamma$ & 2.66 & 6.73 \\
TOI 643 & 2020-02-09 & LDT & Speckle & 692 nm & 3.06 & 4.08 \\
 & 2020-02-09 & LDT & Speckle & 880 nm & 2.90 & 3.56 \\
TOI 654 & 2019-06-10 & Keck & AO & $J-cont$ & 5.06 & 5.89 \\
 & 2019-06-10 & Keck & AO & Br-$\gamma$ & 6.09 & 8.06 \\
 & 2019-06-13 & Hale & AO & $H-cont$ & 2.36 & 7.98 \\
 & 2019-06-13 & Hale & AO & Br-$\gamma$ & 2.21 & 8.14 \\
 & 2020-02-09 & LDT & Speckle & 692 nm & 3.99 & 4.50 \\
 & 2020-02-09 & LDT & Speckle & 880 nm & 3.33 & 4.50 \\
TOI 674 & 2020-01-14 & Gemini South & Speckle & 562 nm & 4.65 & 5.62 \\
 & 2020-01-14 & Gemini South & Speckle & 832 nm & 5.14 & 6.99 \\
TOI 716 & 2020-03-13 & Gemini South & Speckle & 562 nm & 4.83 & 5.40 \\
 & 2020-03-13 & Gemini South & Speckle & 832 nm & 5.15 & 6.27 \\
TOI 717 & 2020-03-16 & Gemini South & Speckle & 562 nm & 4.04 & 4.42 \\
 & 2020-03-16 & Gemini South & Speckle & 832 nm & 5.54 & 7.14 \\
TOI 727 & 2019-11-10 & Hale & AO & $H-cont$ & 3.58 & 8.44 \\
 & 2019-11-10 & Hale & AO & Br-$\gamma$ & 3.49 & 8.47 \\
 & 2019-11-18 & WIYN & Speckle & 562 nm & 3.68 & 4.07 \\
 & 2019-11-18 & WIYN & Speckle & 832 nm & 4.14 & 5.80 \\
TOI 734 & 2020-03-12 & Gemini South & Speckle & 562 nm & 4.91 & 5.02 \\
 & 2020-03-12 & Gemini South & Speckle & 832 nm & 4.45 & 4.76 \\
TOI 736 & 2020-02-09 & LDT & Speckle & 692 nm & 3.12 & 3.52 \\
 & 2020-02-09 & LDT & Speckle & 880 nm & 3.77 & 4.57 \\
TOI 737 & 2019-06-10 & Keck & AO & $J-cont$ & 4.27 & 5.40 \\
 & 2019-06-10 & Keck & AO & Br-$\gamma$ & 5.45 & 6.98 \\
 & 2020-02-09 & LDT & Speckle & 692 nm & 4.29 & 4.67 \\
 & 2020-02-09 & LDT & Speckle & 880 nm & 2.50 & 5.25 \\
TOI 749 & 2020-02-09 & LDT & Speckle & 692 nm & 3.88 & 4.26 \\
 & 2020-02-09 & LDT & Speckle & 880 nm & 4.50 & 4.64 \\
TOI 756 & 2020-03-12 & Gemini South & Speckle & 562 nm & 4.86 & 5.19 \\
 & 2020-03-12 & Gemini South & Speckle & 832 nm & 5.13 & 6.01 \\
TOI 782 & 2020-02-09 & LDT & Speckle & 692 nm & 3.64 & 4.56 \\
 & 2020-02-09 & LDT & Speckle & 880 nm & 4.15 & 4.65 \\
TOI 797 & 2020-03-12 & Gemini South & Speckle & 562 nm & 4.49 & 4.93 \\
 & 2020-03-12 & Gemini South & Speckle & 832 nm & 5.21 & 7.21 \\
TOI 821 & 2020-02-09 & LDT & Speckle & 692 nm & 2.86 & 3.15 \\
 & 2020-02-09 & LDT & Speckle & 880 nm & 3.19 & 3.77 \\
TOI 876 & 2020-02-10 & LDT & Speckle & 692 nm & 2.97 & 3.17 \\
 & 2020-02-10 & LDT & Speckle & 880 nm & 2.37 & 3.05 \\
TOI 900 & 2020-03-13 & Gemini South & Speckle & 562 nm & 4.88 & 5.82 \\
 & 2020-03-13 & Gemini South & Speckle & 832 nm & 4.92 & 6.37 \\
TOI 912 & 2020-03-14 & Gemini South & Speckle & 562 nm & 4.04 & 4.36 \\
 & 2020-03-14 & Gemini South & Speckle & 832 nm & 5.47 & 6.91 \\
TOI 1201 & 2019-11-09 & Hale & AO & $H-cont$ & 3.95 & 7.87 \\
 & 2019-11-09 & Hale & AO & Br-$\gamma$ & 3.58 & 7.90 \\
 & 2019-11-10 & WIYN & Speckle & 562 nm & 3.81 & 4.40 \\
 & 2019-11-10 & WIYN & Speckle & 832 nm & 3.86 & 5.91 \\
TOI 1235 & 2020-02-09 & LDT & Speckle & 692 nm & 3.64 & 4.84 \\
 & 2020-02-09 & LDT & Speckle & 880 nm & 2.40 & 5.25 \\
TOI 1238 & 2020-02-09 & LDT & Speckle & 692 nm & 3.96 & 4.62 \\
 & 2020-02-09 & LDT & Speckle & 880 nm & NaN & NaN \\
TOI 1266 & 2020-02-09 & LDT & Speckle & 692 nm & 3.31 & 4.53 \\
 & 2020-02-09 & LDT & Speckle & 880 nm & 3.80 & 4.29 \\
TOI 1467 & 2020-08-09 & Gemini North & Speckle & 562 nm & 4.66 & 5.13 \\
 & 2020-08-09 & Gemini North & Speckle & 832 nm & 5.33 & 7.16 \\
 & 2020-12-04 & Gemini North & Speckle & 562 nm & 4.66 & 5.24 \\
 & 2020-12-04 & Gemini North & Speckle & 832 nm & 5.25 & 7.76 \\
TOI 1468 & 2020-08-04 & Gemini North & Speckle & 562 nm & 4.92 & 5.38 \\
 & 2020-08-04 & Gemini North & Speckle & 832 nm & 5.07 & 7.02 \\
 & 2021-08-08 & Hale & AO & Br-$\gamma$ & 3.63 & 8.76 \\
TOI 1634 & 2020-12-02 & Gemini North & Speckle & 562 nm & 4.75 & 5.66 \\
 & 2020-12-02 & Gemini North & Speckle & 832 nm & 5.38 & 7.91 \\
TOI 1635 & 2020-02-09 & LDT & Speckle & 692 nm & 4.32 & 4.71 \\
 & 2020-02-09 & LDT & Speckle & 880 nm & 4.49 & 4.79 \\
TOI 1638 & 2020-12-03 & Gemini North & Speckle & 562 nm & 4.61 & 4.77 \\
 & 2020-12-03 & Gemini North & Speckle & 832 nm & 4.75 & 6.41 \\
TOI 1639 & 2020-11-04 & Hale & AO & Br-$\gamma$ & 3.78 & 7.46 \\
 & 2020-12-02 & Gemini North & Speckle & 562 nm & 4.21 & 4.49 \\
 & 2020-12-02 & Gemini North & Speckle & 832 nm & 5.04 & 6.67 \\
 \hline
 \label{table:observations}
\end{longtable}
\twocolumngrid

\subsection{Data Reduction}

Through the comparison of speckle patterns observed simultaneously at two wavelengths, speckle cameras that use EMCCDs can even resolve stellar companions below the diffraction limit \citep{Horch2011bAJ....141..180H}. We leveraged this concept to reduce our speckle observations, and to produce reconstructed images in which the resolved stellar companions are revealed. The reconstructed images were then used to produce contrast curves.

The raw FITS files from our observations were passed through our standard speckle analysis pipeline \citep{Horch2009AJ....137.5057H, Horch2011bAJ....141..180H, Howell2011AJ....142...19H}, which makes use of bispectral analysis \citep{Lohmann1983ApOpt..22.4028L} to compute a reconstructed image. Resolved binary systems produce a fringe pattern in the Fourier plane. The fringe pattern is a product of the data reduction. A 2-D autocorrelation function is calculated for each speckle frame, and then these are summed over all frames. Next, the Fourier transform of the autocorrelation function is found and squared, yielding the power spectrum. The power spectrum of the science target is normalized using a similarly-processed data cube from the point source calibration object. The residual 2-D power spectrum appears as a set of fringes for each pair of stars in the field. This is fit using a cosine-squared function (fringe model) in order to determine the relative astrometry and photometry of any pairs of stars in the field, and to generate a reconstructed image for each target.

We examined annuli in the reconstructed images centered on the primary star, and found all local maxima and minima in each annulus, in order to derive their mean values and standard deviations. We then estimated the detection limit as the mean value of the maxima, plus five times the average of the standard deviations of the maxima and the minima. In doing so, we calculated the detection limit as a function of angular separation for each observation, and generated a contrast curve for each target.

The AO data were processed and analyzed using a custom set of IDL tools. The science frames were flat-fielded and sky-subtracted. The flat fields were generated from a median average of dark subtracted flats taken on-sky, and were normalized such that the median value of the flats was unity. The sky frames were generated from the median average of the dithered science frames. The reduced science frames were combined into a single combined image using an intra-pixel interpolation that conserves flux, shifts the individual dithered frames by the appropriate fractional pixels, and median-coadds the frames. The final resolution of the combined dithers was determined from the full-width half-maximum (FWHM) of the point spread function; typically 0.1\arcsec\ and 0.050\arcsec\ for Palomar and Keck observations, respectively.

The sensitivities of the final combined AO images were determined by injecting simulated sources azimuthally around the primary target every $20^\circ$ at separations of integer multiples of the central source's FWHM \citep{furlan2017, LundCiardi2020AAS...23524906L}. The brightness of each injected source was scaled until standard aperture photometry detected it with $5\sigma$ significance. The resulting brightness of the injected sources relative to the primary target set the contrast limits at that injection location. The final $5\sigma$ limit at each separation was determined from the average of all of the determined limits at that separation, and the uncertainty on the limit was set by the rms dispersion of the azimuthal slices at a given radial distance. For the one detected AO companion, aperture photometry was used to determine the relative brightness of the detected companion.

All of our final data products are now available through the ExoFOP.

\section{Stellar Companions} \label{sec:stellarcompanions}

Though we observed \TOIs M-dwarf TOIs with speckle imaging, we detected only one likely bound stellar companion. We find an additional stellar companion in the AO data. We find three stellar companions that are known to the literature. We also find 12 CPM companions: nine with Gaia EDR3, two with SUPERWIDE, and one with \citet{El-Badry2021MNRAS.506.2269E}.

\subsection{Speckle Companion} \label{subsec:speckle_companion}

We observed TOI 482 on UT 2019 October 11 with NESSI at the 3.5-m WIYN telescope. TOI 482 is located at a distance of 173.8 pc, with a TESS magnitude of 13.1, and a $\rm T_{\rm eff}$ of 3692 K. 

\begin{figure*}
    \centering
    \includegraphics[width=0.49\textwidth]{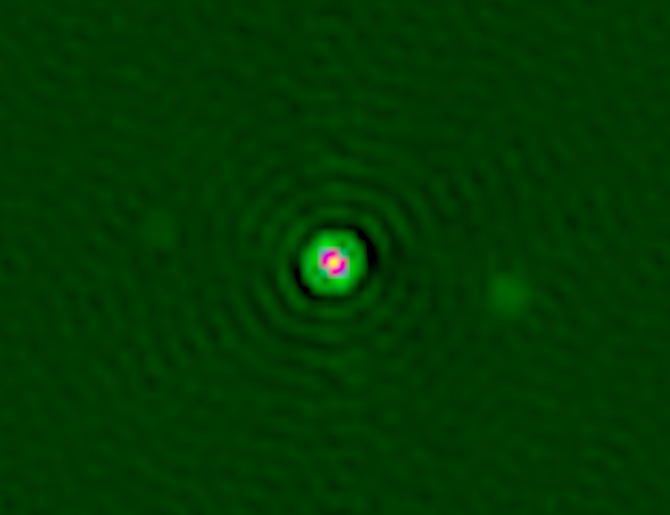}
    \includegraphics[width=0.49\textwidth]{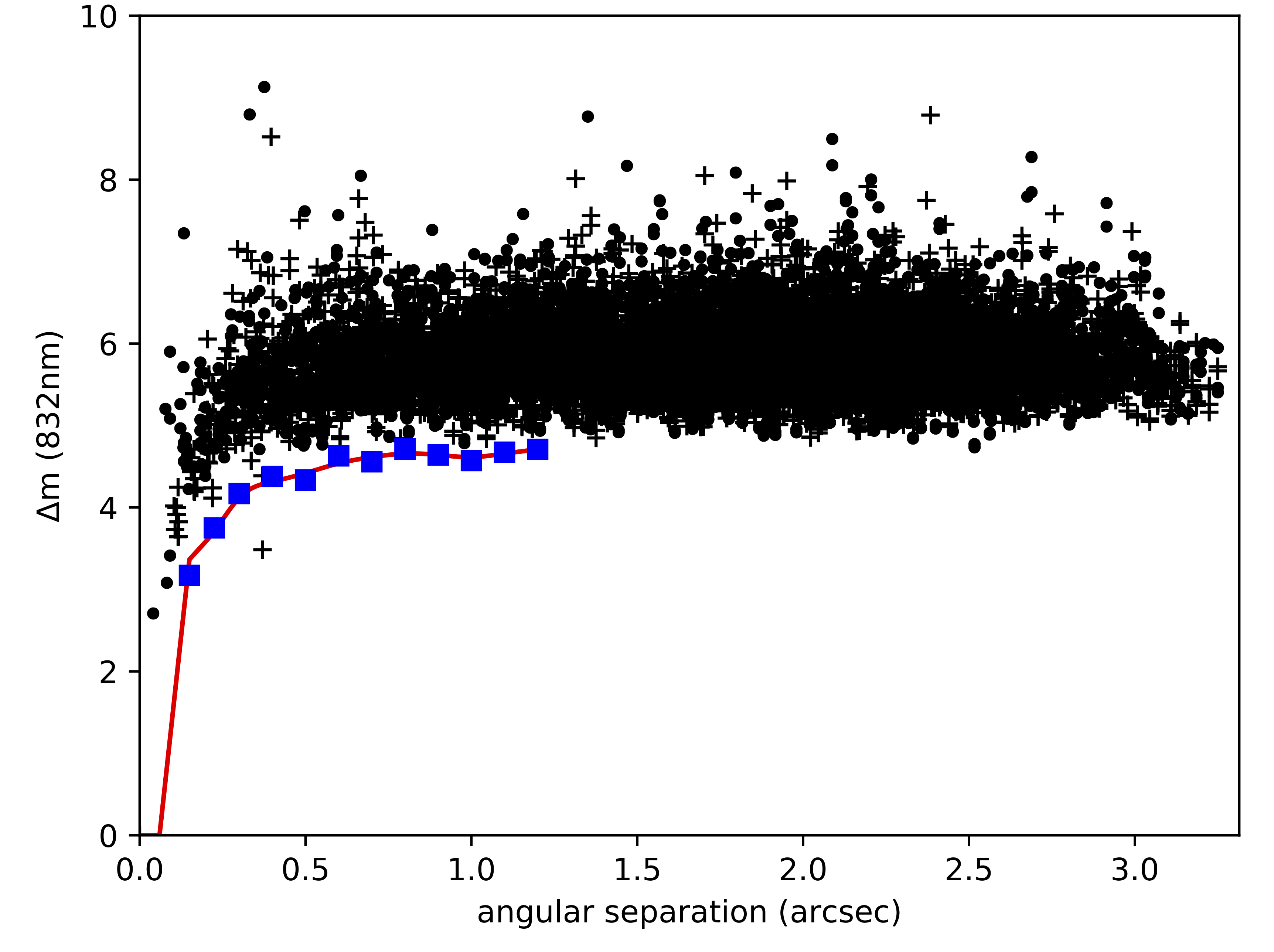}
    \caption{The only stellar companion revealed by our speckle observations, shown in the reconstructed image on the left, and marked with a cross beneath the contrast curve on the right. The stellar companion to TOI 482 was detected at 0.398\arcsec\ (57 au), at $267^{\circ}$, and with a delta magnitude of 2.43 at 832 nm.}
    \label{fig:TOI482}
\end{figure*}

The reconstructed image and the contrast curve for this object are shown in Figure \ref{fig:TOI482}. The stellar companion was detected at an angular separation of $0.398\arcsec\pm0.004$, a position angle of $267\pm1^{\circ}$, and a delta magnitude of $2.43\pm0.20$ at 832 nm.

We used this observational data to estimate additional astrophysical parameters for the stellar companion to TOI 482. We calculated a projected separation of $57\pm2$ au using the Gaia EDR3 parallax and parallax error for this object. As we only detected one stellar companion with our speckle observations, we assigned an angular separation uncertainty based on a similar analysis of a large data set; we assumed comparable astrometric precision to Figure 2 of \citet{Colton2021AJ....161...21C}.

To calculate the mass ratio for this system, we used the mass-luminosity relationship that was calibrated for late-type stars by \citet{Mann2019ApJ...871...63M}. We estimate the mass of the primary to be $0.16\pm0.01$ $M_{\odot}$ and the mass of the secondary to be $0.09\pm0.01$ $M_{\odot}$ using their code that is publicly available on $\tt github$\footnote{\url{https://github.com/awmann/M_-M_K-}}. These values give a mass ratio of $0.56\pm0.04$.

We calculated an orbital period of roughly 859 years by assuming that the instantaneous spatial separation detected in our imaging is approximately the orbital semi-major axis.

\subsection{AO Companion}

TOI 737 was observed on UT 2019 June 10 with NIRC2 at the 10 m Keck II telescope. TOI 737 is located at a distance of 299.4 pc, with a TESS magnitude of 14.8, and a $\rm T_{\rm eff}$ of 3394 K. 

\begin{figure*}
    \centering
    \includegraphics[width=0.49\textwidth]{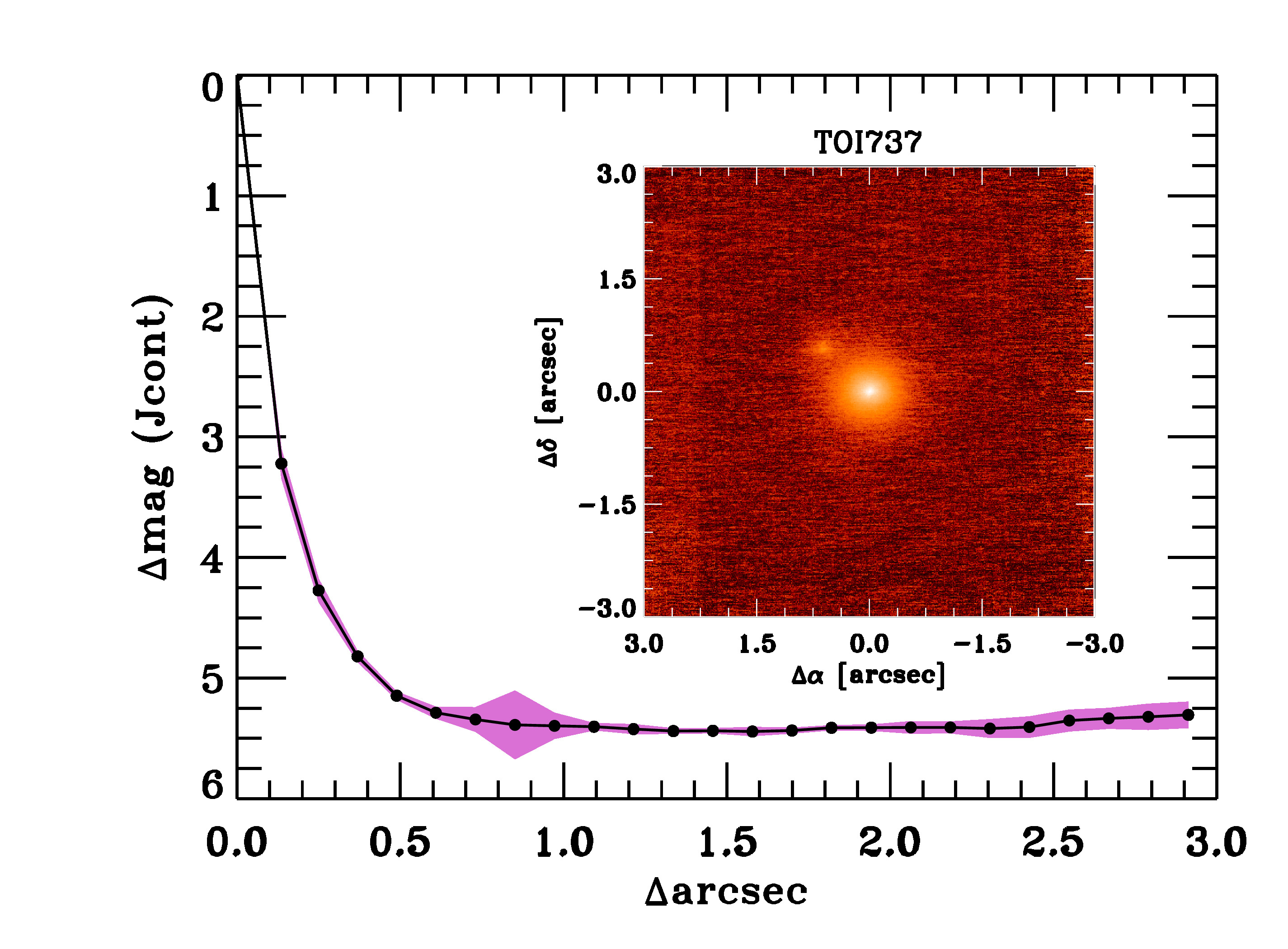}
    \includegraphics[width=0.49\textwidth]{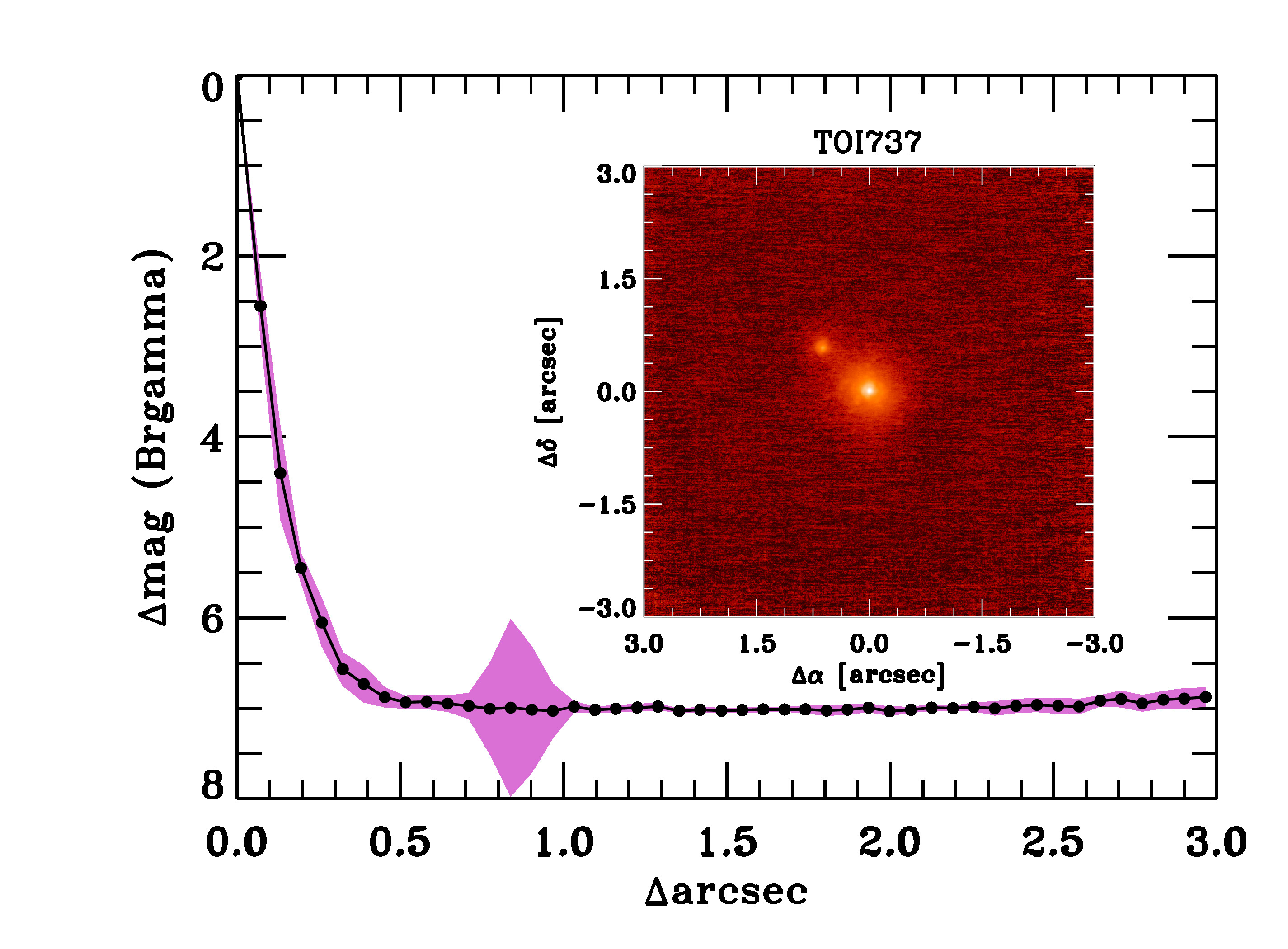}
    \caption{The stellar companion revealed by Keck AO data, shown in the the $J-cont$ filter on the left, and in the Br-$\gamma$ filter on the right. The stellar companion to TOI 737 was detected at 0.845\arcsec\ (247 au), at $46^{\circ}$, and with deblended relative magnitudes of $\Delta J = 2.64$ mag and $\Delta K=2.61$ mag.}
    \label{fig:TOI737}
\end{figure*}

The final combined images and contrast curves for this object are shown in Figure \ref{fig:TOI737}. The stellar companion was detected at an angular separation of $0.842\arcsec\pm0.001$ and a position angle of $46\pm1^{\circ}$. The deblended magnitudes correspond to relative brightnesses of $\Delta J = 2.64\pm0.03$ mag and $\Delta K=2.62\pm0.01$ mag.

We used the same methods from Section \ref{subsec:speckle_companion} to estimate the projected separation, mass ratio, orbital period for TOI 737 and its stellar companion. We find a projected separation of $246\pm3$ au. We estimate the component masses to be $0.58\pm0.02$ $M_{\odot}$ for the primary and $0.23\pm0.01$ $M_{\odot}$ for the secondary. These values correspond to a mass ratio of $0.40\pm0.02$, yielding an orbital period of roughly 4.28\sci{e3} years.

It should be noted that two TOIs (256 and 654) are listed on the ExoFOP as having stellar companions that were detected in their AO observations; however, these companions were determined to be background stars based on their Gaia EDR3 parallaxes.

\subsection{Companions Known to the Literature}

Upon searching the literature for stellar companions to these \TOIs M-dwarf TOIs, we find three previously known companions to these objects.

TOI 455 is known to be in a hierarchical triple star system based on visual micrometry by \citet{Rossiter1955POMic..11....1R}, \textit{Hubble Space Telescope} images from \citet{Dieterich2012AJ....144...64D}, and, most recently, data obtained in 2017 that is available in the Washington Double Star (WDS) catalog\footnote{\raggedright\url{https://www.usno.navy.mil/USNO/astrometry/optical-IR-prod/wds/WDS}} \citep{Mason2001AJ....122.3466M}. Furthermore, a planet orbiting TOI 455 was discovered by \citet{Winters2019AJ....158..152W}. We use the angular separation and position angle values listed in \citet{Dieterich2012AJ....144...64D}, and the parallax and magnitude values listed in \citet{Winters2019AJ....158..152W}, to calculate the A-BC estimated astrophysical parameters for Table \ref{table:summary}.

TOI 529 was listed as a candidate eclipsing binary by \citet{Armstrong2015A&A...579A..19A}. As they did not list observational data for this system, we do not include this stellar companion in Table \ref{table:summary} or Figure \ref{fig:periods}.

TOI 1635, or LSPM J1626+6608E, was listed as a CPM companion to LSPM J1626+6608W in the Luyten proper motion catalog \citep{Luyten(1997)}. Additionally, \citet{Lamman2020AJ....159..139L} reported a companion to LSPM J1626+6608W at $0.195\arcsec$, making the system trinary. We use the angular separation and position angle values listed in \citet{Luyten(1997)}, and the parallax and magnitude values listed in Gaia EDR3, to calculate the A-B estimated astrophysical parameters for Table \ref{table:summary}. Errors on separation were not provided in \citet{Luyten(1997)}, so we assume a 0.5\arcsec error to be a generous estimate for the catalog and for 66 years of proper motion. Furthermore, LSPM J1626+6608W does not have a parallax listed in Gaia EDR3, likely due to its sub-arcsecond companion. We have reason to believe that the A-B components have similar parallax values because of their similar proper motions, and no other parallax values were forthcoming from the literature. We therefore used the LSPM J1626+6608E parallax value for LSPM J1626+6608W as well.

\clearpage

\subsection{Common Proper Motion Companions}

We also searched for widely-separated CPM companions using three catalogs. We found nine CPM companions with Gaia EDR3 \citep{Gaia2021A&A...649A...1G} using a search radius of $60\arcsec$. We considered stars to be in a CPM pair if their parallaxes and proper motions agreed with one another to within $3\sigma$. In order to probe the most widely-separated binaries, we cross-matched our sample with the SUPERWIDE catalog \citep{HartmanLepine2020ApJS..247...66H}, which consists of nearly 100,000 high probability wide binaries, and is the result of a two-part Bayesian analysis of the high proper motion stars in Gaia DR2. We identified two additional CPM companions with this catalog. Finally, to probe widely-separated binaries with proper motions smaller than 40 milli-arcseconds per year, we cross-matched our sample with \citet{El-Badry2021MNRAS.506.2269E}, and identified one additional CPM companion. The SUPERWIDE catalog and \citet{El-Badry2021MNRAS.506.2269E} allowed us to search more precisely for companions outside of $60\arcsec$. In all cases, both the primaries and associated CPM secondaries appear in Gaia EDR3.

It should be noted that the completeness of Gaia EDR3 to wide binaries is beyond the scope of this paper and, in general, does not affect the primary results of this work, which is the characterization of close-in binaries. If there are more wide binaries that have gone undetected in this work, they do not affect the results of the high-resolution imaging, nor the conclusions drawn regarding the presence of close-in companions. The sensitivity and completeness of the SUPERWIDE and \citet{El-Badry2021MNRAS.506.2269E} surveys are discussed in their respective papers.

The CPM companions are summarized in Table \ref{table:CPMcompanions}. As a note, the speckle companion was not detected by either Gaia EDR3 or the CPM catalogs. The AO companion is listed in Gaia EDR3; however, it has no parallax nor proper motions listed.

\onecolumngrid
\begin{longtable}{c c c c c c c c} 
 \caption{Common proper motion companions}\\
 \hline
 Target & Component & Gaia EDR3 & Parallax & Parallax Error & Proper Motion & Proper Motion Error & RUWE \\
 & & Source ID & (mas) & (mas) & (mas/yr) & (mas/yr) & \\
 \hline
 \endfirsthead
 \multicolumn{8}{c}%
 {\tablename\ \thetable\ -- \textit{Continued from previous page}} \\
 \hline
 Target & Component & Gaia EDR3 & Parallax & Parallax Error & Proper Motion & Proper Motion Error & RUWE \\
 & & Source ID & (mas) & (mas) & (mas/yr) & (mas/yr) & \\
 \hline
 \endhead
 \hline \multicolumn{8}{c}{\textit{Continued on next page}} \\
 \endfoot
 \endlastfoot
 TOI 277 & B & 2353440974955205632 & 15.4 & 0.019 & -115, -249 & 0.026, 0.025 & 1.28 \\
 & A & 2353440974955205504 & 15.4 & 0.018 & -113, -248 & 0.025, 0.023 & 1.25 \\
 TOI 468 & B & 2966680597368750720 & 5.89 & 0.019 & -2.57, 7.37 & 0.015, 0.018 & 1.08 \\
 & A & 2966587448117647872 & 5.91 & 0.014 & -2.49, 7.47 & 0.014, 0.016 & 0.98 \\
 TOI 488 & A & 3094290054327367168 & 36.6 & 0.024 & -403, -381 & 0.025, 0.017 & 1.30 \\
 & B & 3094290019967631360 & 36.5 & 0.065 & -399, -381 & 0.067, 0.047 & 1.14 \\
 TOI 507 & A & 5724250571514167424 & 9.10 & 0.020 & 47.7, -15.3 & 0.019, 0.015 & 1.10 \\
 & B & 5724250880751513728 & 9.12 & 0.024 & 47.7, -15.0 & 0.022, 0.018 & 1.05 \\
 TOI 513 & A & 5713325132487143424 & 5.20 & 0.166 & 68.7, -54.5 & 0.144, 0.138 & 5.30 \\
 & B & 5713325136782172544 & 4.58 & 0.176 & 68.1, -53.7 & 0.226, 0.235 & 1.00 \\
 TOI 573 & A & 5689391929738059648 & 11.3 & 0.018 & -51.7, -92.2 & 0.019, 0.014 & 1.09 \\
 & B & 5689391929738875776 & 11.2 & 0.049 & -47.0, -95.2 & 0.057, 0.042 & 1.09 \\
 TOI 633 & B & 5735743903992405632 & 8.58 & 0.037 & 30.6, 3.15 & 0.030, 0.020 & 1.25 \\
 & A & 5735744144510573696 & 8.51 & 0.024 & 30.8, 2.84 & 0.020, 0.020 & 1.12 \\
 TOI 717 & A & 3846754645112261760 & 29.1 & 0.024 & -25.9, 61.6 & 0.029, 0.032 & 1.18 \\
 & B & 3846848417133797504 & 29.1 & 0.025 & -23.7, 61.7 & 0.030, 0.033 & 1.15 \\
 TOI 749 & A & 3501271053530782080 & 4.46 & 0.094 & -36.4, -34.2 & 0.117, 0.094 & 3.01 \\
 & B & 3501271053530027904 & 5.08 & 0.182 & -40.4, -38.1 & 0.239, 0.224 & 1.02 \\
 TOI 756 & A & 6129327525817451648 & 11.6 & 0.017 & -217, 29.2 & 0.016, 0.013 & 1.24 \\
 & B & 6129327319659021056 & 11.6 & 0.028 & -216, 29.5 & 0.028, 0.023 & 1.15 \\
 TOI 1201 & A & 5157183324996790272 & 26.6 & 0.022 & 164, 46.6 & 0.025, 0.027 & 1.55 \\
 & B & 5157183324996789760 & 26.5 & 0.023 & 174, 45.5 & 0.026, 0.029 & 1.53 \\
 TOI 1634 & A & 223158499179138432 & 28.5 & 0.018 & 81.4, 13.6 & 0.020, 0.015 & 1.23 \\
 & B & 223158499176634112 & 28.6 & 0.113 & 80.6, 14.5 & 0.126, 0.091 & 1.71 \\
  \hline
 \label{table:CPMcompanions}
\end{longtable}
\twocolumngrid

It should be noted that TOIs 277, 468, and 633 are the B components of their respective binary systems. In the case of TOI 633, the A component is TOI 1883, and the same transit event is listed for both stars in the ExoFOP. The TESS Follow-up Observing Program Working Group 1 showed that the transit from TOI 1883 is real, and that the transit from TOI 633 is a false positive. So in the case of TOI 633, the planet does orbit the primary star, which is TOI 1883. However, in the cases of TOIs 277 and 468, the planet orbits the secondary star in the binary system.

\subsection{Potential Close-In Companions} \label{subsec:close-in_companions}

To search for very close companions that may have escaped detection with our high-resolution imaging, we examined the Gaia EDR3 re-normalized unit weight error (RUWE) values associated with the M-dwarf TOIs in our sample. The Gaia RUWE metric acts like a reduced chi-squared, where large values can indicate a poor model fit to the astrometry, assuming that the star is single. Single sources typically have RUWE values of $\sim1$, while sources with RUWE values $>1.4$ are likely non-single or otherwise extended \citep{Ziegler2020AJ....159...19Z, Gaia2021A&A...649A...1G}. It should be noted that large RUWE values only indicate a poor astrometric fit, and do not definitively indicate the presence of a stellar companion. However, in this work, as in previous works, we use the RUWE value as an indication that something is amiss with the single star fit, and that the cause could be a hidden stellar companion. Following \citet{Vrijmoet2020AJ....160..215V}, which surveyed M-dwarfs specifically, we use RUWE $>2$ to distinguish single and (potentially) non-single sources. TOI 531 (RUWE = 21.2) is the only target to  have a large RUWE value but no close-in stellar companion detected by the high-resolution imaging or by Gaia EDR3.  

Another tool that can be used to indicate the presence of an unseen stellar companion is the apparent brightness of the stars in comparison to the Main Sequence in an H-R diagram. Figure~\ref{fig:h-r_diagram} reveals three stars in our sample that sit approximately 1 magnitude above the Main Sequence (TOIs 436, 633, 717). While two of these stars have CPM companions (TOIs 633 and 727) with $\rho>15\arcsec$, none of these targets have a large RUWE value or a known, close-in stellar companion. The excess brightness may indicate a nearly-equal-brightness -- and therefore, nearly-equal-mass -- stellar companion. It is unclear as to why these targets would not exhibit a large RUWE value, but for the purposes of this work, we assume that the excess brightness (like the excess RUWE) indicates an undetected close-in stellar companion. 

As we do not have observational data for these potential close-in stellar companions, nor confirmation that these companions exist, we do not include them in Table \ref{table:summary} or Figure \ref{fig:periods}, but we do discuss their potential properties (mass and period) in Section \ref{sec:missed_companions}.

\subsection{Summary of Companions with Observational Data}

A summary of all 16 stellar companions with observational data are included in Table \ref{table:summary}. We estimate the astrophysical parameters of the systems by assuming that the eccentricity is zero, the inclination is $90^{\circ}$, and the instantaneous spatial separation is approximately the semi-major axis. It should be noted that the mass for the stellar companion to TOI 468 was obtained from \citet{Kervella2019A&A...623A..72K}.

\onecolumngrid
\LTcapwidth=\textwidth
\begin{ThreePartTable}
\begin{TableNotes}
\footnotesize
\item [*] The M-dwarf TOI is the secondary in the system.
\end{TableNotes}
\begin{longtable}{c c c c c c c c c} 
 \caption{Summary of observational data and estimated astrophysical parameters for detected companions} \\
 \hline
 Target & $\rho$ & $\theta$ & $\Delta$m & Bandpass & Projected Separation & Mass & Orbital Period & Source \\
 & (\arcsec) & ($^{\circ}$) & & & (au) & Ratio & (yr) & \\
 \hline
 \endfirsthead
 \multicolumn{9}{c}
 {\tablename\ \thetable\ -- \textit{Continued from previous page}} \\
 \hline
 Target & $\rho$ & $\theta$ & $\Delta$m & Bandpass & Projected & Mass & Orbital & Source \\
 & (\arcsec) & ($^{\circ}$) & & & Separation (au) & Ratio & Period (yr) & \\
 \hline
 \endhead
 \hline \multicolumn{9}{c}{\textit{Continued on next page}} \\
 \endfoot
 \insertTableNotes
 \endlastfoot
 TOI 277\tnote{*} & 17.2 & 352 & $-0.304\pm0.001$ & G & 1.12\sci{e3} & 1.06 & 3.58\sci{e4} & Gaia EDR3 \\
 TOI 455 & 7.71 & 315 & $-0.08\pm0.04$ & $\rm I_{\rm KC}$ & 53.0 & 1.15 & 643 &  \citet{Winters2019AJ....158..152W} \\
 TOI 468\tnote{*} & 330 & 257 & $-7.699\pm0.001$ & G & 5.60\sci{e4} & 4.14 & 7.61\sci{e6} & \citet{El-Badry2021MNRAS.506.2269E} \\
 TOI 482 & 0.398 & 267 & $2.43\pm0.20$ & 832 nm & 57.0 & 0.56 & 859 & Speckle \\
 TOI 488 & 49.3 & 223 & $3.482\pm0.001$ & G & 1.35\sci{e3} & 0.32 & 7.38\sci{e4} & Gaia EDR3 \\
 TOI 507 & 73.5 & 339 & $0.469\pm0.001$ & G & 8.08\sci{e3} & 0.78 & 7.66\sci{e5} & SUPERWIDE \\
 TOI 513 & 0.718 & 42.7 & $1.695\pm0.004$ & G & 138 & 0.56 & 2.04\sci{e3} & Gaia EDR3 \\
 TOI 573 & 1.71 & 122 & $0.879\pm0.001$ & G & 152 & 0.76 & 2.10\sci{e3} & Gaia EDR3 \\
 TOI 633\tnote{*} & 15.8 & 15.3 & $-0.696\pm0.001$ & G & 1.84\sci{e3} & 1.17 & 8.18\sci{e4} & Gaia EDR3 \\
 TOI 717 & 65.5 & 88.6 & $0.098\pm0.001$ & G & 2.25\sci{e3} & 0.97 & 1.24\sci{e5} & SUPERWIDE \\
 TOI 737 & 0.842 & 46 & $2.615\pm0.010$ & Br-$\gamma$ & 246 & 0.40 & 4.28\sci{e3} & AO \\
 TOI 749 & 1.32 & 252 & $2.326\pm0.004$ & G & 296 & 0.62 & 5.01\sci{e3} & Gaia EDR3 \\
 TOI 756 & 11.1 & 157 & $1.353\pm0.001$ & G & 956 & 0.66 & 3.21\sci{e4} & Gaia EDR3 \\
 TOI 1201 & 8.35 & 98.7 & $0.288\pm0.001$ & G & 314 & 0.92 & 5.79\sci{e3} & Gaia EDR3 \\
 TOI 1634 & 2.69 & 89.5 & $3.387\pm0.003$ & G & 94.3 & 0.32 & 1.19\sci{e3} & Gaia EDR3 \\
 TOI 1635\tnote{*} & 78.0 & 107 & $-2.504\pm0.001$ & G & 6.70\sci{e3} & 1.77 & 5.69\sci{e5} & \citet{Luyten(1997)} \\
  \hline
 \label{table:summary}
\end{longtable}
\end{ThreePartTable}
\twocolumngrid

\section{Assessment of Potentially Missed Companions}
\label{sec:missed_companions}

Here we analyze the completeness of our speckle observations and the AO data using our derived contrast curves and the code adapted from \citet{LundCiardi2020AAS...23524906L}. We also explore properties of the potential close-in stellar companions indicated by a RUWE $>2$ or excess brightness.

\subsection{Completeness of the High-Resolution Imaging} \label{subsec: completeness}

While high-resolution imaging is a powerful resource for finding stellar companions to exoplanet hosts, there are companions that could be missing from our observations: companions with a large magnitude difference from their primaries, very red companions that are faint in the optical bandpass, companions at separations outside the fields of view of the instruments, and close-in companions that are inside the resolution limits of the instruments. We have addressed each of these populations in the following ways:

\begin{enumerate}
    \item To find faint companions with large magnitude differences from their primaries, we enlisted speckle cameras at larger telescopes (Gemini North and Gemini South) to gather more photons (Figure \ref{fig:median_contrast_curves}).
    \item To capture very red companions that are faint in the optical bandpass, we included AO data in our analysis. The AO data were obtained with the PHARO and NIRC2 instruments, which observe at NIR wavelengths; our data were observed with the $J-cont$, $H-cont$, and Br-$\gamma$ filters.
    \item To reveal companions at separations outside the fields of view of the instruments we used, we cross-matched our sample with Gaia EDR3, SUPERWIDE, and the catalog from \citet{El-Badry2021MNRAS.506.2269E} to search for CPM companions.
    \item To account for close-in companions, we examined the Gaia EDR3 RUWE values associated with the M-dwarf TOIs in our sample.
\end{enumerate}

Even after combining our speckle observations, the AO data, and the CPM catalogs, there are still likely stellar companions to the M-dwarf TOIs in our sample that we missed. In order to identify the population of stellar companions that were not detected by our high-resolution imaging, we used existing code that was originally developed for Palomar and Keck AO observations \citep{LundCiardi2020AAS...23524906L}, and adapted it to the specific needs of our speckle survey. The purpose of these simulations is to estimate the fraction of stellar companions that would be detectable within our high-resolution images, assuming that the star has a companion. The code makes the assumption that if the companion is outside the field of view of the instrument, then it would be revealed by other methods.

\begin{figure*}
    \centering
    \includegraphics[width=0.49\textwidth]{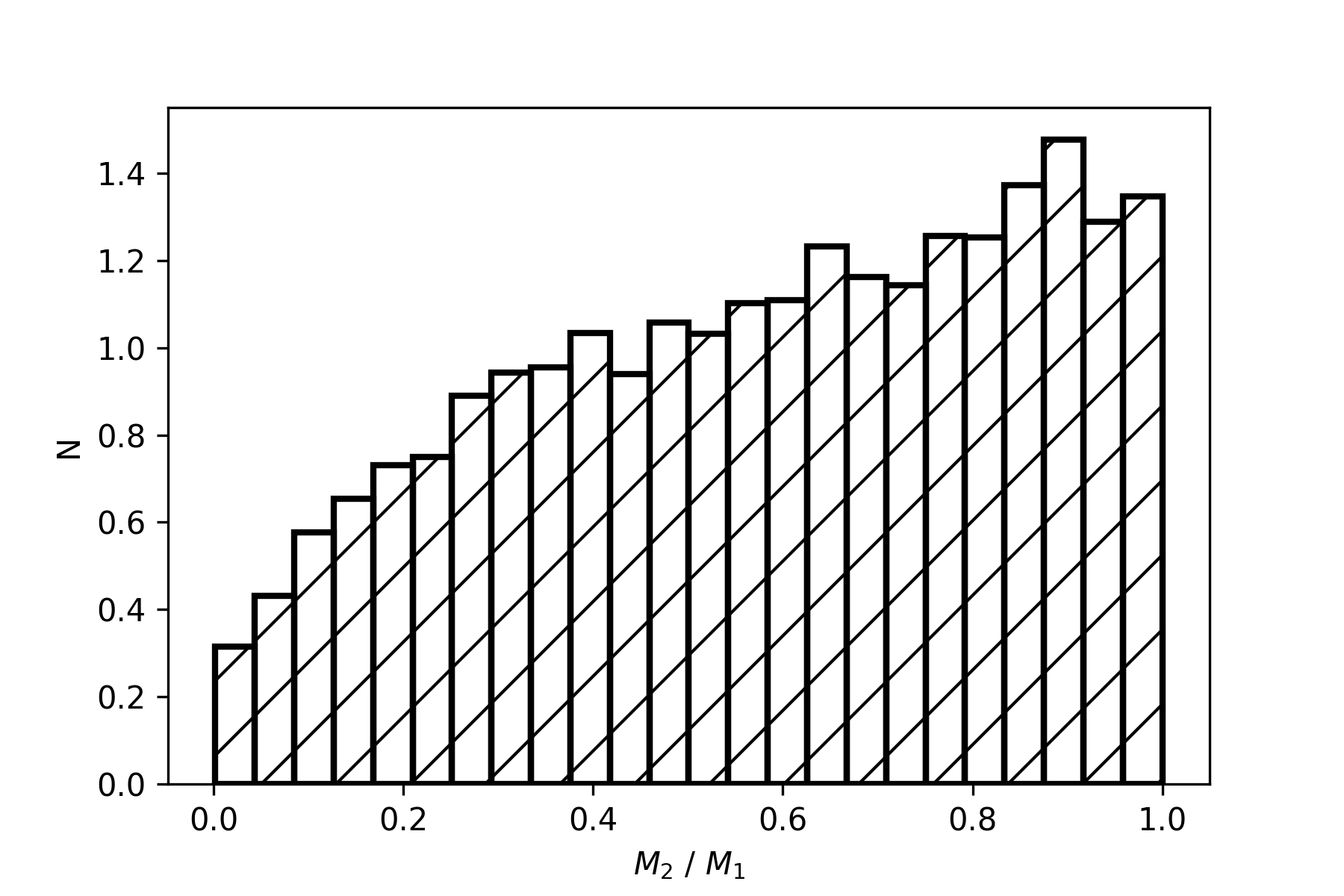}
    \includegraphics[width=0.49\textwidth]{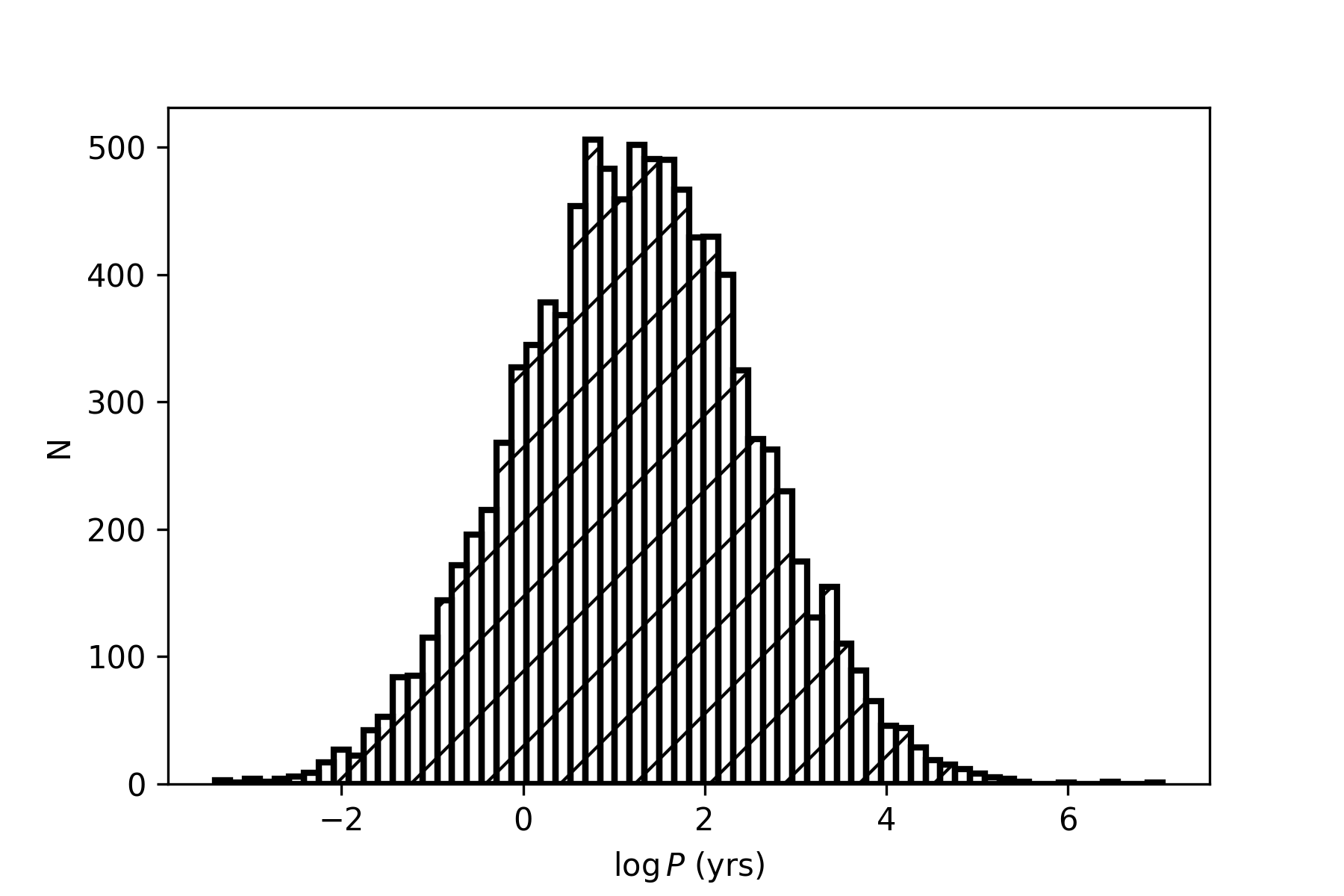}
    \caption{The mass ratio (left) and orbital period (right) distributions used to simulate stellar companions to the M-dwarf TOIs in our sample, from \citet{DucheneKraus2013ARA&A..51..269D}. Values for these figures were drawn assuming a primary mass of 0.44~$M_{\odot}$, which is the median primary mass for the M-dwarf TOIs in our sample.}
    \label{fig:distributions_duchenekraus}
\end{figure*}

The code works by first identifying the population of stellar companions that could orbit each star, and then uses the derived contrast curves to evaluate the sensitivity of each observation to these stellar companions. The code identifies the populations of stellar companions by matching the star to a best-fit stellar isochrone from the Dartmouth isochrones \citep{Dotter2008ApJ...687L..21D}. These isochrones were chosen to enable a more complete comparison to the results of \citet{Ciardi2015ApJ...805...16C}, which used these models. The code then independently draws from the mass ratio and orbital period distributions for M dwarfs from the \citet{DucheneKraus2013ARA&A..51..269D} review paper on stellar multiplicity (Figure \ref{fig:distributions_duchenekraus}). The mass ratios from \citet{DucheneKraus2013ARA&A..51..269D} follow a relatively flat distribution, with an upper limit at 1, and a lower limit for the primary star at $0.1 M_{\odot}$, near the hydrogen burning limit at $0.08 M_{\odot}$ \citep{BaraffeChabrier1996ApJ...461L..51B}. The orbital period distribution follows a log-normal distribution.

The simulated companions to and derived contrast curves for each M-dwarf TOI in our sample are shown in the Appendix. The separations for the simulated companions are calculated assuming a circular orbit, and using the period and stellar masses to compute a semi-major axis. The companion is then given a random position on that orbit. This simulation is performed 10,000 times for each star. It should be noted that the TOI 532 z' band observational data are not included in this analysis because the code from \citet{LundCiardi2020AAS...23524906L} is unable to account for the z' bandpass. Additionally, the TOI 1238 880 nm observational data are not included because the innermost annulus of the reconstructed image did not have enough maxima or minima to do the calculations required to generate a contrast curve.

In general, the simulated companions that were not detected by our technique are either very close-in, or are much fainter than their primary stars. It should be noted that \citet{DucheneKraus2013ARA&A..51..269D} considered companions significantly above the substellar limit when generating their mass ratio and orbital period distributions for low-mass stars. Although faint companions have less of an impact on measured planet properties, these biases influence the stellar companions that are simulated for each M-dwarf TOI, and therefore influence the fraction of simulated companions detectable by our technique as well. Table \ref{table:fraction_detectable} lists the fraction of simulated companions detectable in our observations for each M-dwarf TOI, based not only on the mass ratio and orbital period distributions from \citet{DucheneKraus2013ARA&A..51..269D}, but also from \citet{Raghavan2010ApJS..190....1R} and \citet{Winters2019AJ....157..216W}. Although the distributions from \citet{DucheneKraus2013ARA&A..51..269D} are biased to earlier M dwarfs and more massive stellar companions, the distributions from \citet{Raghavan2010ApJS..190....1R} were formulated based on solar-type stars, and the mass ratio distribution from \citet{Winters2019AJ....157..216W} is incomplete, as the sample had not been surveyed for stellar companions at small separations ($<2\arcsec$), so companions are likely missing that would fill the large unequal mass ratio portion of the distribution. For these reasons, we determined that \citet{DucheneKraus2013ARA&A..51..269D} produced the most appropriate distributions from which to draw populations of stellar companions to assess the completeness of our observations. However, these biases do point to the necessity of a comprehensive M-dwarf multiplicity survey that examines the distribution of stellar companions to M dwarfs at all mass ratios and orbital periods.

\onecolumngrid
\begin{longtable}{c c c c} 
 \caption{Fraction of simulated stellar companions detectable} \\
 \hline
 & & Fraction Detectable & \\
 Target & Duchêne \& Kraus (2013) & Raghavan et al. (2010) & Winters et al. (2019) \\
 \hline
 \endfirsthead
 \multicolumn{4}{c}%
 {\tablename\ \thetable\ -- \textit{Continued from previous page}} \\
 \hline
 & & Fraction Detectable & \\
 Target & Duchêne \& Kraus (2013) & Raghavan et al. (2010) & Winters et al. (2019) \\
 \hline
 \endhead
 \hline \multicolumn{4}{c}{\textit{Continued on next page}} \\
 \endfoot
 \endlastfoot
TOI 198 & 80.3 & 87.1 & 88.1 \\
TOI 244 & 77.7 & 85.3 & 86.0 \\
TOI 256 & 90.6 & 93.4 & 93.6 \\
TOI 277 & 56.6 & 74.8 & 72.2 \\
TOI 278 & 63.0 & 77.5 & 73.8 \\
TOI 436 & 55.6 & 73.2 & 70.1 \\
TOI 442 & 57.1 & 76.6 & 74.3 \\
TOI 455 & 91.6 & 92.6 & 94.4 \\
TOI 468 & 29.7 & 58.1 & 51.2 \\
TOI 482 & 27.5 & 58.3 & 48.2 \\
TOI 488 & 74.2 & 84.0 & 83.2 \\
TOI 497 & 52.0 & 71.5 & 66.8 \\
TOI 507 & 40.3 & 64.9 & 60.1 \\
TOI 513 & 25.5 & 55.4 & 46.0 \\
TOI 516 & 33.8 & 61.7 & 55.1 \\
TOI 519 & 36.6 & 60.5 & 55.3 \\
TOI 521 & 70.8 & 81.1 & 81.4 \\
TOI 526 & 50.5 & 71.8 & 67.5 \\
TOI 528 & 56.9 & 57.4 & 49.5 \\
TOI 529 & 53.7 & 70.4 & 67.2 \\
TOI 530 & 29.6 & 58.6 & 49.4 \\
TOI 531 & 30.3 & 59.4 & 52.0 \\
TOI 532 & 61.1 & 62.6 & 54.4 \\
TOI 538 & 31.9 & 61.1 & 53.0 \\
TOI 549 & 43.0 & 65.7 & 60.8 \\
TOI 552 & 24.2 & 53.1 & 43.8 \\
TOI 557 & 38.7 & 65.2 & 59.5 \\
TOI 562 & 89.0 & 91.6 & 92.8 \\
TOI 573 & 58.2 & 75.4 & 71.7 \\
TOI 620 & 68.3 & 80.2 & 80.5 \\
TOI 633 & 54.0 & 73.6 & 68.8 \\
TOI 643 & 48.0 & 69.4 & 64.0 \\
TOI 654 & 69.3 & 80.7 & 79.7 \\
TOI 674 & 65.1 & 78.5 & 77.3 \\
TOI 716 & 48.5 & 70.5 & 67.3 \\
TOI 717 & 67.0 & 78.9 & 79.1 \\
TOI 727 & 63.6 & 78.3 & 77.3 \\
TOI 734 & 86.9 & 90.5 & 90.9 \\
TOI 736 & 74.0 & 82.4 & 80.6 \\
TOI 737 & 61.7 & 62.4 & 53.2 \\
TOI 749 & 26.5 & 57.0 & 48.3 \\
TOI 756 & 50.2 & 71.9 & 68.8 \\
TOI 782 & 56.7 & 73.6 & 71.1 \\
TOI 797 & 59.0 & 76.5 & 74.9 \\
TOI 821 & 59.0 & 75.5 & 73.6 \\
TOI 876 & 38.0 & 55.4 & 53.5 \\
TOI 900 & 65.7 & 65.5 & 60.8 \\
TOI 912 & 74.0 & 83.8 & 84.4 \\
TOI 1201 & 61.5 & 76.9 & 75.1 \\
TOI 1235 & 76.8 & 75.9 & 76.2 \\
TOI 1238 & 66.0 & 65.8 & 63.2 \\
TOI 1266 & 67.1 & 80.5 & 78.7 \\
TOI 1467 & 65.6 & 78.9 & 78.2 \\
TOI 1468 & 88.8 & 92.0 & 92.6 \\
TOI 1634 & 67.4 & 79.5 & 80.2 \\
TOI 1635 & 46.8 & 68.8 & 62.9 \\
TOI 1638 & 65.7 & 65.2 & 60.1 \\
TOI 1639 & 43.0 & 68.2 & 62.9 \\
  \hline
 \label{table:fraction_detectable}
\end{longtable}
\twocolumngrid

\begin{figure*}
    \centering
    \includegraphics[width=\textwidth]{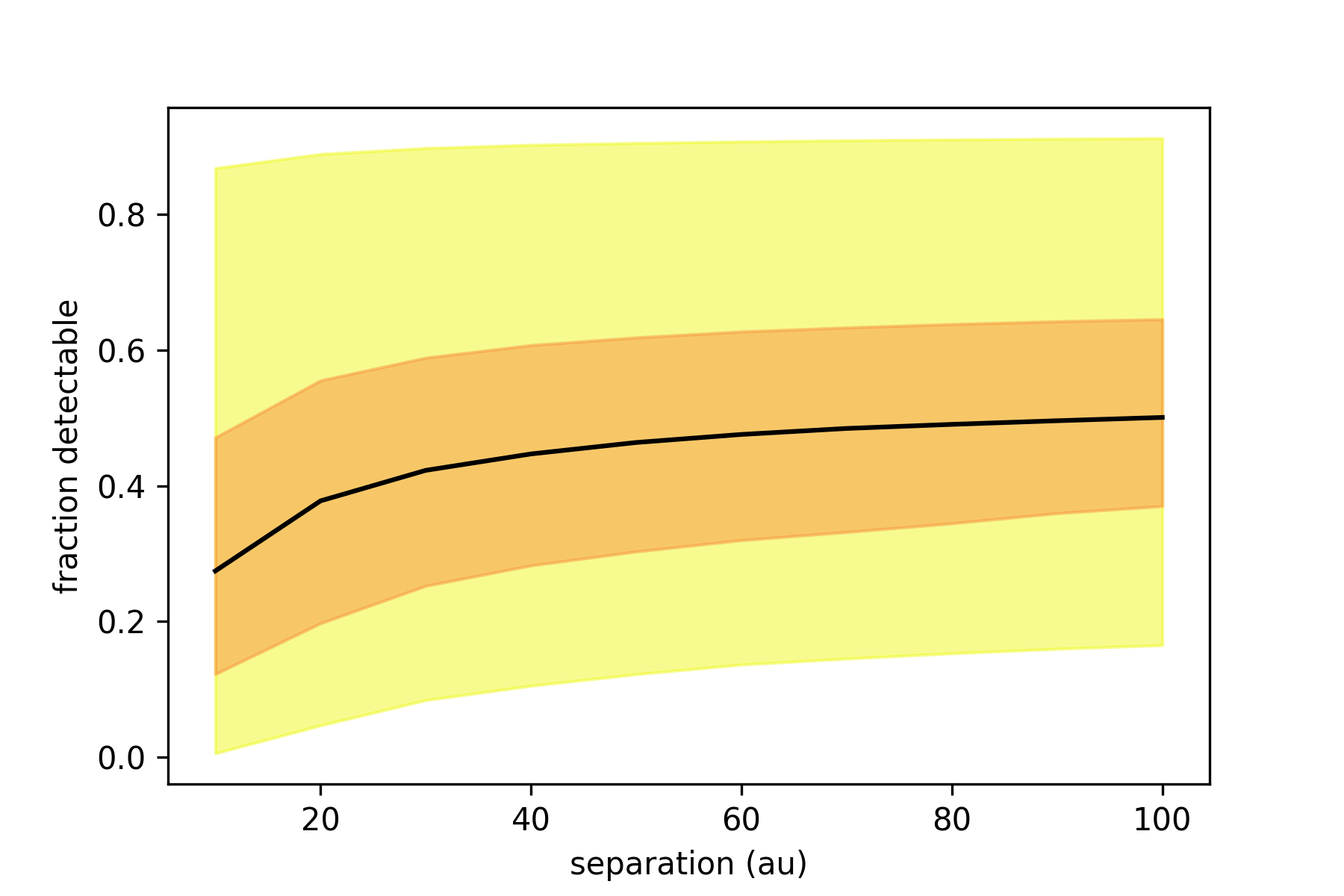}
    \caption{The fraction of simulated companions detectable by the high resolution imaging observations within 100 au for all targets in the sample. As a function of projected separation, the black line shows the median value, the orange shaded area shows the interquartile region, and the yellow shaded area shows the region between the maxima and minima. This plot shows that we are able to reliably survey the region around exoplanet hosts, even at close separations where we have found a dearth of stellar companions.}
    \label{fig:fraction_detectable}
\end{figure*}

We also assess the completeness of our observations within 100 au specifically in order to draw more robust conclusions on the dearth of \textit{close-in} stellar companions, and to survey the parameter space where the number of stellar companions begins to fall off for solar-type stars \citep[e.g.,][]{Hirsch2021AJ....161..134H}. We use the distributions from \citet{DucheneKraus2013ARA&A..51..269D} to plot the fraction of simulated companions detectable within 100 au as a function of projected separation for all targets in the sample in Figure \ref{fig:fraction_detectable}.

\subsection{Properties of the Potential Close-In Companions} \label{subsec: properties}

Using this code, we also calculate a mass and orbital period for the potential close-in stellar companions indicated by a RUWE $>2$ or excess brightness on the Main Sequence (Table \ref{table:missed_companions}). These calculations are based on the median and 1$\sigma$ dispersion of the simulated stellar companions that remain undetected by our high-resolution imaging of these objects. As noted above, we cannot confirm the existence of these potential companions, as the higher RUWE and brightness values may be caused by effects unrelated to binarity (e.g., Gaia instrument artifacts). Additionally, as not all simulated stellar companions are equally likely to contribute to a high RUWE value or excess brightness, these distributions represent the prior probability without taking into account the RUWE value or excess brightness. Furthermore, as the high RUWE value or excess brightness suggests that the likelihood of the star being single is lower than in the canonical probability distribution we implement, the true fraction of simulated companions detectable by our technique may also change in these cases. Finally, we note that this analysis does not contribute to the fraction of simulated stellar companions detectable from Table \ref{table:fraction_detectable}.

\onecolumngrid
\LTcapwidth=\textwidth
\begin{longtable}{c c c c c c} 
 \caption{Potential median masses and orbital periods for undetected companions
 } \\
 \hline
 Target & $\langle$Mass$\rangle$ & $\sigma_{\rm mass}$ & $\langle \rm log P \rangle$ & $\sigma_{\rm log P}$ & Source \\
 & ($M_{\odot}$) & ($M_{\odot}$) & (yr) & (yr) \\
 \hline
 \endfirsthead
 \multicolumn{5}{c}%
 {\tablename\ \thetable\ -- \textit{Continued from previous page}} \\
 \hline
 Target & $\langle$Mass$\rangle$ & $\sigma_{\rm mass}$ & $\langle \rm log P \rangle$ & $\sigma_{\rm log P}$ & Source \\
 & ($M_{\odot}$) & ($M_{\odot}$) & (yr) & (yr) \\
 \hline
 \endhead
 \hline \multicolumn{5}{c}{\textit{Continued on next page}} \\
 \endfoot
 \endlastfoot
 TOI 436 & 0.236 & 0.08 & 0.390 & 0.8 & Excess Brightness \\
 TOI 531 & 0.245 & 0.08 & 0.778 & 0.9 & RUWE $>2$ \\
 TOI 633 & 0.251 & 0.08 & 0.475 & 0.9 & Excess Brightness \\
 TOI 717 & 0.236 & 0.07 & 0.055 & 0.8 & Excess Brightness \\
   \hline
 \label{table:missed_companions}
\end{longtable}
\twocolumngrid

\section{Implications for the Orbital Period Distribution for Planet-Hosting M dwarfs} \label{sec:implications}

Following our analysis in Section \ref{subsec: completeness}, if we assume that the stellar companions have the mass ratio and orbital period distributions for low-mass stars from \citet{DucheneKraus2013ARA&A..51..269D}, then our speckle observations and the AO data revealed 59\% of the simulated companions. The high-resolution imaging revealed 73\% of the stellar companions using the distributions from \citet{Raghavan2010ApJS..190....1R}, or 69\% of the stellar companions if the distributions follow \citet{Winters2019AJ....157..216W}. Overall, our high-resolution imaging is 59-73\% complete. Our speckle and AO images revealed two companions, so it is possible that one or two additional inner companions were missed, potentially those with a high RUWE value or excess brightness (Sections \ref{subsec:close-in_companions} and \ref{subsec: properties}).

Based on the stellar multiplicity rate for field M dwarfs \citep[$\sim27\%$;][]{Winters2019AJ....157..216W}, we expected to detect approximately 16 stellar companions to the \TOIs M-dwarf TOIs in our sample across all techniques. In total, we found 17 stellar companions to the M-dwarf TOIs in our sample: two were revealed by the high-resolution imaging, three were known to the literature, and 12 were shown to be in wide CPM pairs. However, the \citet{Winters2019AJ....157..216W} stellar multiplicity rate is relevant specifically for M dwarfs with M-dwarf companions. TOI 633 is actually the lower mass companion to a more massive star; thus we cannot include it in our comparison to the \citet{Winters2019AJ....157..216W} stellar multiplicity rate. Our detected companion rate for our M-dwarf TOI sample is therefore in complete agreement with the companion rate for M dwarfs in general from \citet{Winters2019AJ....157..216W}.

These results indicate that if the stellar companions to planet-hosting M dwarfs do have an orbital period distribution that is similar to that of the general field sample, then the fraction of M-dwarf TOIs with stellar companions would be \textit{larger} than the fraction of field M dwarfs with stellar companions, and that those companions would be \textit{closer} to their primary stars than our observations could detect. These outcomes would be inconsistent with the results found for both solar-type Kepler stars \citep[e.g.,][]{Kraus2016AJ....152....8K} and for solar-type TESS stars \citep[e.g.,][]{Lester2021AJ....162...75L}.

\begin{figure*}
    \centering
    \includegraphics{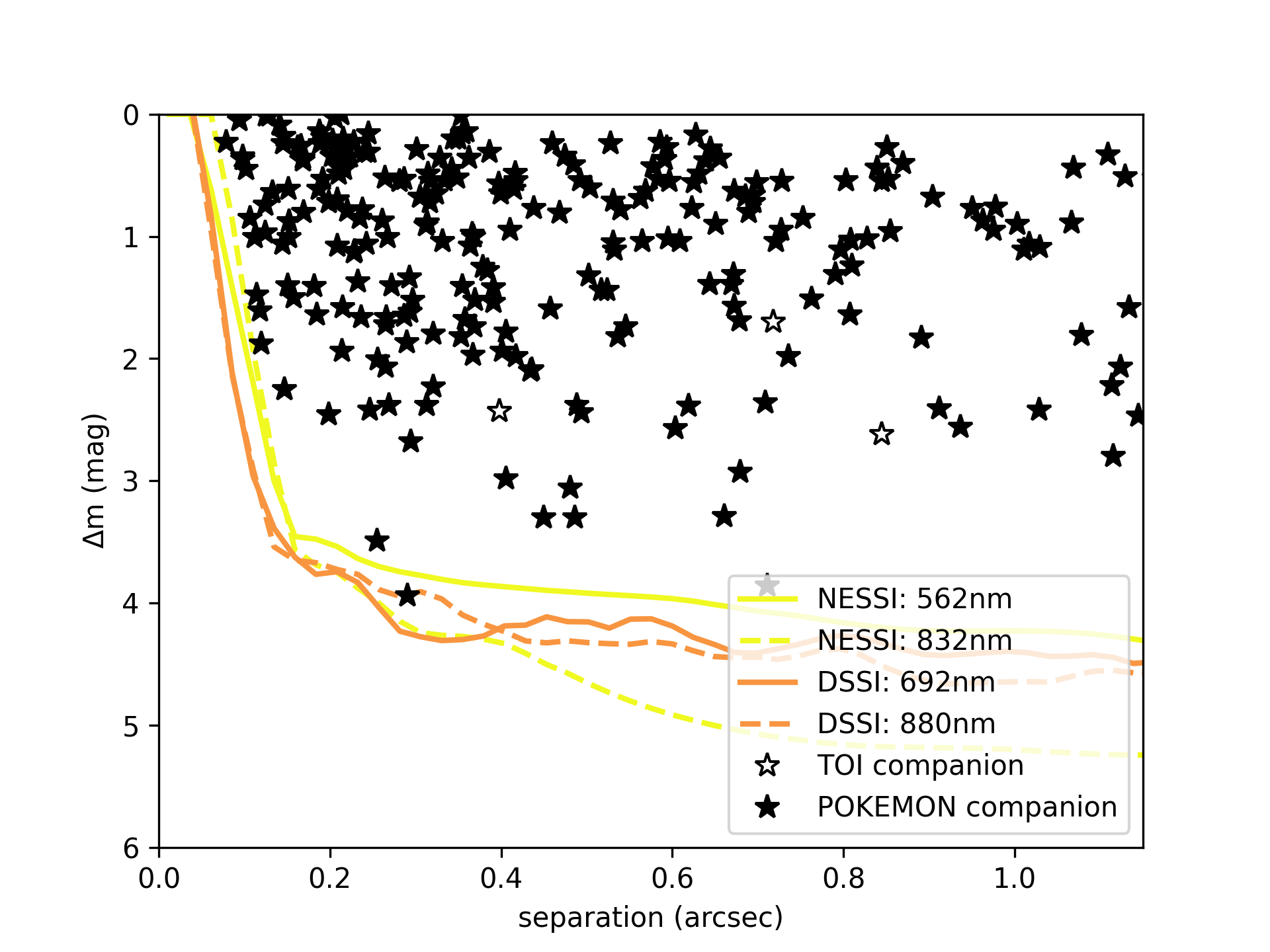}
    \caption{Stellar companions detected throughout the POKEMON survey of nearby field M dwarfs (black stars) compared with the stellar companions detected throughout this survey of planet-hosting M dwarfs (white stars), overlaid on the median contrast curves from NESSI (yellow lines) and DSSI (orange lines), which were used for the POKEMON survey. The range of angular separations for these companions demonstrates that we typically detect companions at the resolution limits of our instruments. However, this was not the case for our survey of M-dwarf TOIs; the closest stellar companion revealed by our high-resolution images was at an angular separation of $0.398\arcsec$.}
    \label{fig:pokemon}
\end{figure*}

These outcomes would also be inconsistent with our findings. The majority of the M-dwarf TOIs in our sample have RUWE values and brightnesses that suggest these stars are single, indicating that the fraction of M-dwarf TOIs with stellar companions is consistent with the fraction of field M dwarfs with stellar companions. Furthermore, the angular separations of the stellar companions detected throughout the upcoming POKEMON speckle survey of 1070 nearby field M dwarfs (Clark et al. in prep) demonstrate that we typically detect companions at the resolution limits of our instruments (Figure \ref{fig:pokemon}); this was not the case for this survey of planet-hosting M dwarfs, indicating that the M-dwarf TOIs in our sample likely do not have unseen stellar companions at the resolution limits of these instruments.

Therefore, a more likely explanation is that the stellar companions to planet-hosting M dwarfs have a different orbital period distribution than the general field sample; they are further separated from their primary stars than the stellar companions to non-planet-hosting M dwarfs. This conclusion is in agreement with both the observed lack of planets in close-in binary systems, and the lack of close-in stellar companions to exoplanet hosts \citep[e.g.,][]{Kraus2012ApJ...745...19K, Bergfors2013MNRAS.428..182B, Wang2014ApJ...791..111W, Kraus2016AJ....152....8K, Fontanive2019MNRAS.485.4967F, FontaniveBardalezGagliuffi2021FrASS...8...16F, Hirsch2021AJ....161..134H, Mustill2021arXiv210315823M, Su2021AJ....162..272S}. This conclusion is also in agreement with what has been shown for nearby solar-type stars with combined radial velocity and imaging data \citep{Hirsch2021AJ....161..134H}, and with what has been seen with speckle imaging of planet-hosting solar-type stars \citep[e.g.,][]{Ziegler2018AJ....156..259Z, Howell2021AJ....161..164H, Lester2021AJ....162...75L, MoeKratter2021MNRAS.507.3593M}. Additionally, \citet{Winters2019AJ....157..216W} found a multiplicity rate of 20.2\% for M dwarfs with stellar companions within 50 au, and therefore only a 6.4\% multiplicity rate for companions at separations beyond 50 au --- the region where we find all of the companions to the M-dwarf TOIs in our sample, and where we find a companion fraction of $\sim27\%$. These findings further substantiate the claim that planet-hosting M dwarfs have a different orbital period distribution than the general field sample.

\begin{figure*}
    \centering
    \includegraphics[width=\textwidth]{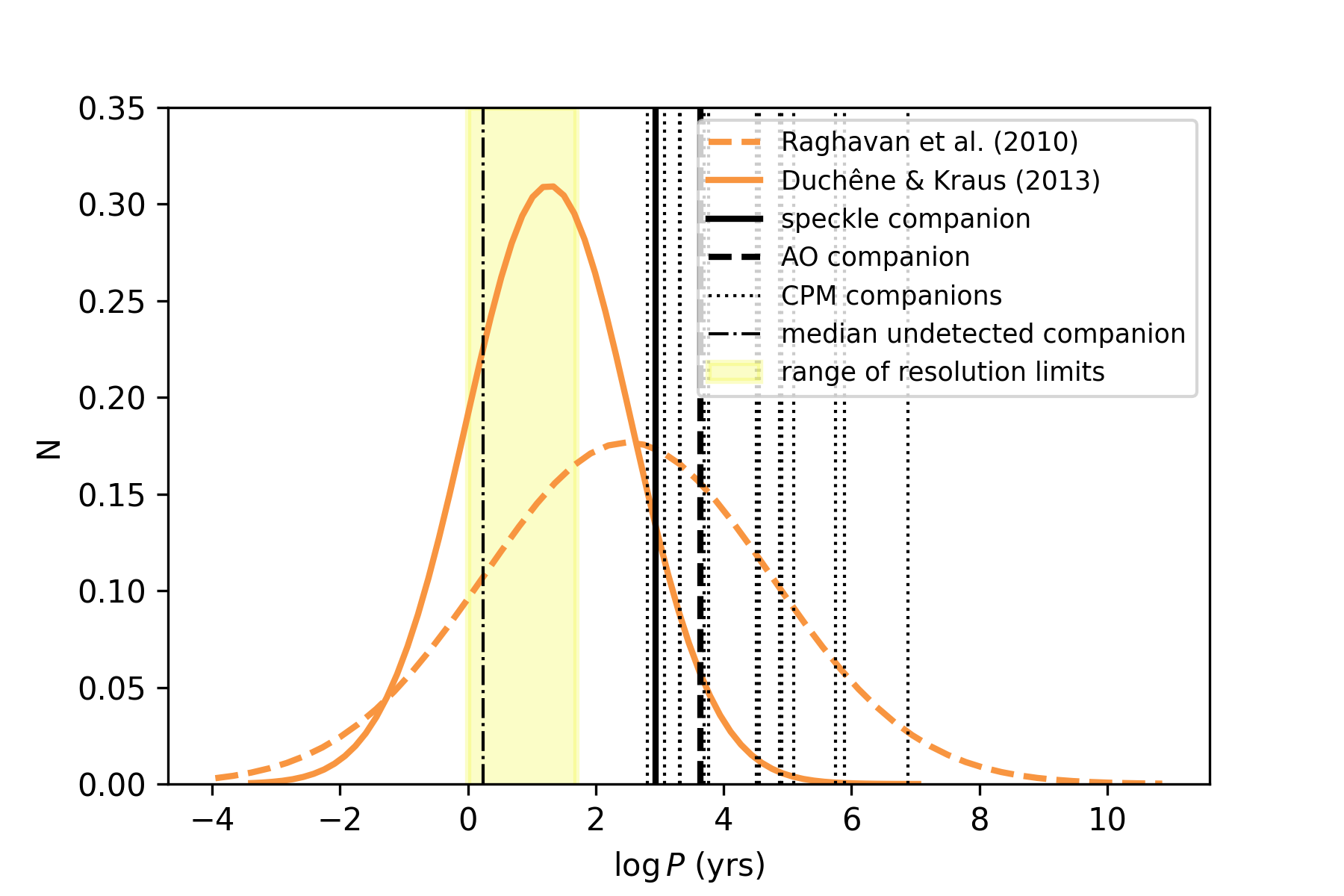}
    \caption{The orbital periods of the stellar companions to the M-dwarf TOIs in our sample, compared with the orbital period distributions for field M dwarfs and solar-type field stars. The orbital period of the speckle companion is marked with a solid black line. The orbital period of the AO companion is marked with a dashed black line. The orbital periods of the CPM companions are marked with dotted black lines. The median orbital period of the potential close-in companions is marked with a black dashdot line. These orbital periods were calculated assuming that the eccentricity is zero, the inclination is $90^{\circ}$, and the instantaneous spatial separation is approximately the orbital semi-major axis. The yellow box demonstrates our range of resolution limits for the speckle and AO observations, assuming primary and secondary masses of our median value of 0.44~$M_{\odot}$. These limits demonstrate that we could detect stellar companions at shorter orbital periods than the speckle, AO, and CPM companions, if they existed. The dashed orange line shows a Gaussian fit to the \citet{Raghavan2010ApJS..190....1R} orbital period distribution, with a peak at $\log P = 2.44$ and a standard deviation of $\sigma_{\log P} = 2.3$. The solid orange line shows a Gaussian fit to the \citet{DucheneKraus2013ARA&A..51..269D} orbital period distribution, with a peak at $\log P = 1.27$ and a standard deviation of $\sigma_{\log P} = 1.3$. The peak of the orbital period distribution for the multiples in our sample is at $\log P = 4.32$ with a standard deviation of $\sigma_{\log P} = 1.1$. Therefore, the orbital period distribution of the stellar companions to the M-dwarf TOIs in our sample is shifted to longer periods compared to the distributions for solar-type field stars and field M dwarfs.}
    \label{fig:periods}
\end{figure*}

In general, we find that the fraction of M-dwarf TOIs with stellar companions is in complete agreement with the fraction for field M-dwarfs in general \citep[$\sim27\%$;][]{Winters2019AJ....157..216W}; the orbital period distribution, however, is skewed to longer periods (Figure \ref{fig:periods}). There are too few companions to determine a meaningful orbital period distribution for our sample, but it appears to have a peak at $\log P = 4.32$, as compared to $\log P = 1.27$ for the field M dwarfs \citep{DucheneKraus2013ARA&A..51..269D}, where P is measured in years.

\section{Summary and Future Work} \label{sec:summary}

In an effort to characterize the rate at which planet-hosting M dwarfs (TOIs in this case) host stellar companions as well, we used high-resolution imaging, a literature search, Gaia EDR3, and two wide binary catalogs to search for stellar companions to \TOIs M-dwarf TOIs. These methods revealed 17 stellar companions to these TOIs; two were uncovered by the high-resolution imaging, three were known to the literature, and 12 were detected as CPM companions by Gaia EDR3, SUPERWIDE, and \citet{El-Badry2021MNRAS.506.2269E}. Additionally, we evaluated the RUWE value and excess brightness of the M-dwarf TOIs in our sample to account for close-in stellar companions.

With these techniques, we find a companion rate that is in complete agreement with the expected $\sim27\%$ stellar multiplicity rate for field M dwarfs. However, we also find that the orbital period distribution of stellar companions to planet-hosting M dwarfs is shifted to longer periods compared to the expected distribution for field M dwarfs. 

A control sample of non-planet-hosting M dwarfs is needed to further vet observational biases. The upcoming volume-limited, diffraction-limited POKEMON speckle survey (Clark et al. in prep), which consists of multiplicity measurements for 1070 nearby M dwarfs for all subtypes M0 to M9, will provide this critical sample of M dwarfs not seen to show a transit event. We used DSSI, NESSI, and the newly commissioned Quad-camera Wavefront-sensing Six-channel Speckle Interferometer \citep[QWSSI;][]{Clark2020SPIE11446E..2AC} to carry out this survey. This control sample will provide the crucial multiplicity measurements needed to detect Earth’s cousins among the low-mass stars.

\acknowledgments

We thank our anonymous reviewer for their thoughtful contributions. We also thank Elliott Horch, Zachary Hartman, Joe Llama, Andrew Richardson, Schuyler Borges, and all the students of the spring 2021 semester of INF 604 for their contributions to and feedback on this manuscript.

This research was supported by NSF Grant No.~AST-1616084 and JPL RSA No.~1610345.

JGW is supported by the National Aeronautics and Space Administration under Grant No.~80NSSC18K0476 issued through the XRP Program.

These results made use of the Lowell Discovery Telescope at Lowell Observatory. Lowell is a private, non-profit institution dedicated to astrophysical research and public appreciation of astronomy and operates the LDT in partnership with Boston University, the University of Maryland, the University of Toledo, Northern Arizona University and Yale University. Lowell Observatory sits at the base of mountains sacred to tribes throughout the region. We honor their past, present, and future generations, who have lived here for millennia and will forever call this place home.

These results are also based on observations from Kitt Peak National Observatory, the NSF's National Optical-Infrared Astronomy Research Laboratory (PI: S. Howell), which is operated by the Association of Universities for Research in Astronomy (AURA) under a cooperative agreement with the National Science Foundation. Data presented herein were obtained at the WIYN Observatory from telescope time allocated to NN-EXPLORE through the scientific partnership of the National Aeronautics and Space Administration, the National Science Foundation, and the NSF's National Optical-Infrared Astronomy Research Laboratory.

Some of the observations in the paper made use of the High-Resolution Imaging instruments ‘Alopeke and Zorro. ‘Alopeke and Zorro were funded by the NASA Exoplanet Exploration Program and built at the NASA Ames Research Center by Steve B. Howell, Nic Scott, Elliott P. Horch, and Emmett Quigley. ‘Alopeke and Zorro were mounted on the Gemini North and South telescope of the international Gemini Observatory, a program of NSF’s NOIRLab, which is managed by the Association of Universities for Research in Astronomy (AURA) under a cooperative agreement with the National Science Foundation. on behalf of the Gemini partnership: the National Science Foundation (United States), National Research Council (Canada), Agencia Nacional de Investigación y Desarrollo (Chile), Ministerio de Ciencia, Tecnología e Innovación (Argentina), Ministério da Ciência, Tecnologia, Inovações e Comunicações (Brazil), and Korea Astronomy and Space Science Institute (Republic of Korea).

This research has made use of the Exoplanet Follow-up Observation Program website \citep{https://doi.org/10.26134/exofop3}, which is operated by the California Institute of Technology, under contract with the National Aeronautics and Space Administration under the Exoplanet Exploration Program. This work has used data products from the Two Micron All Sky Survey \citep{https://doi.org/10.26131/irsa2}, which is a joint project of the University of Massachusetts and the Infrared Processing and Analysis Center at the California Institute of Technology, funded by NASA and NSF.

This work presents results from the European Space Agency (ESA) space mission Gaia. Gaia data are being processed by the Gaia Data Processing and Analysis Consortium (DPAC). Funding for the DPAC is provided by national institutions, in particular the institutions participating in the Gaia MultiLateral Agreement (MLA). The Gaia mission website is \url{https://www.cosmos.esa.int/gaia}. The Gaia archive website is \url{https://archives.esac.esa.int/gaia}.

Information was collected from several additional large database efforts: the Simbad database and the VizieR catalogue access tool, operated at CDS, Strasbourg, France; NASA's Astrophysics Data System; and the Washington Double Star Catalog maintained at the U.S. Naval Observatory.

%

\vspace{5mm}
\facilities{WIYN(NESSI), LDT(DSSI), Gemini North('Alopeke), Gemini South(Zorro)}


\software{IPython \citep{IPython2007}, Matplotlib \citep{Matplotlib2007}, NumPy \citep{NumPy2020}, SciPy \citep{SciPy2020}}

\clearpage




\bibliographystyle{aasjournal}
\bibliography{references}



\appendix

\begin{figure*}[!htb]
  \begin{center}
  	  \includegraphics[width=0.3\textwidth]{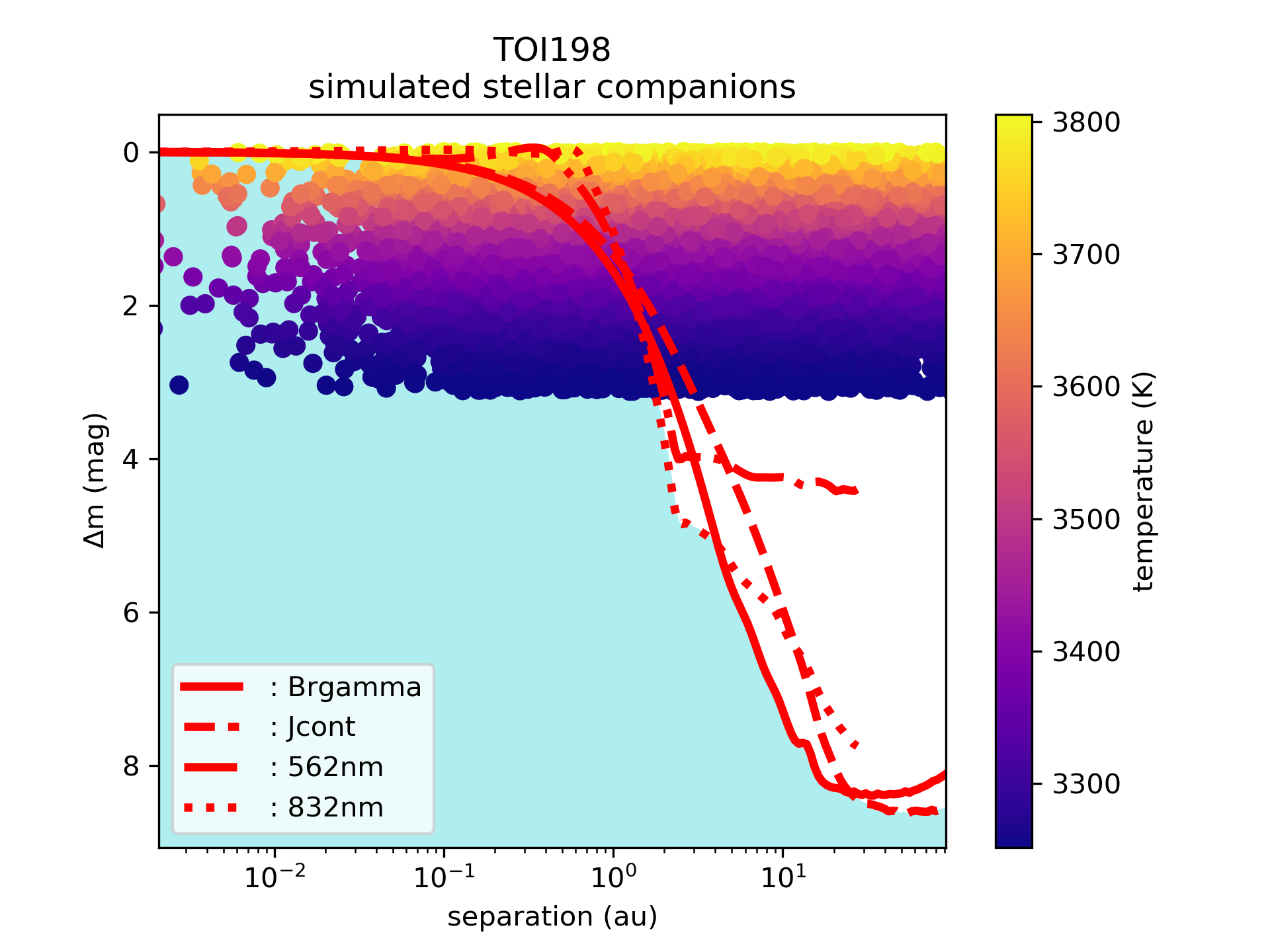}
	  \includegraphics[width=0.3\textwidth]{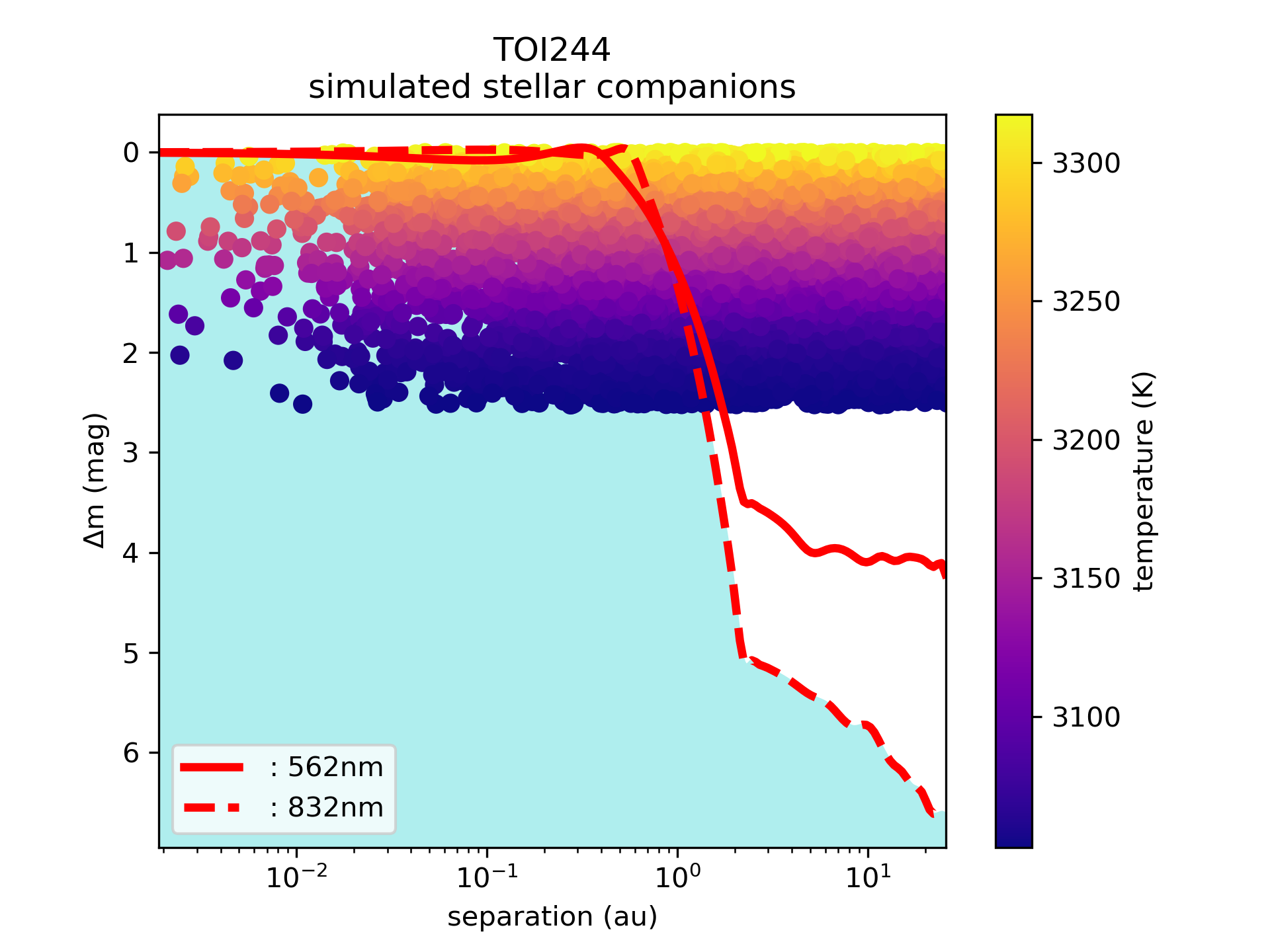}
	  \includegraphics[width=0.3\textwidth]{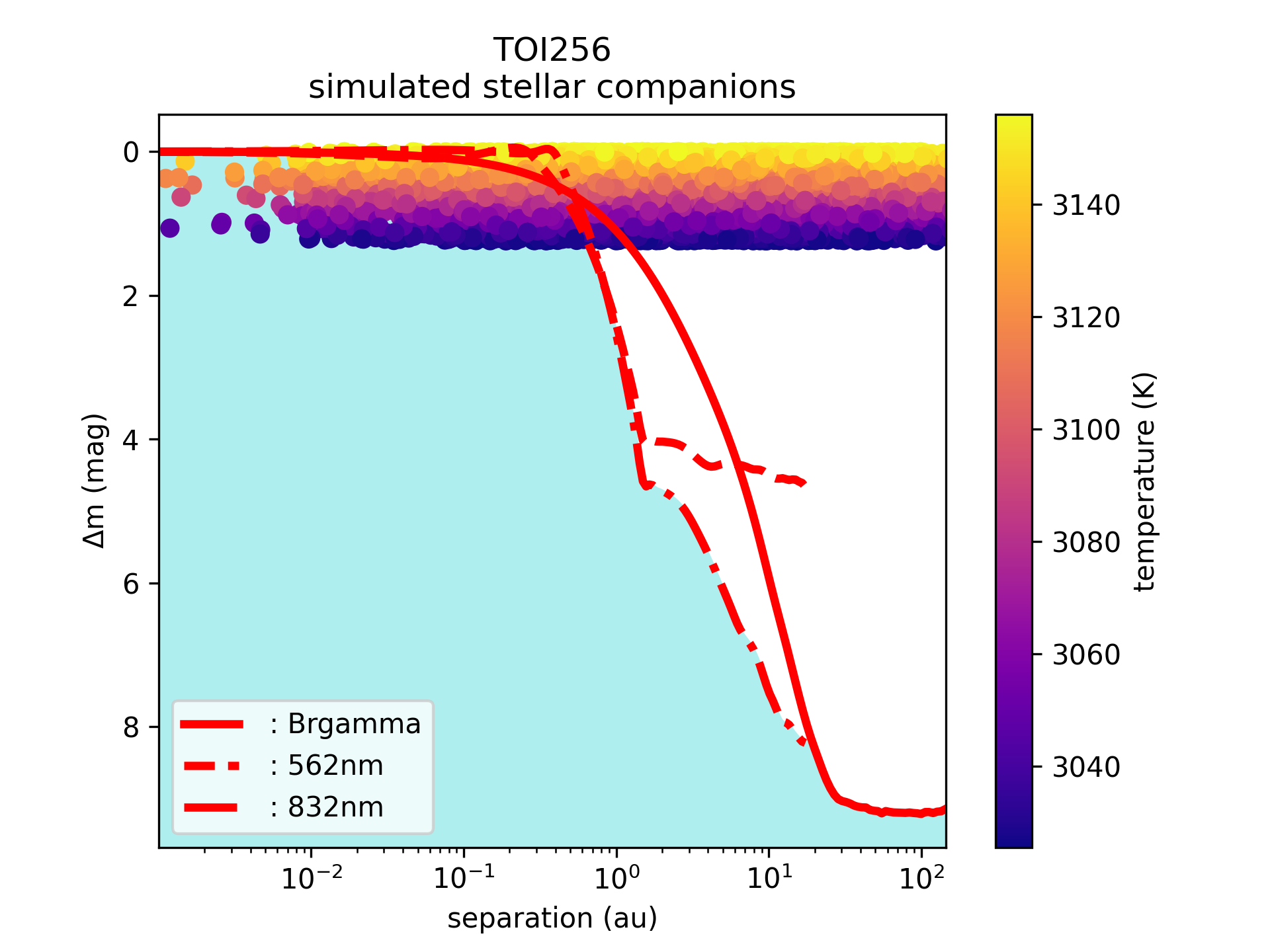}
	  \includegraphics[width=0.3\textwidth]{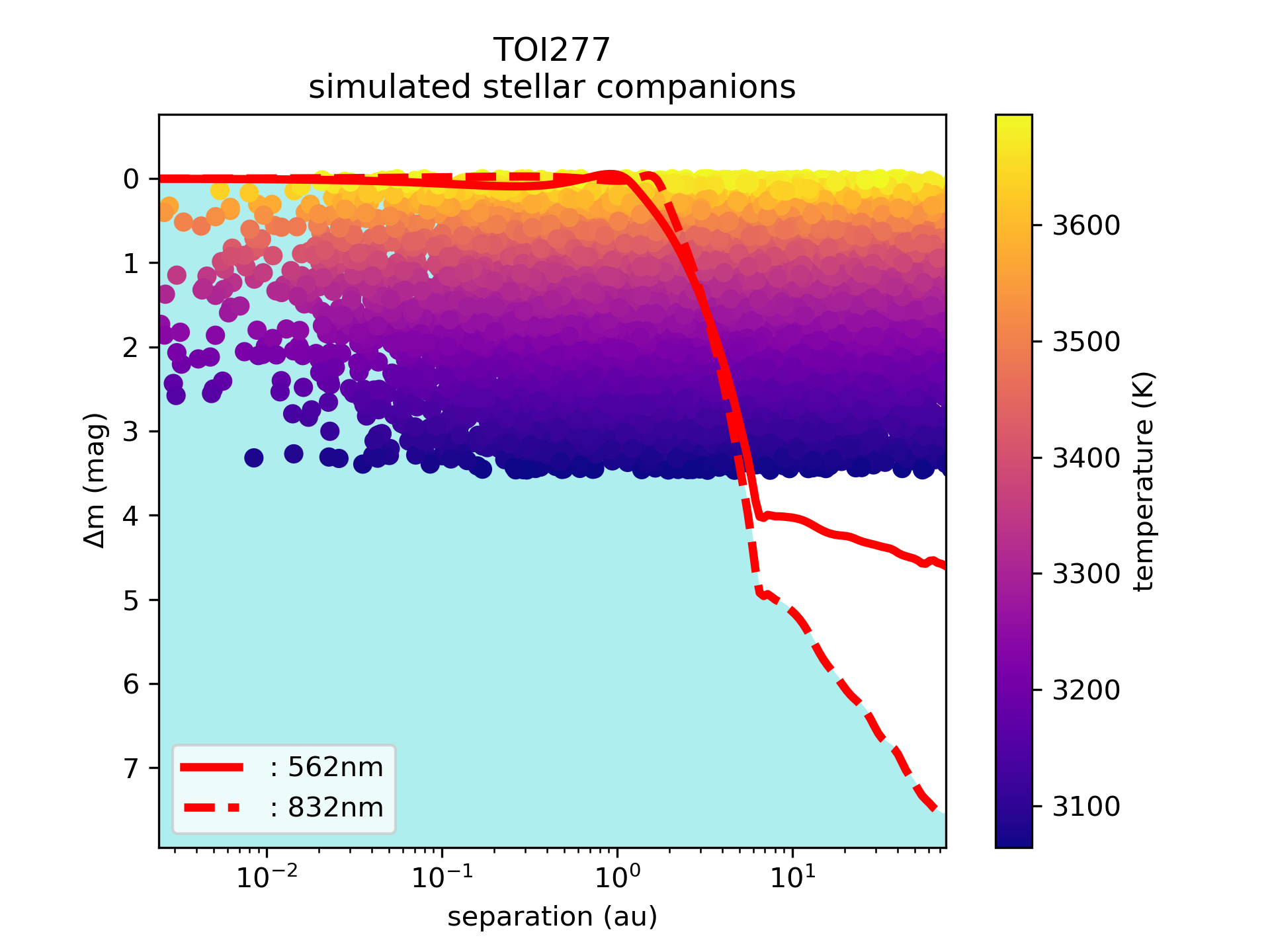}
	  \includegraphics[width=0.3\textwidth]{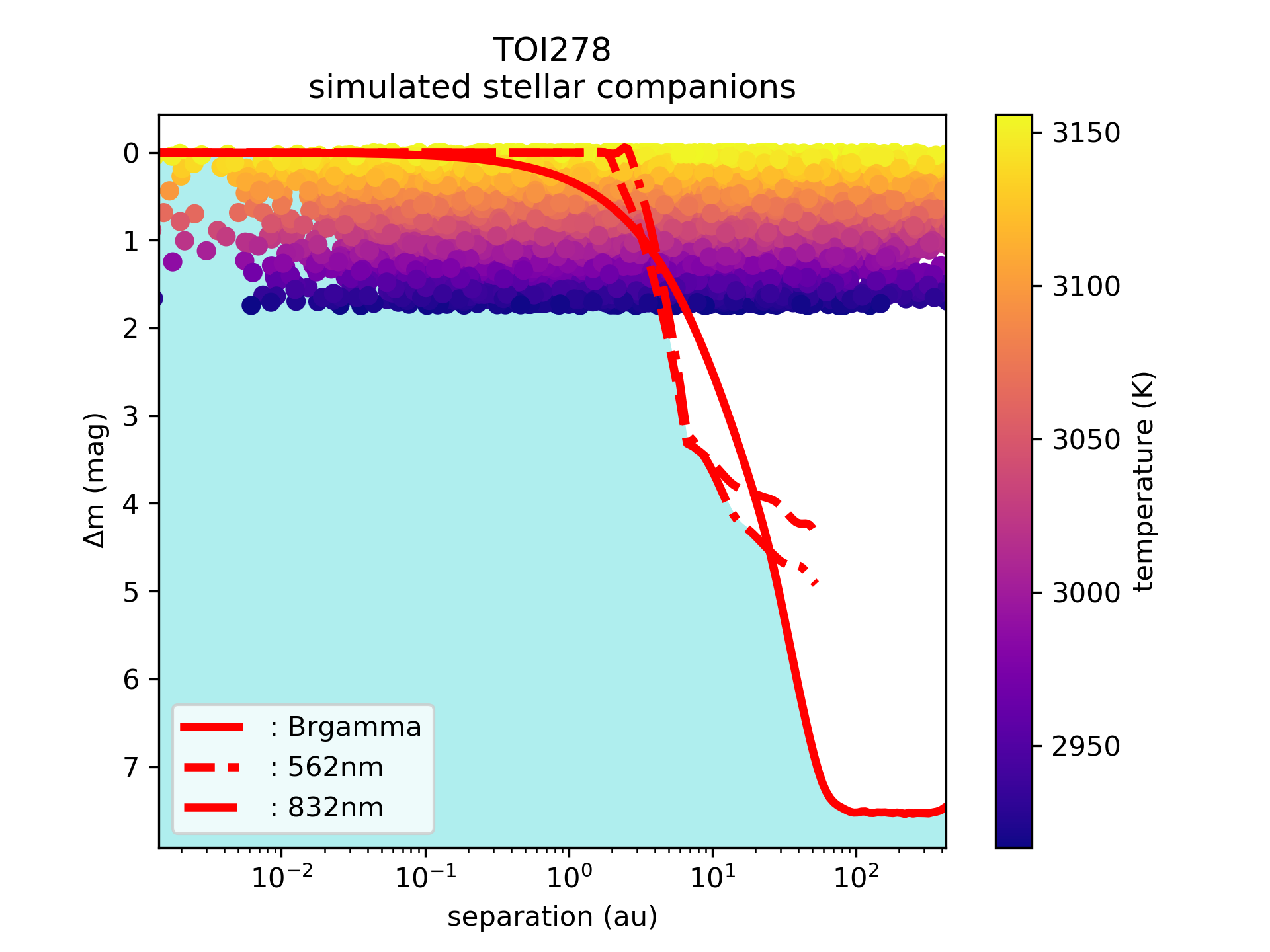}
	  \includegraphics[width=0.3\textwidth]{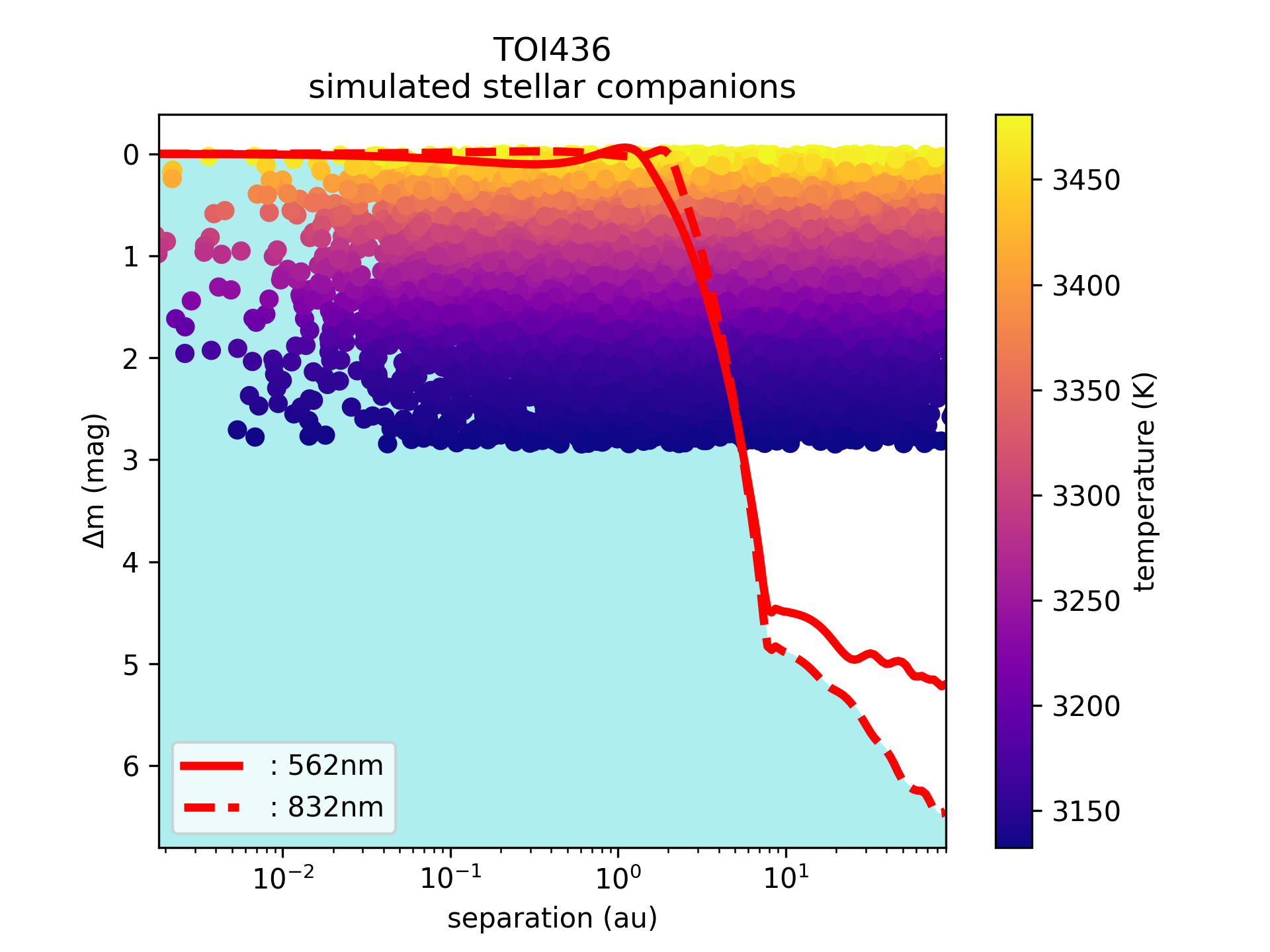}
	  \includegraphics[width=0.3\textwidth]{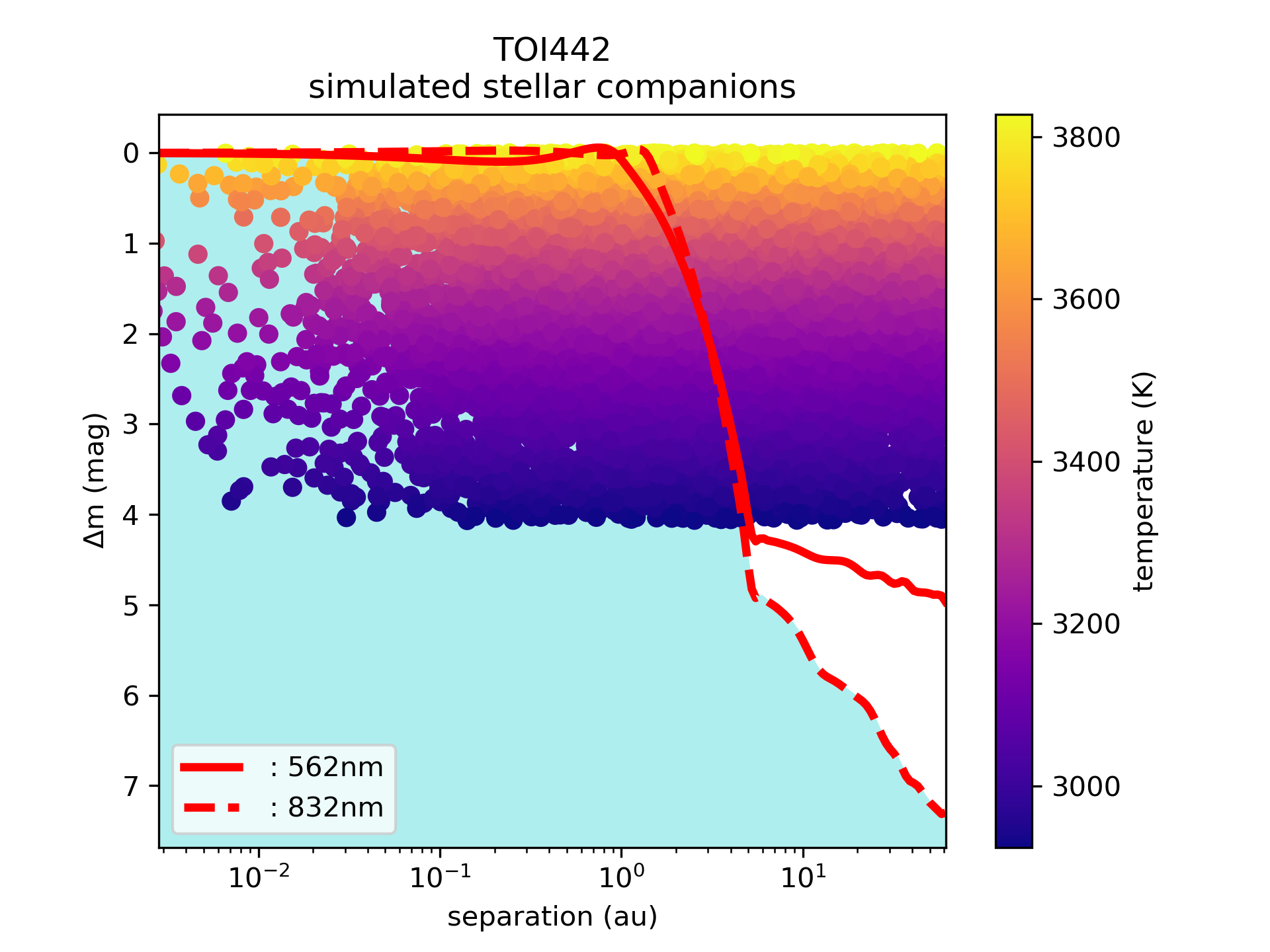}
	  \includegraphics[width=0.3\textwidth]{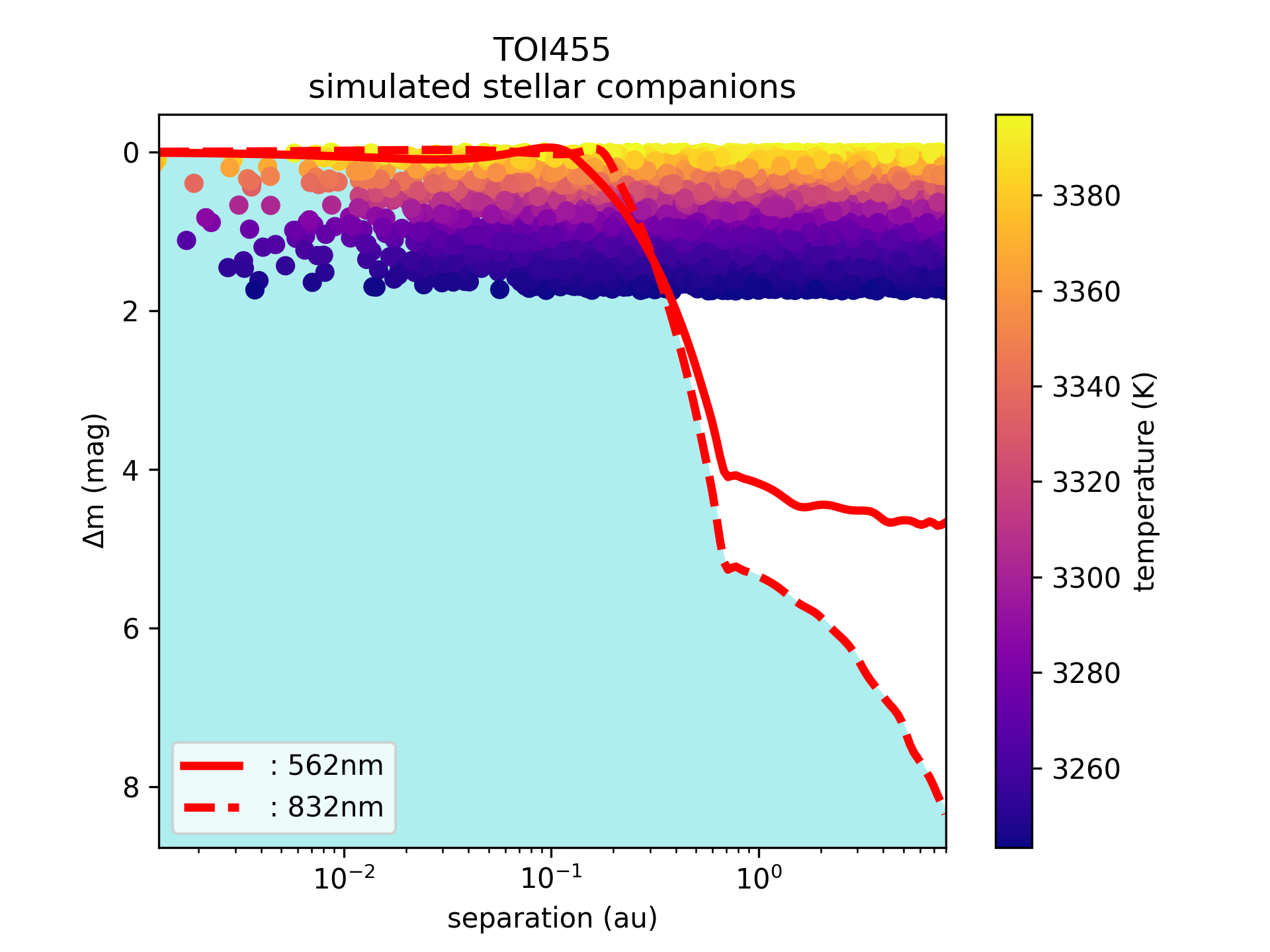}
	  \includegraphics[width=0.3\textwidth]{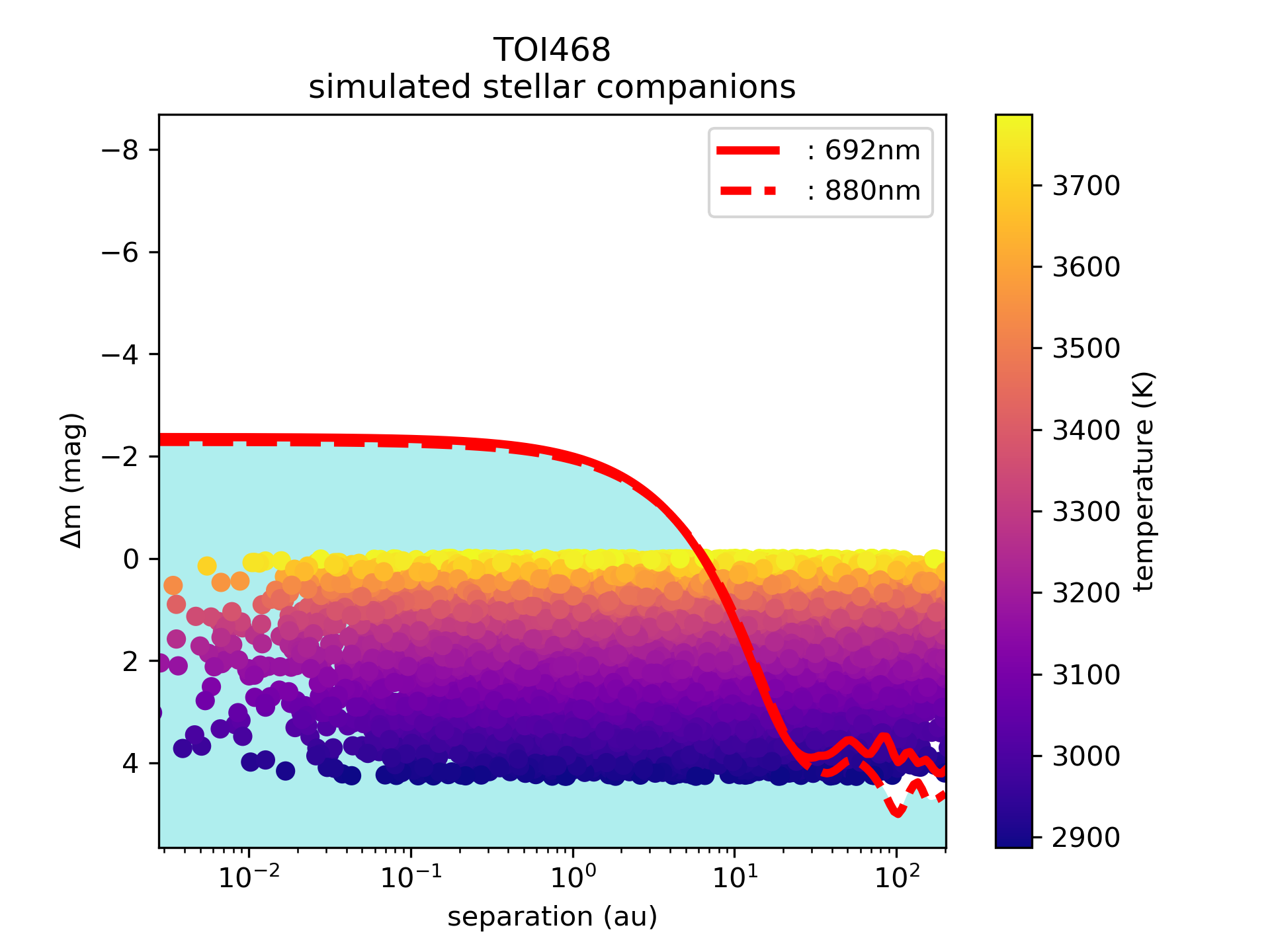}
	  \includegraphics[width=0.3\textwidth]{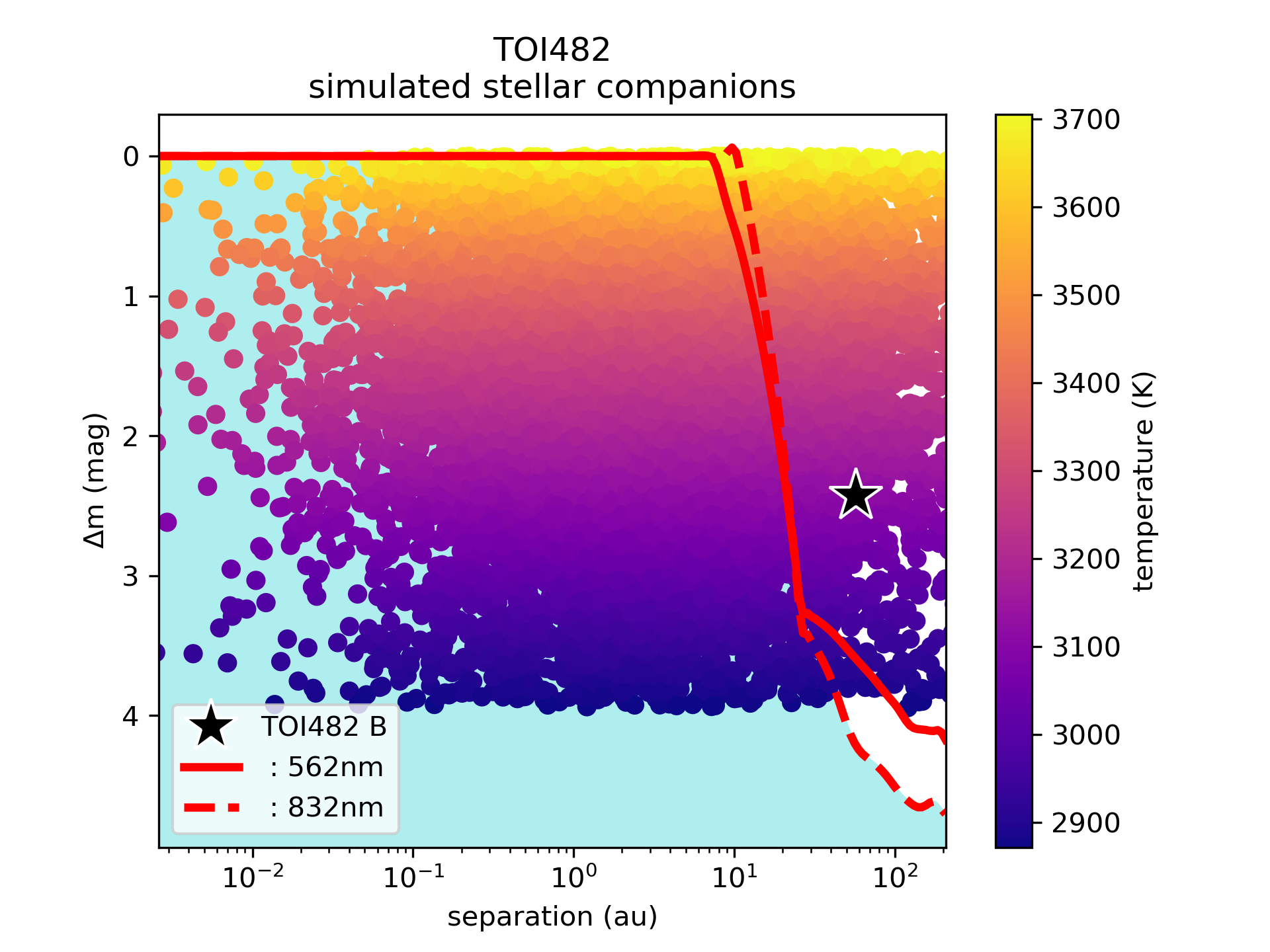}
	  \includegraphics[width=0.3\textwidth]{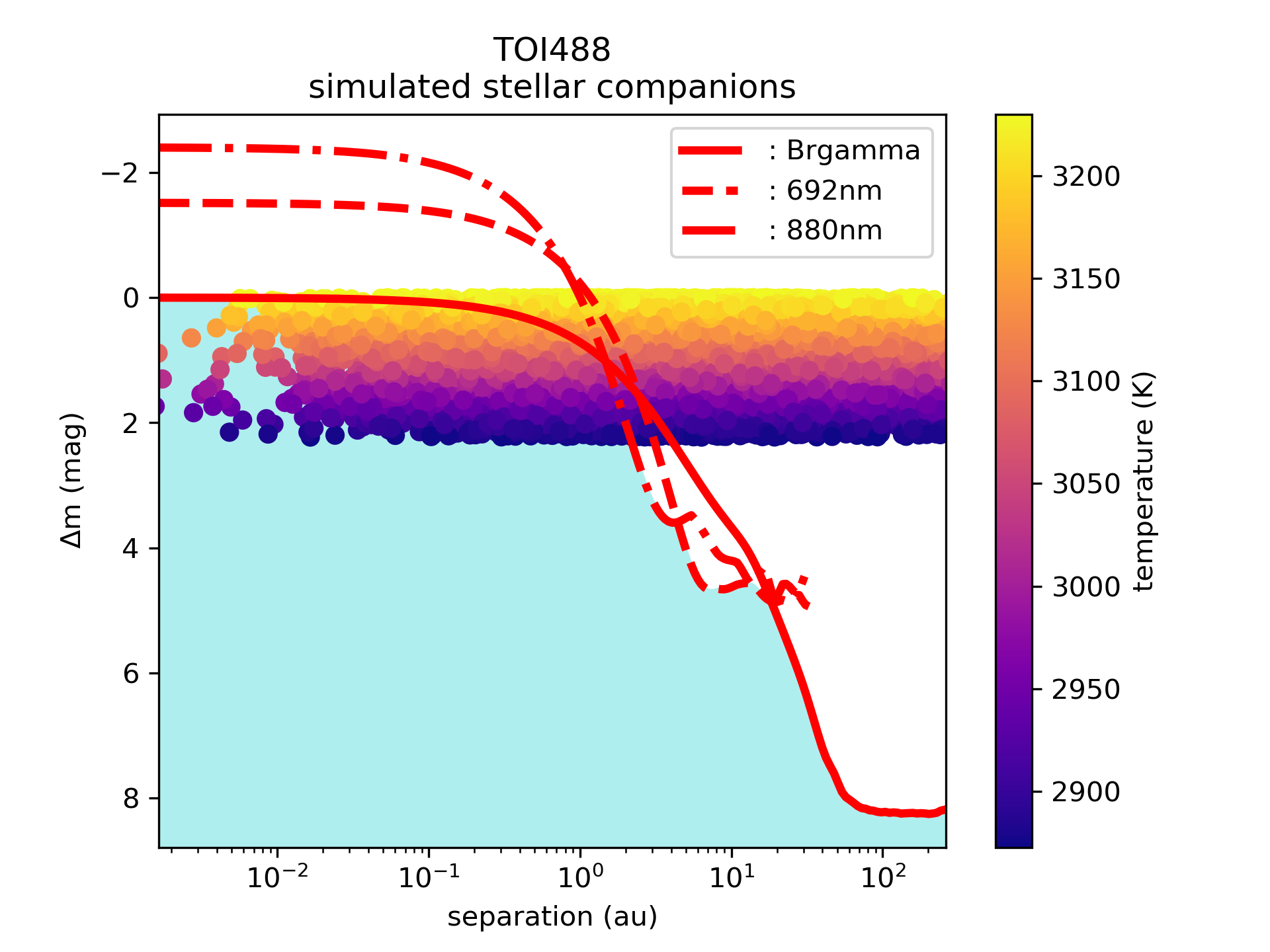}
	  \includegraphics[width=0.3\textwidth]{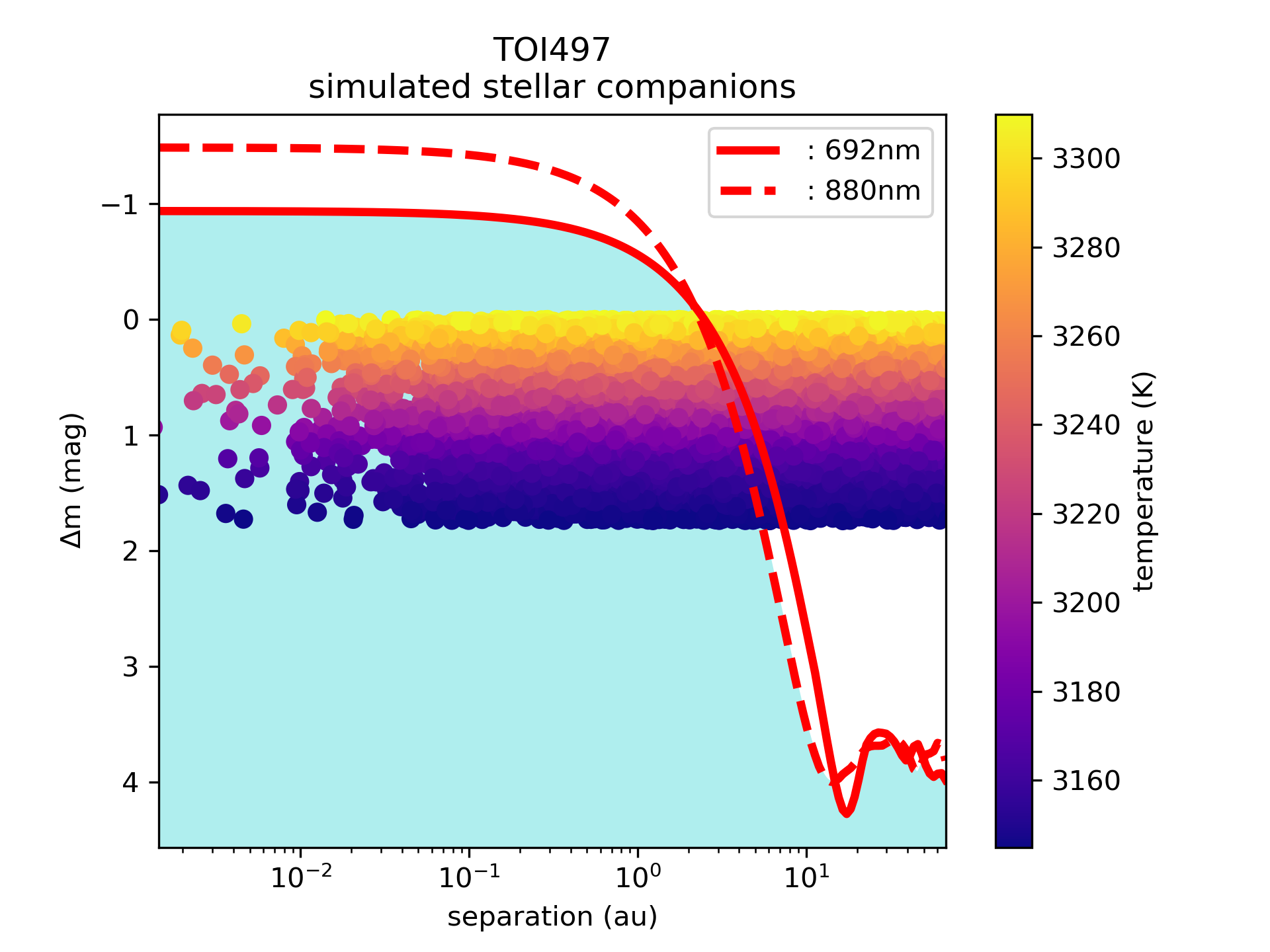}
	  \includegraphics[width=0.3\textwidth]{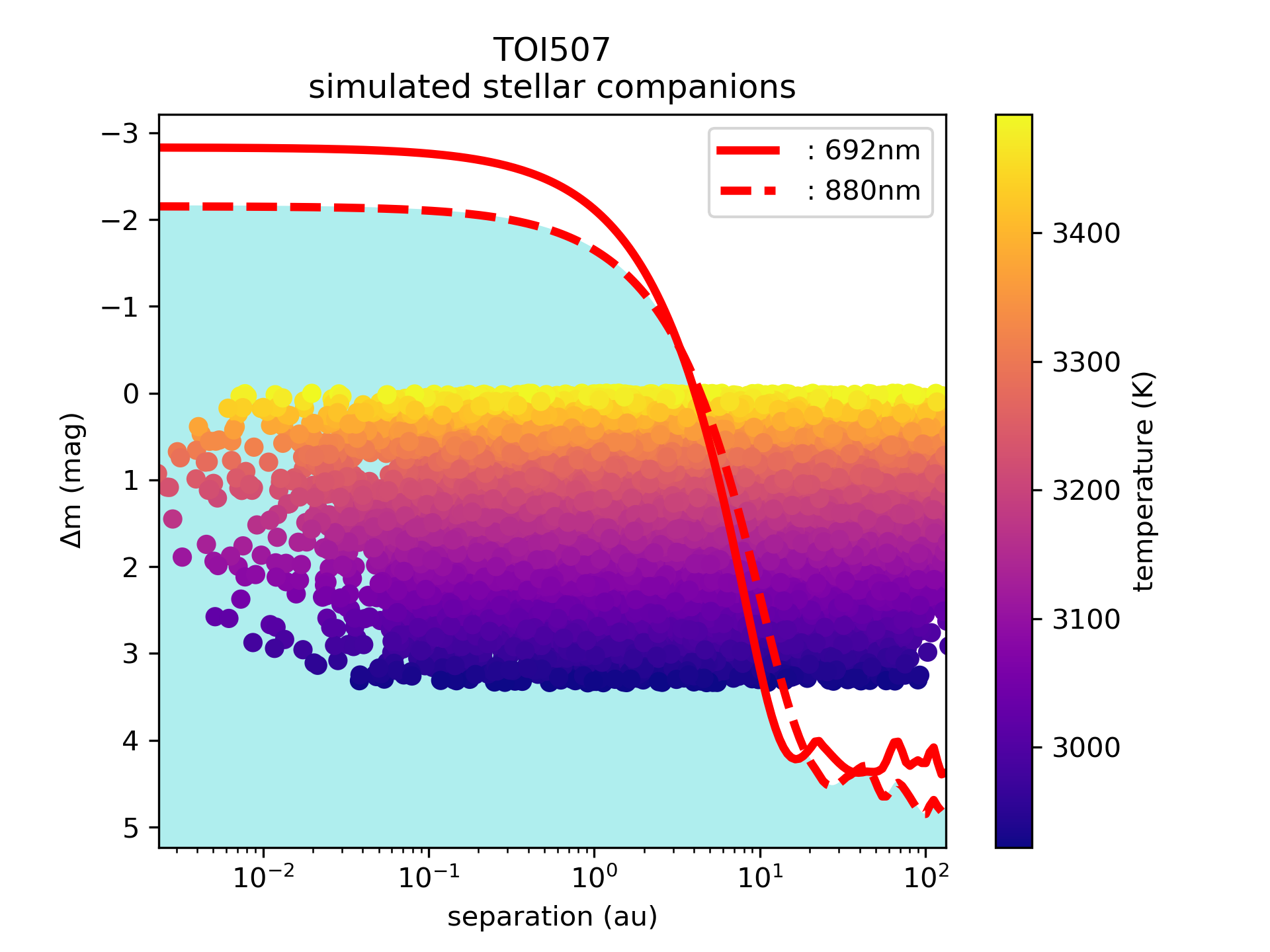}
	  \includegraphics[width=0.3\textwidth]{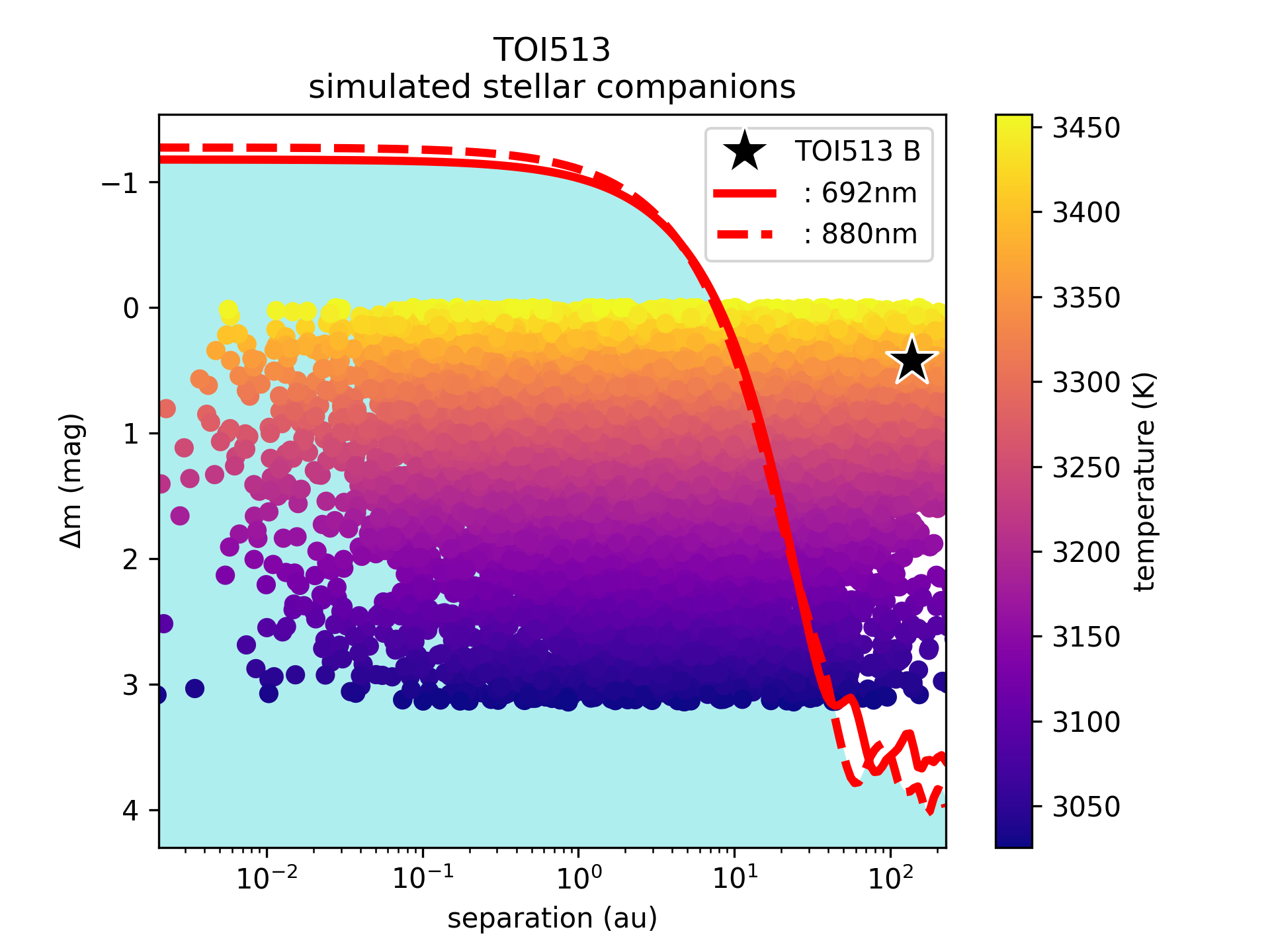}
	  \includegraphics[width=0.3\textwidth]{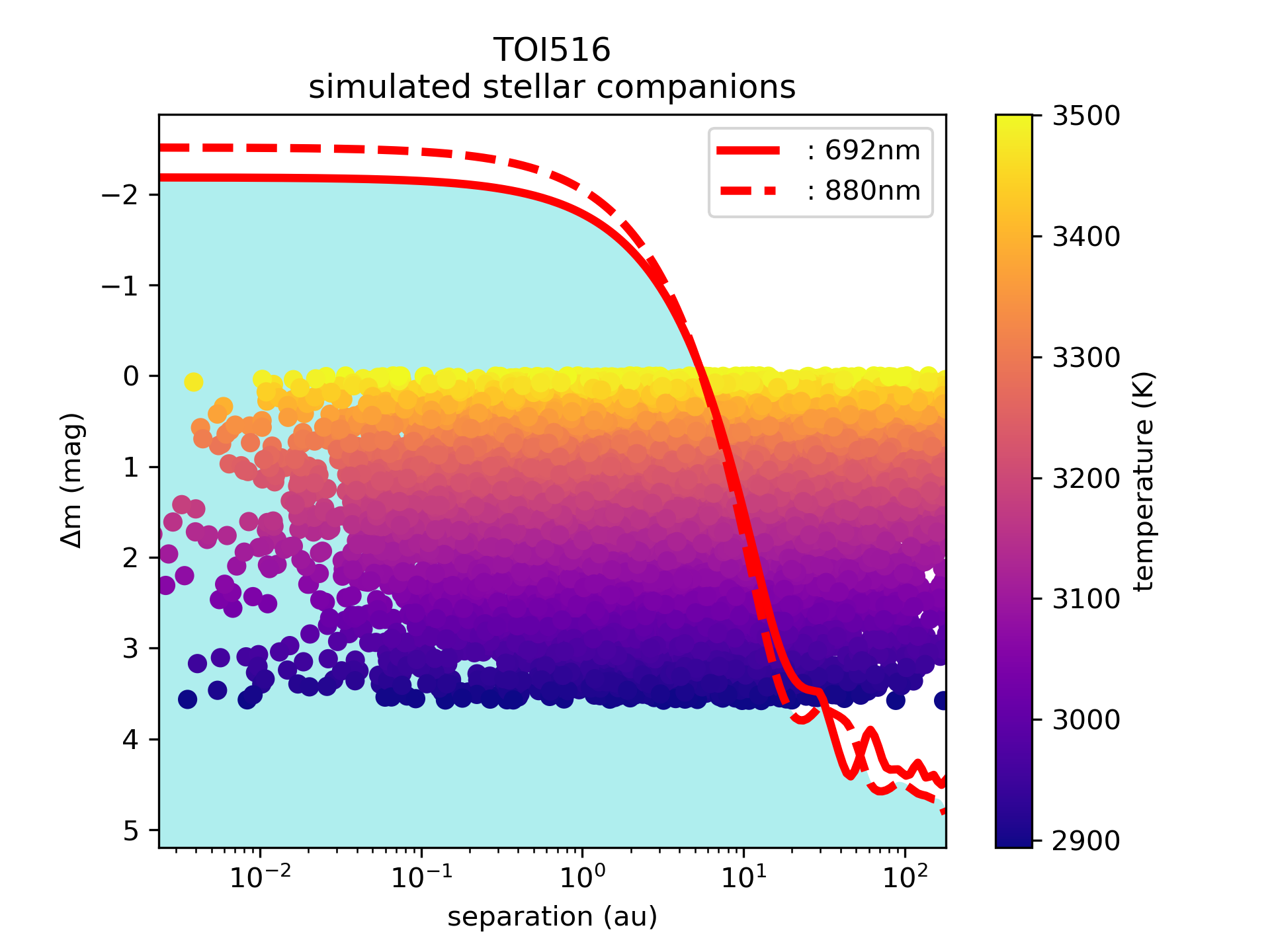}
  \end{center}
  \caption{Simulated Companions: TOI198 to TOI516}
  \label{fig:Sim_Comp_1}
\end{figure*}
\begin{figure*}[!htb]
  \begin{center}
  	  \includegraphics[width=0.3\textwidth]{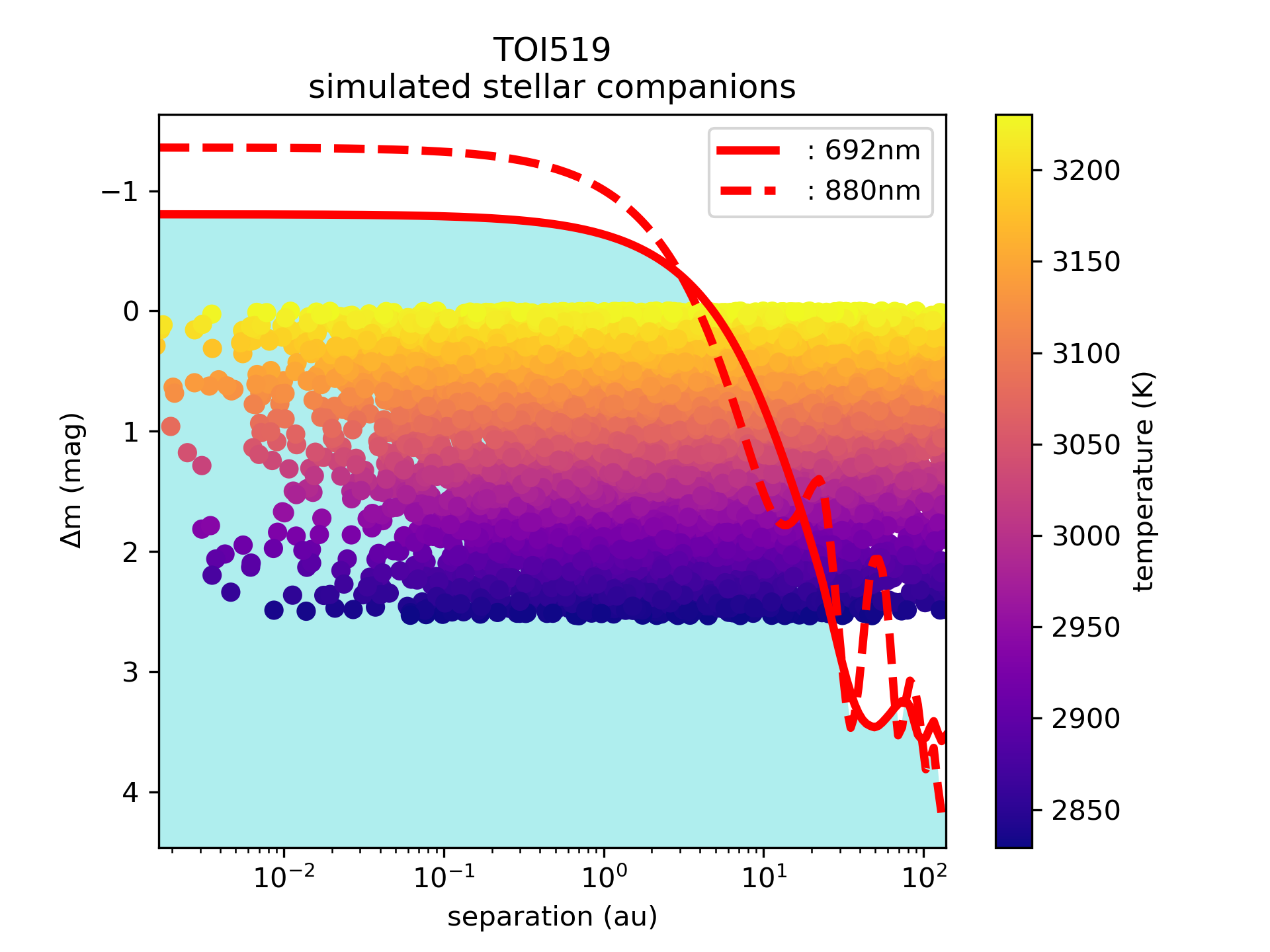}
	  \includegraphics[width=0.3\textwidth]{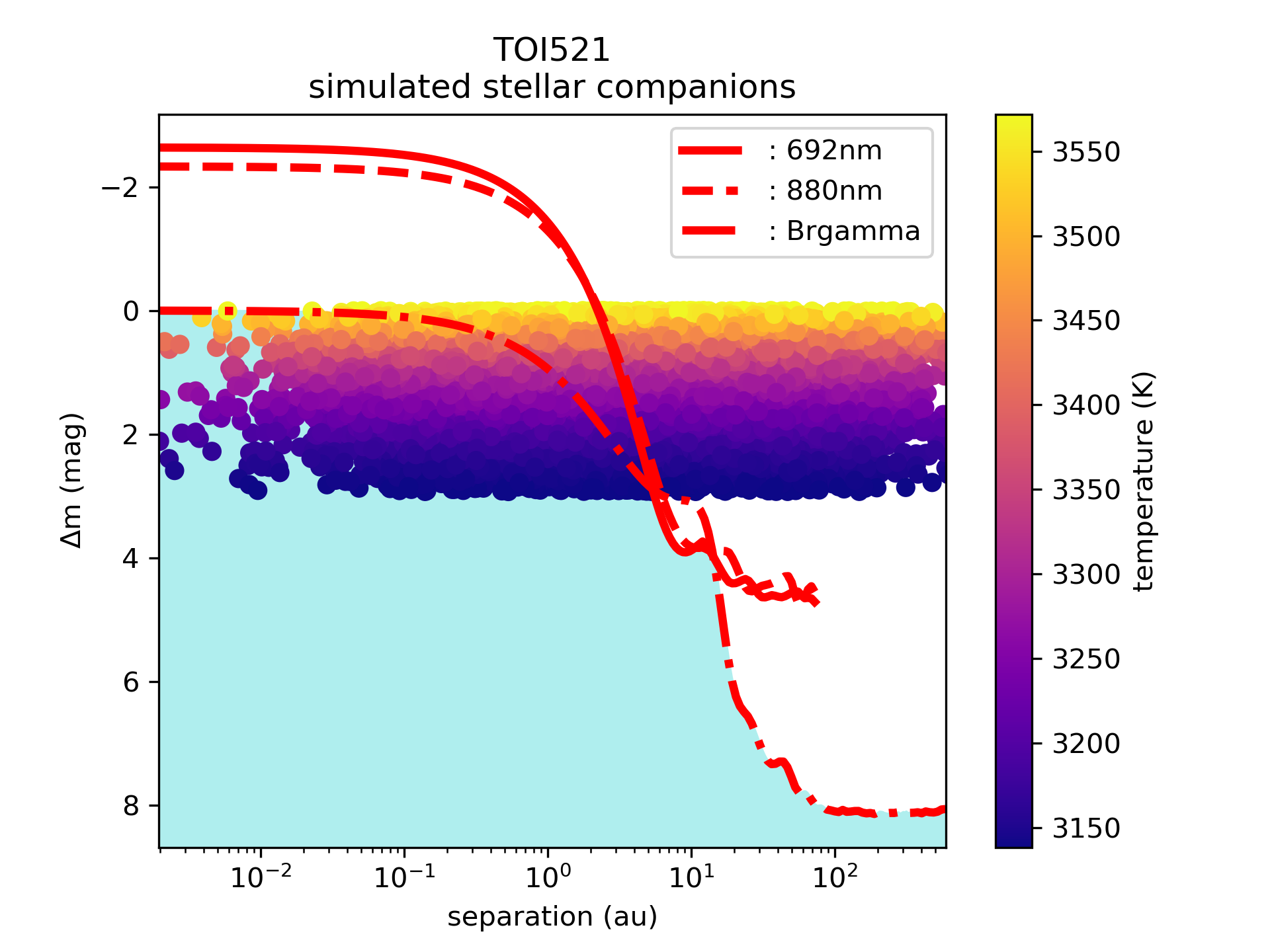}
	  \includegraphics[width=0.3\textwidth]{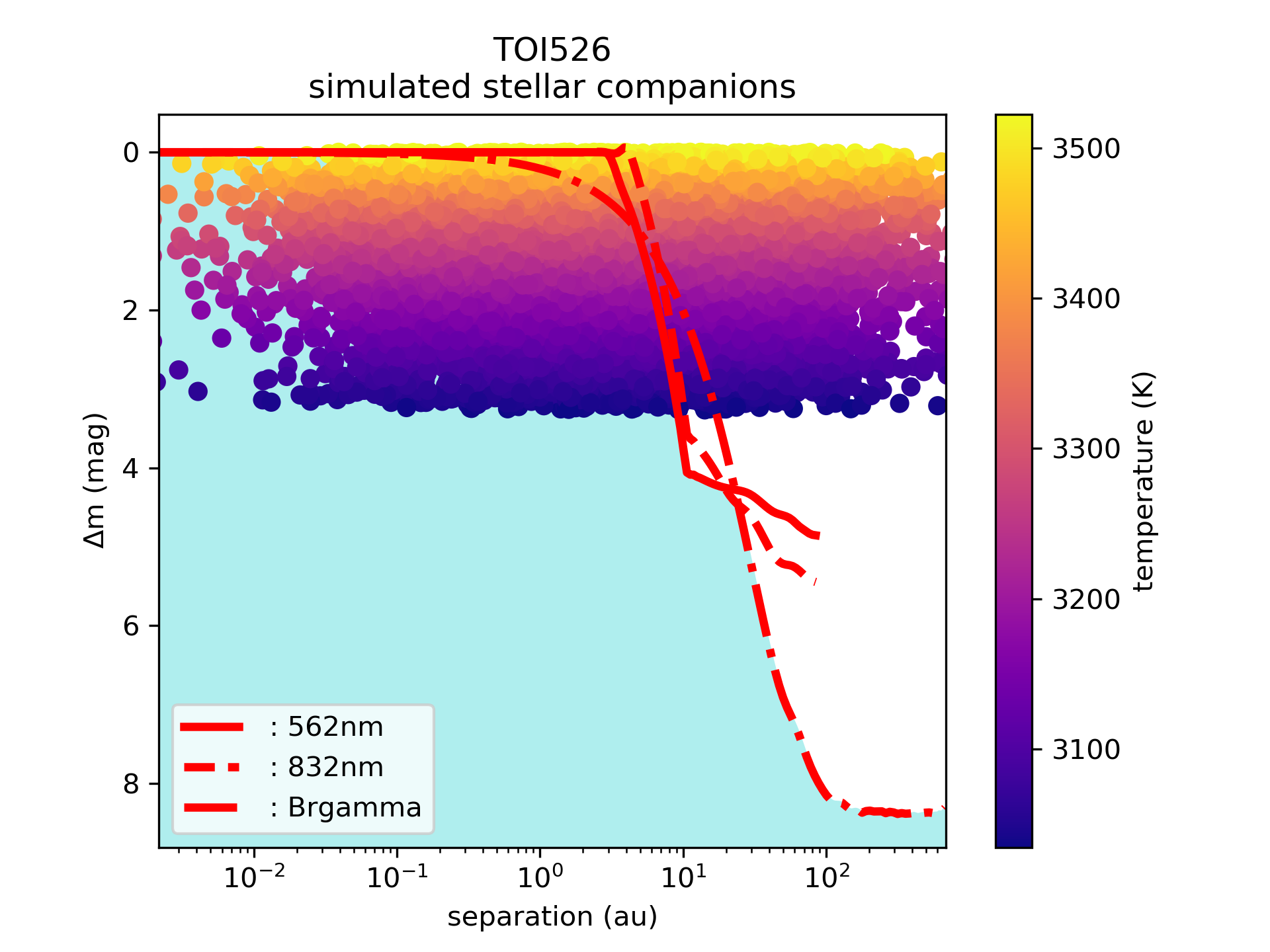}
	  \includegraphics[width=0.3\textwidth]{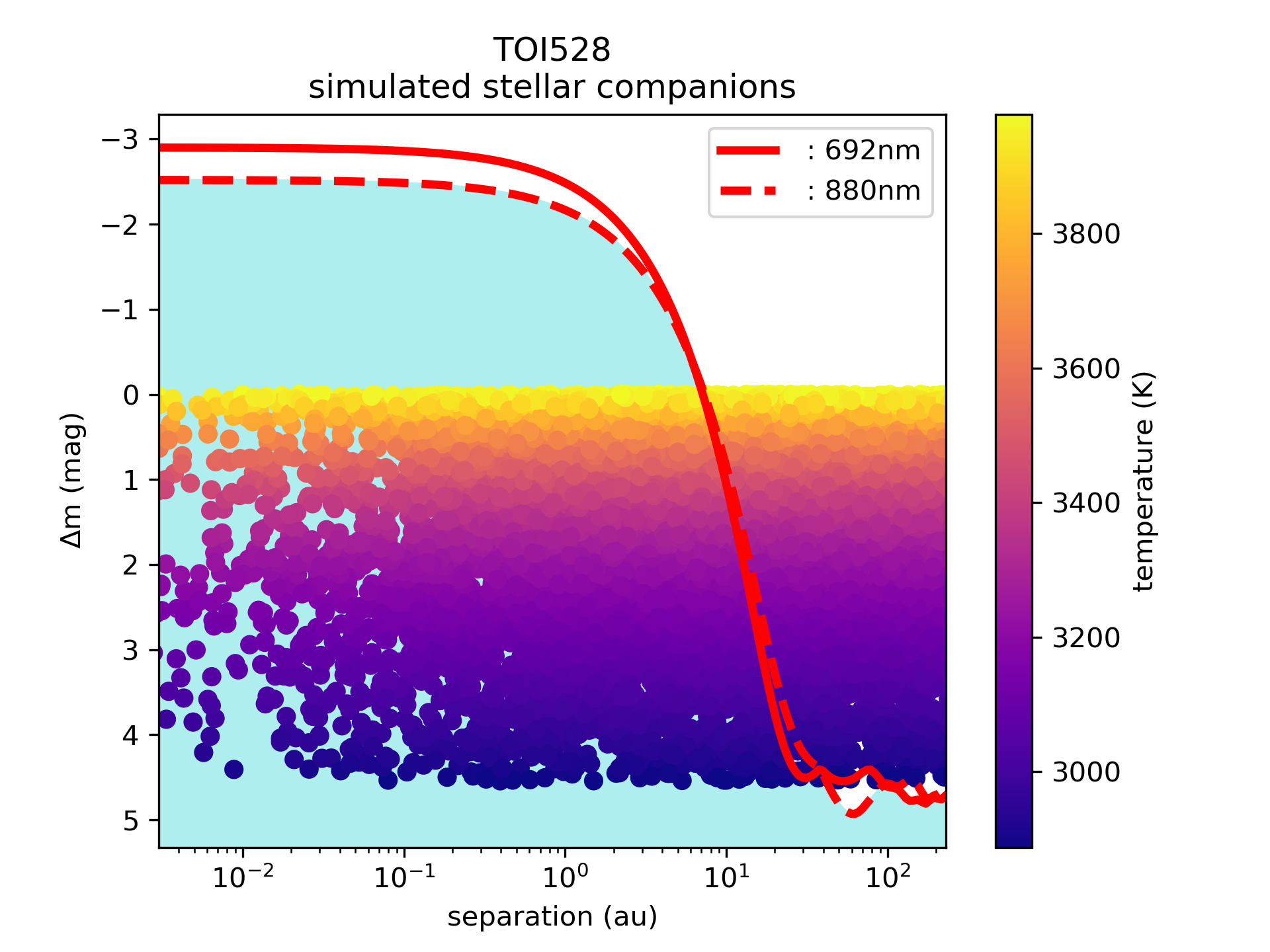}
	  \includegraphics[width=0.3\textwidth]{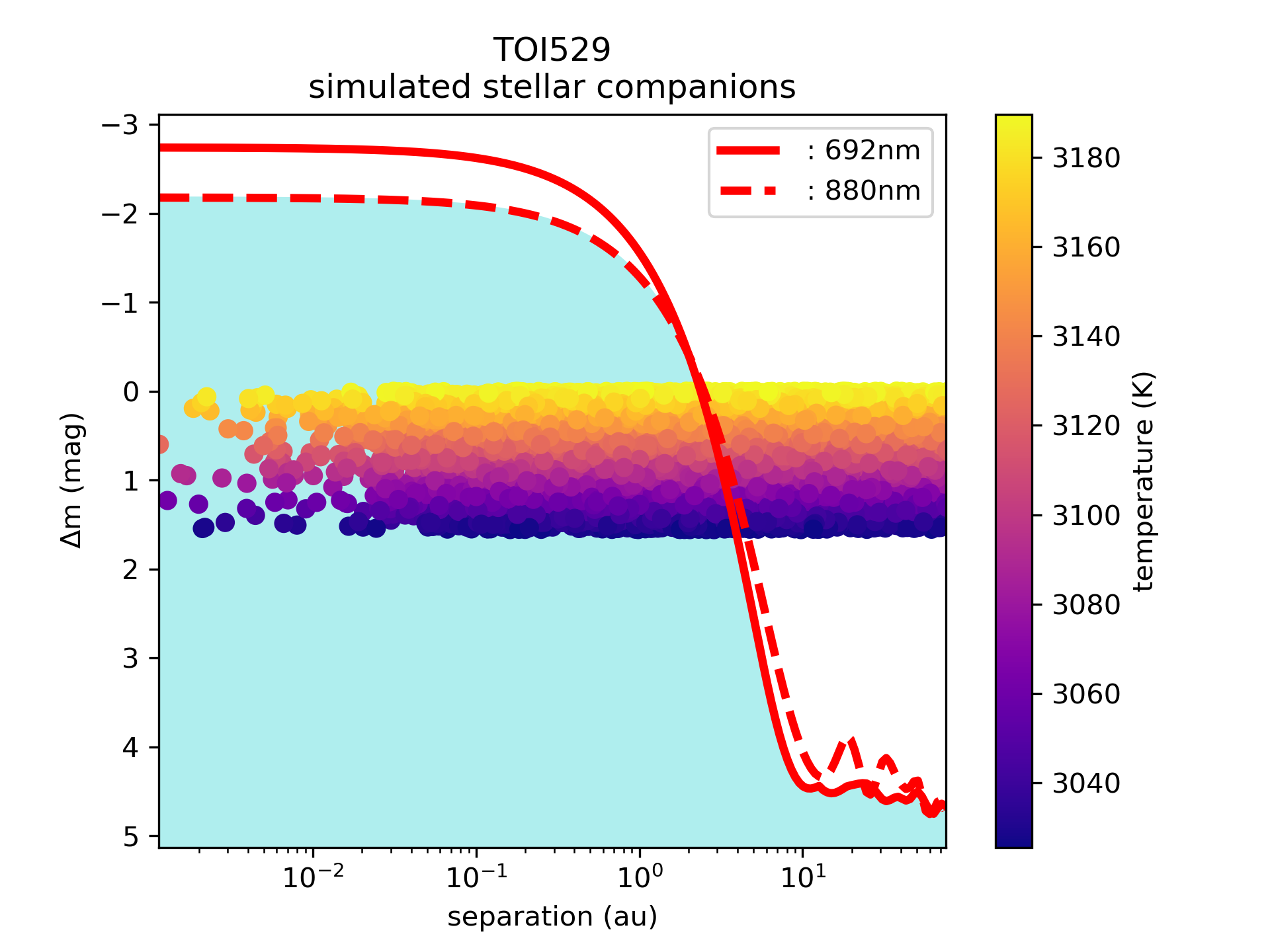}
	  \includegraphics[width=0.3\textwidth]{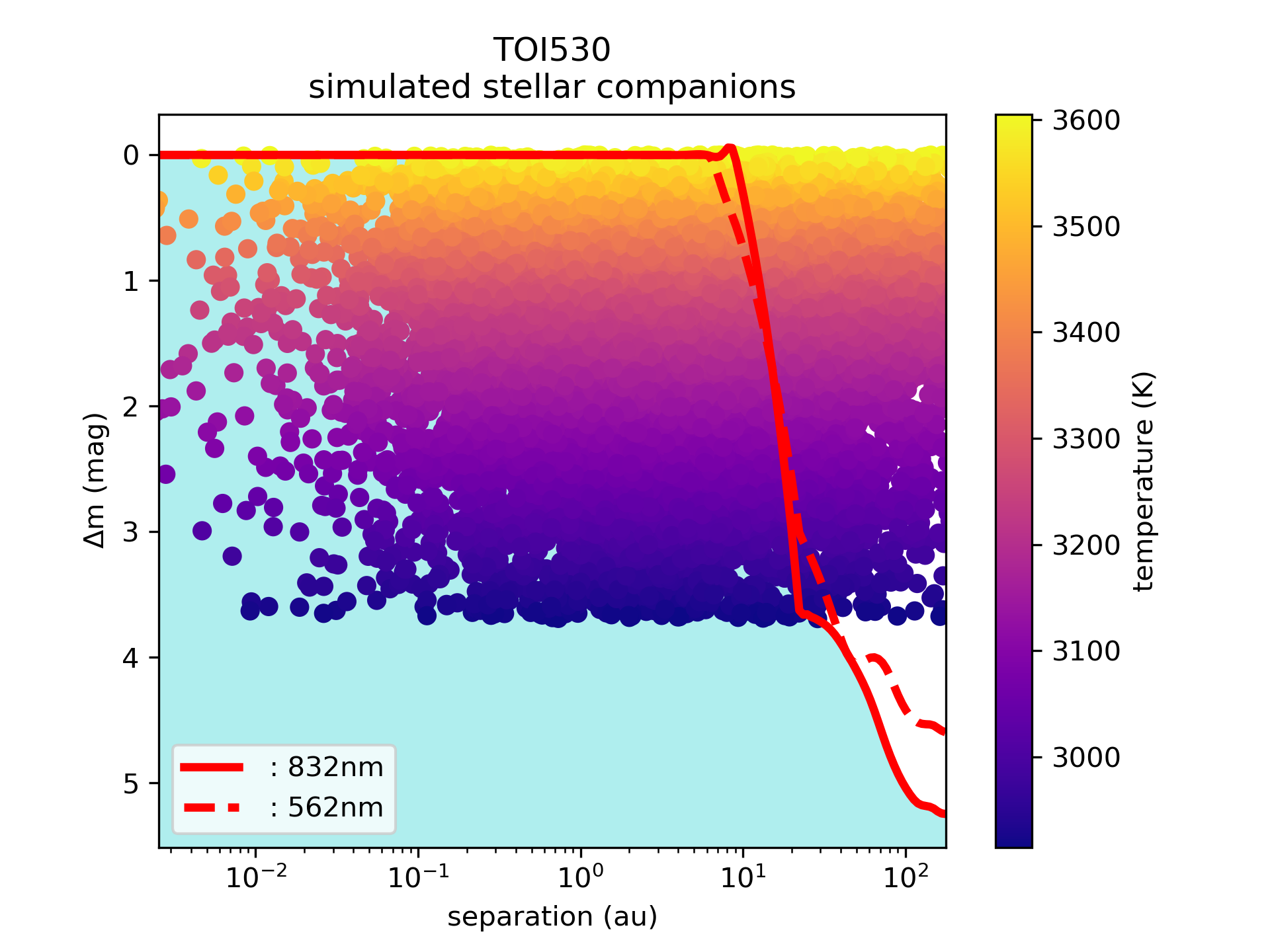}
	  \includegraphics[width=0.3\textwidth]{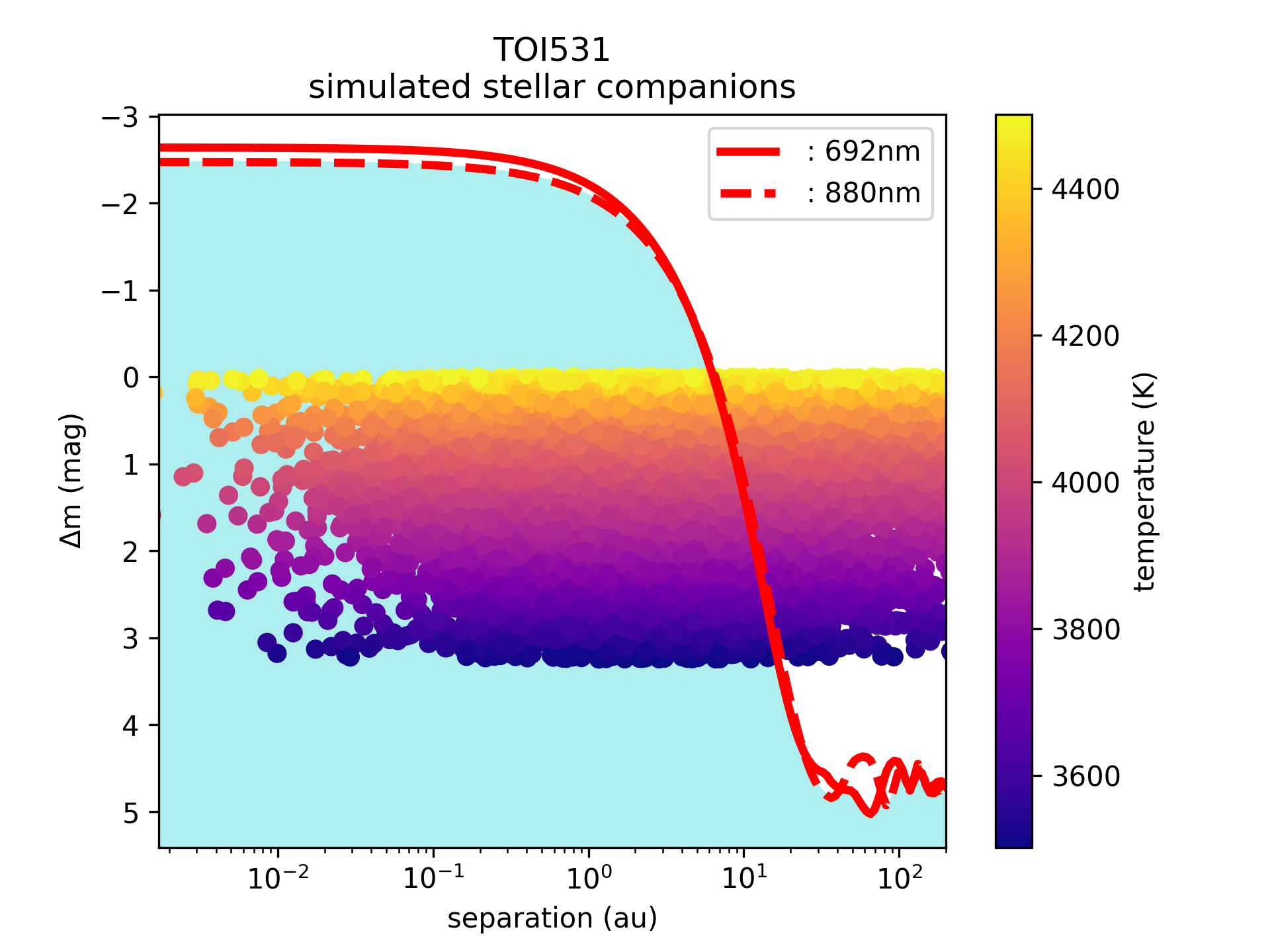}
	  \includegraphics[width=0.3\textwidth]{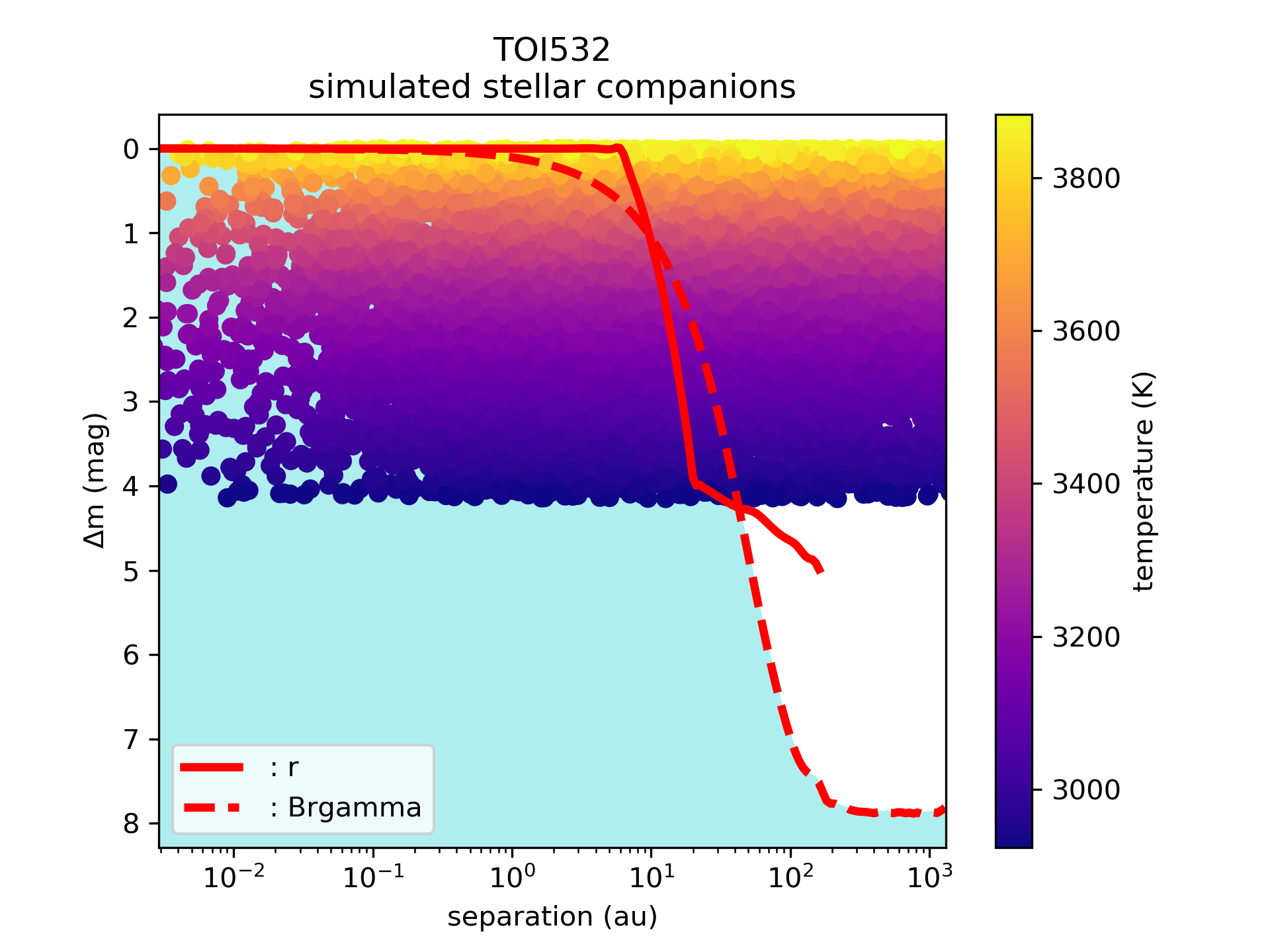}
	  \includegraphics[width=0.3\textwidth]{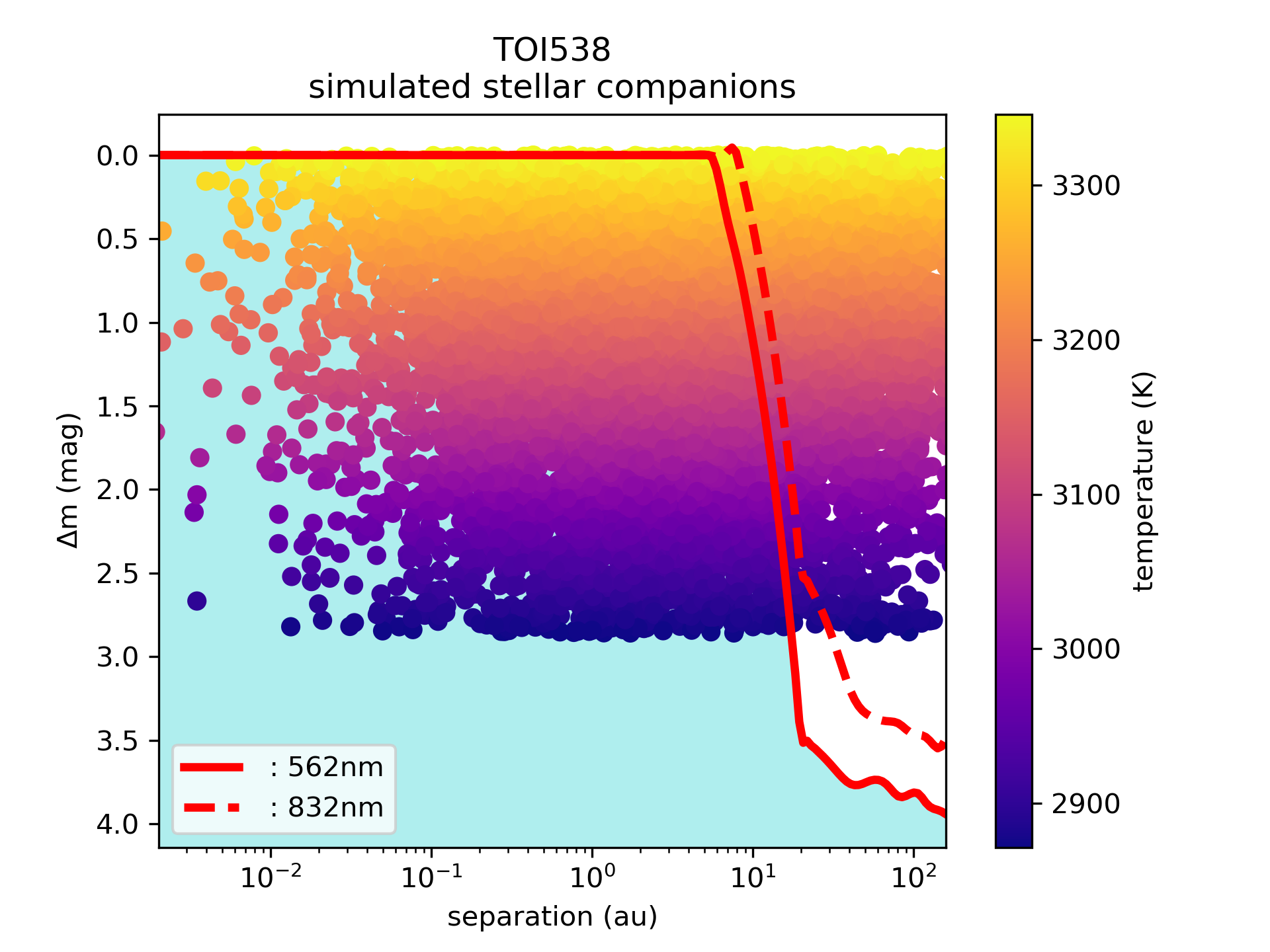}
	  \includegraphics[width=0.3\textwidth]{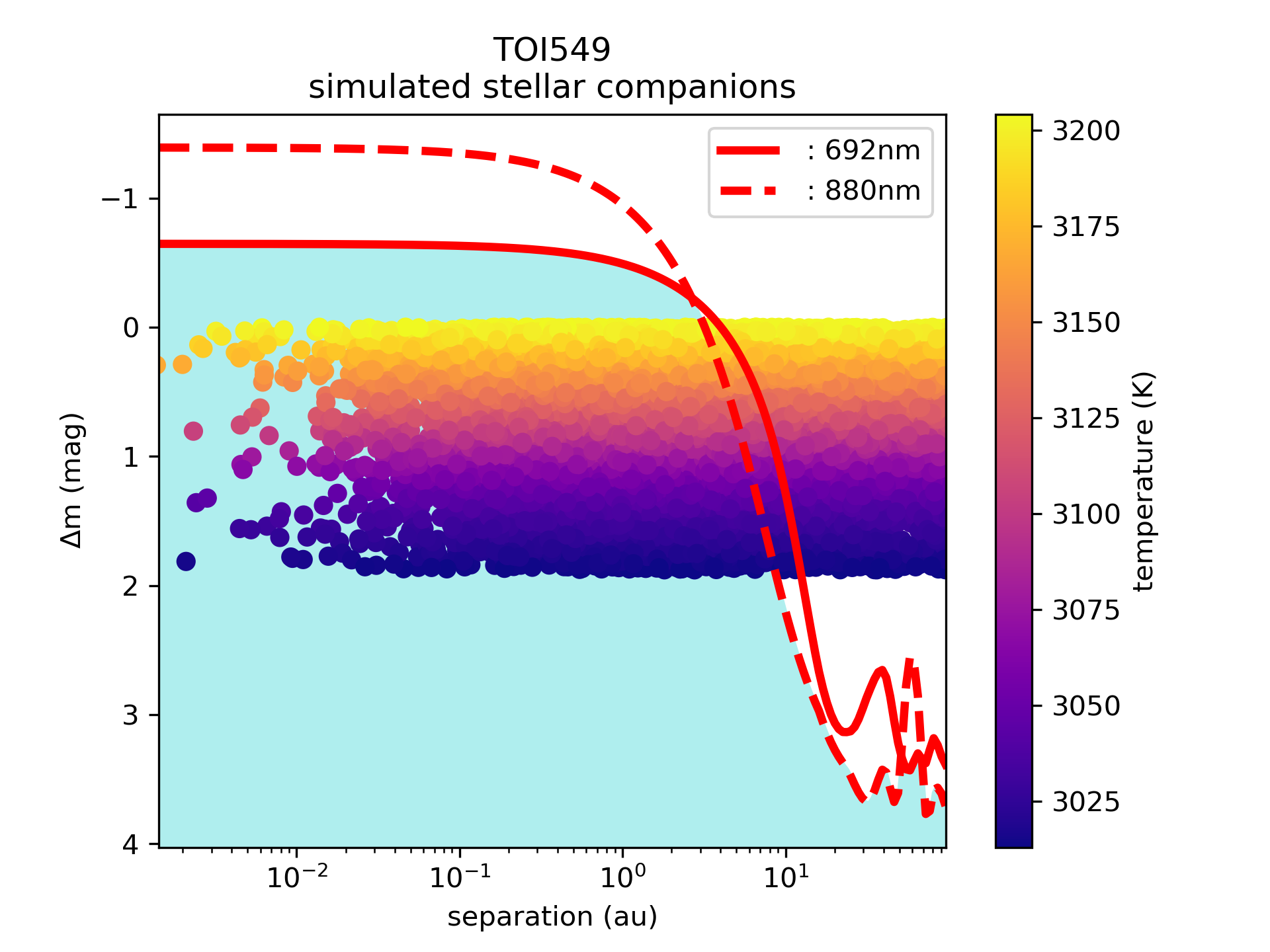}
	  \includegraphics[width=0.3\textwidth]{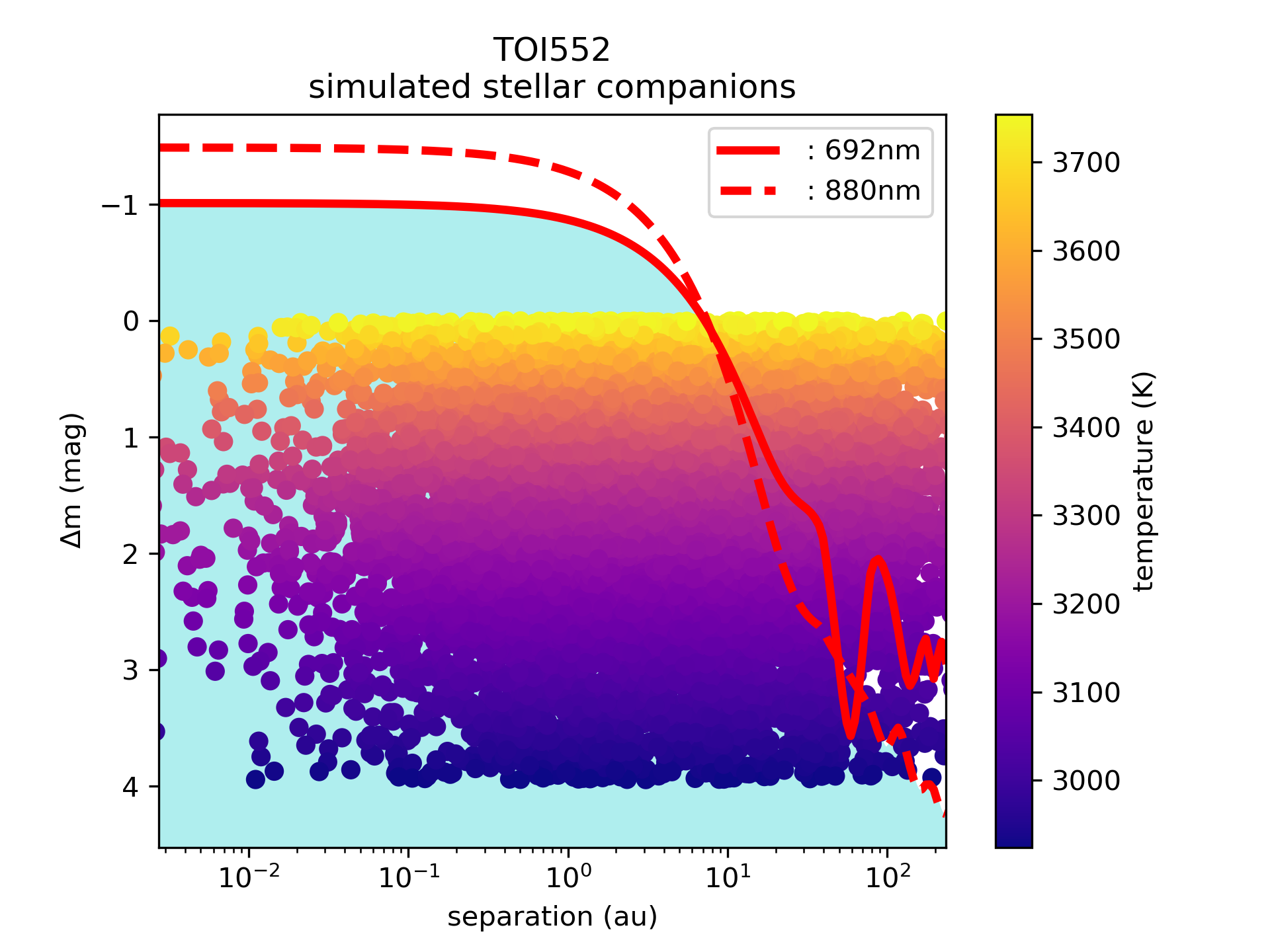}
	  \includegraphics[width=0.3\textwidth]{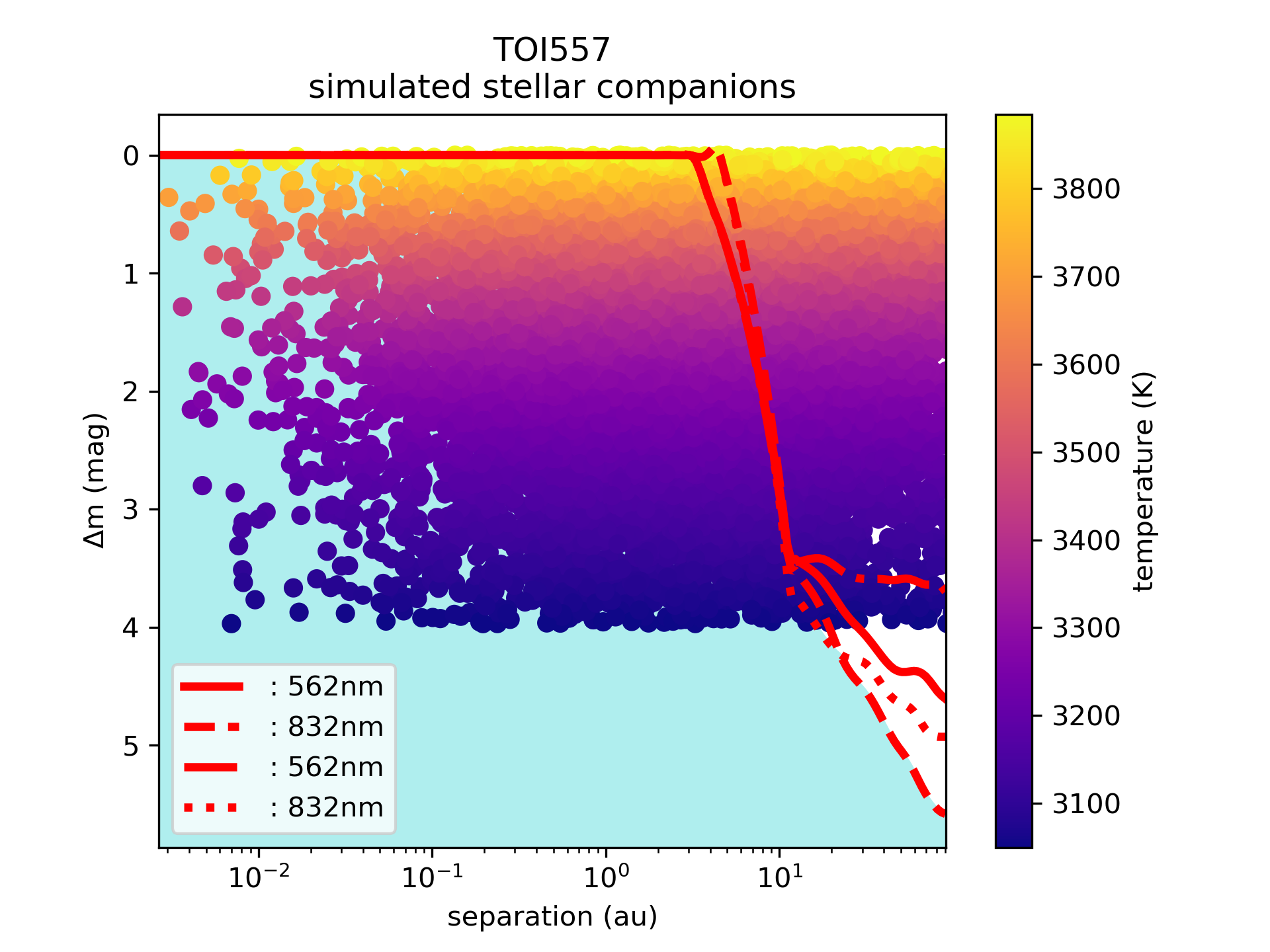}
	  \includegraphics[width=0.3\textwidth]{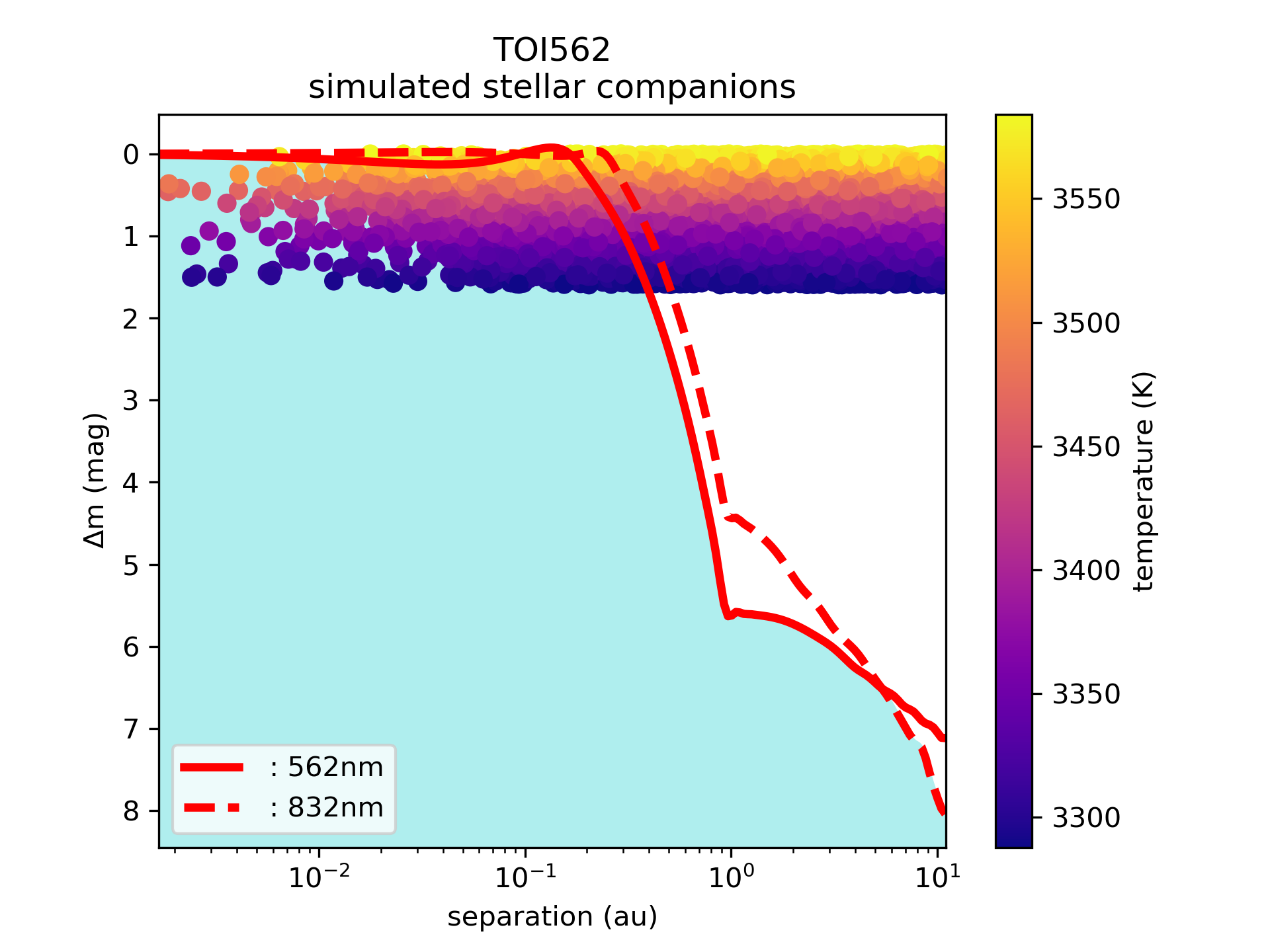}
	  \includegraphics[width=0.3\textwidth]{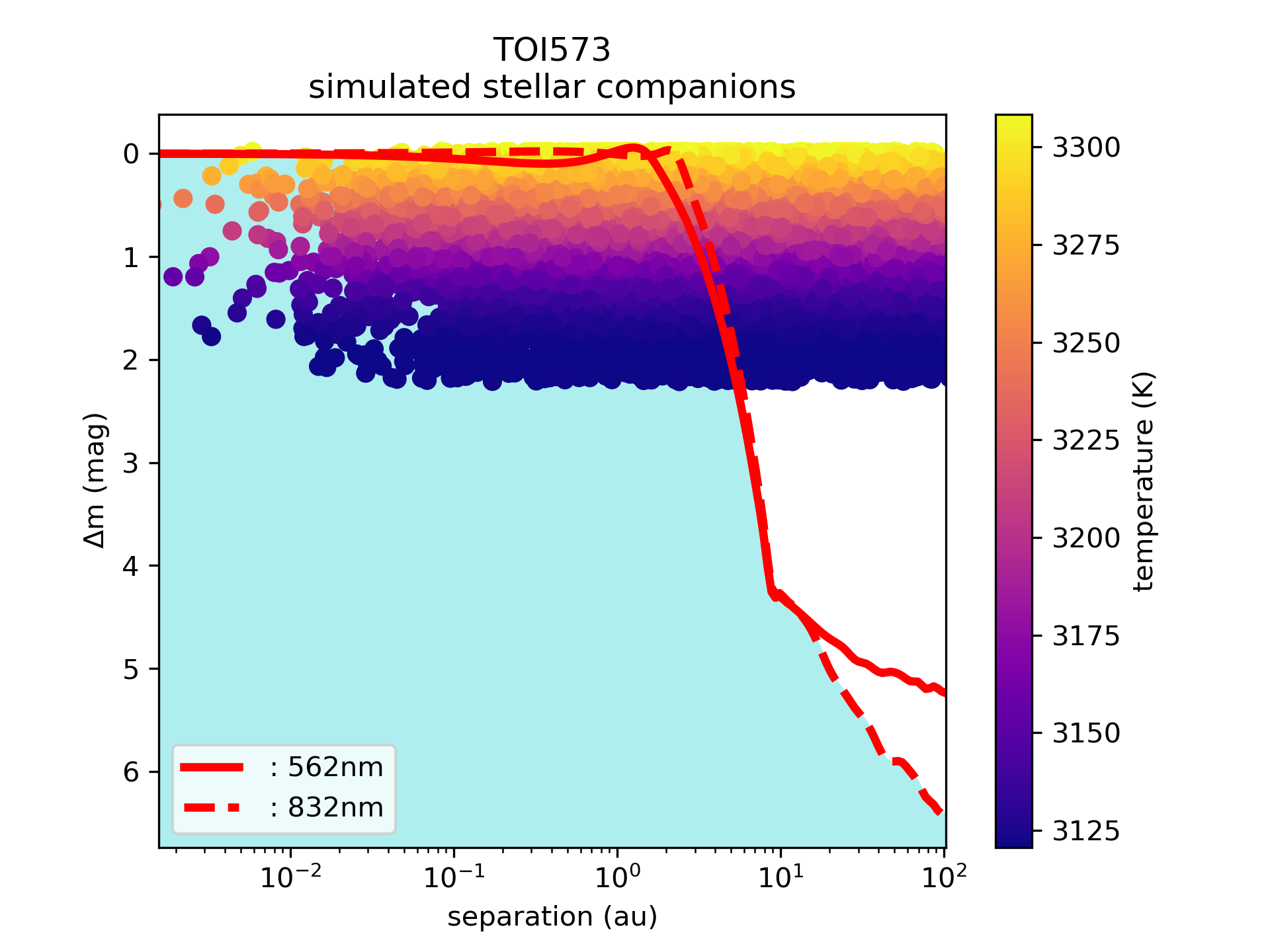}
	  \includegraphics[width=0.3\textwidth]{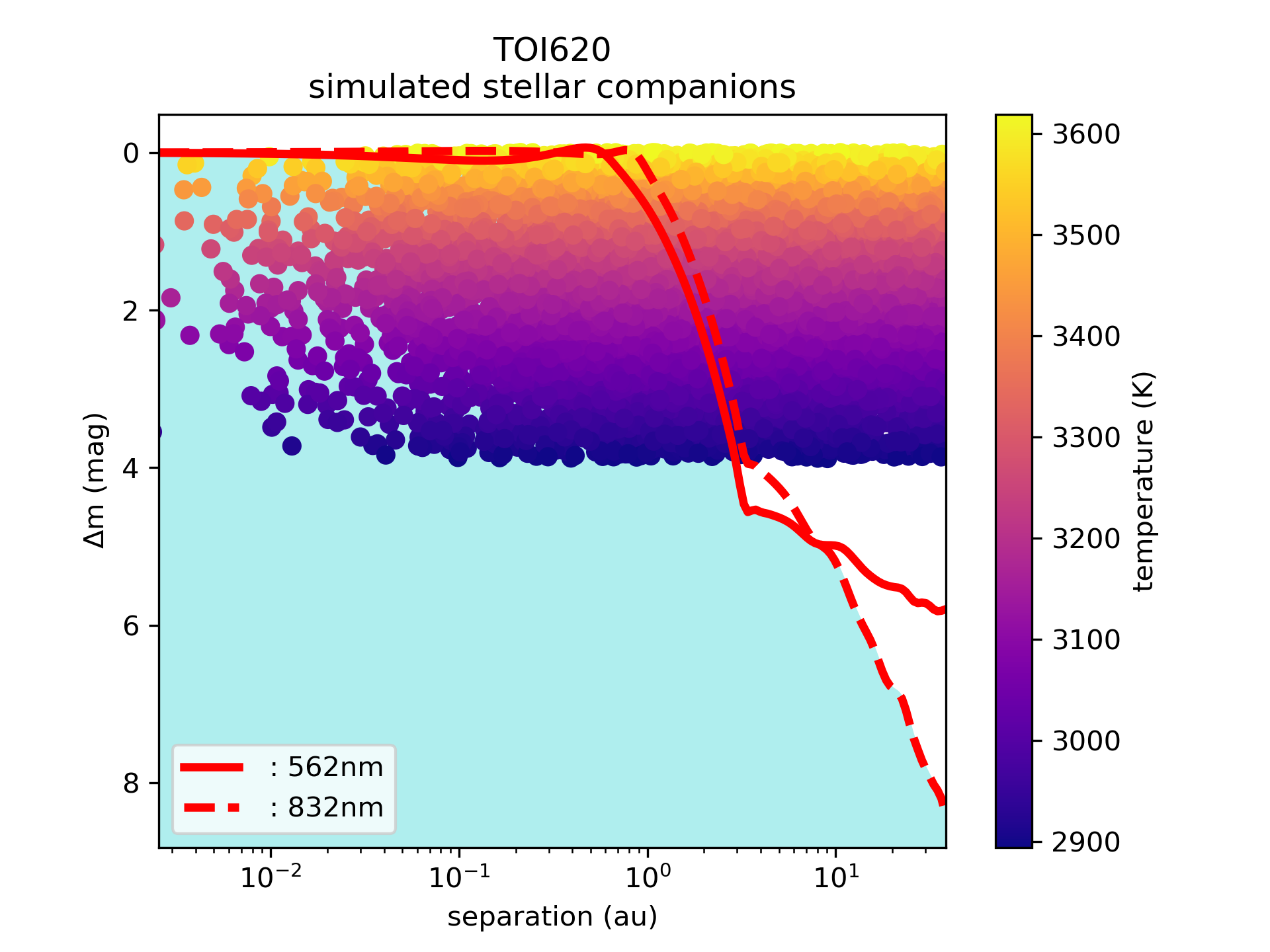}
  \end{center}
  \caption{Simulated Companions: TOI519 to TOI620}
  \label{fig:Sim_Comp_2}
\end{figure*}
\begin{figure*}[!htb]
  \begin{center}
  	  \includegraphics[width=0.3\textwidth]{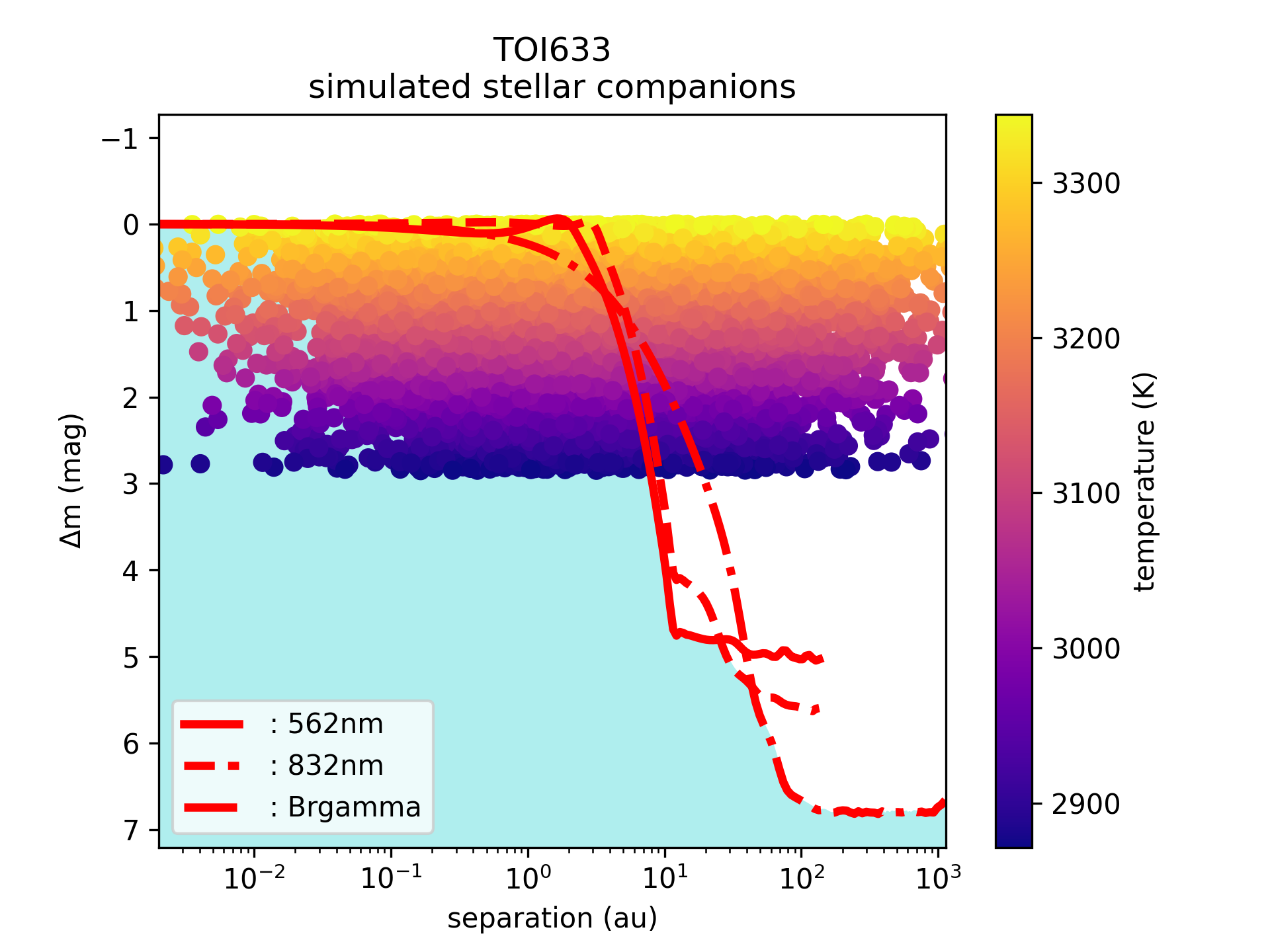}
	  \includegraphics[width=0.3\textwidth]{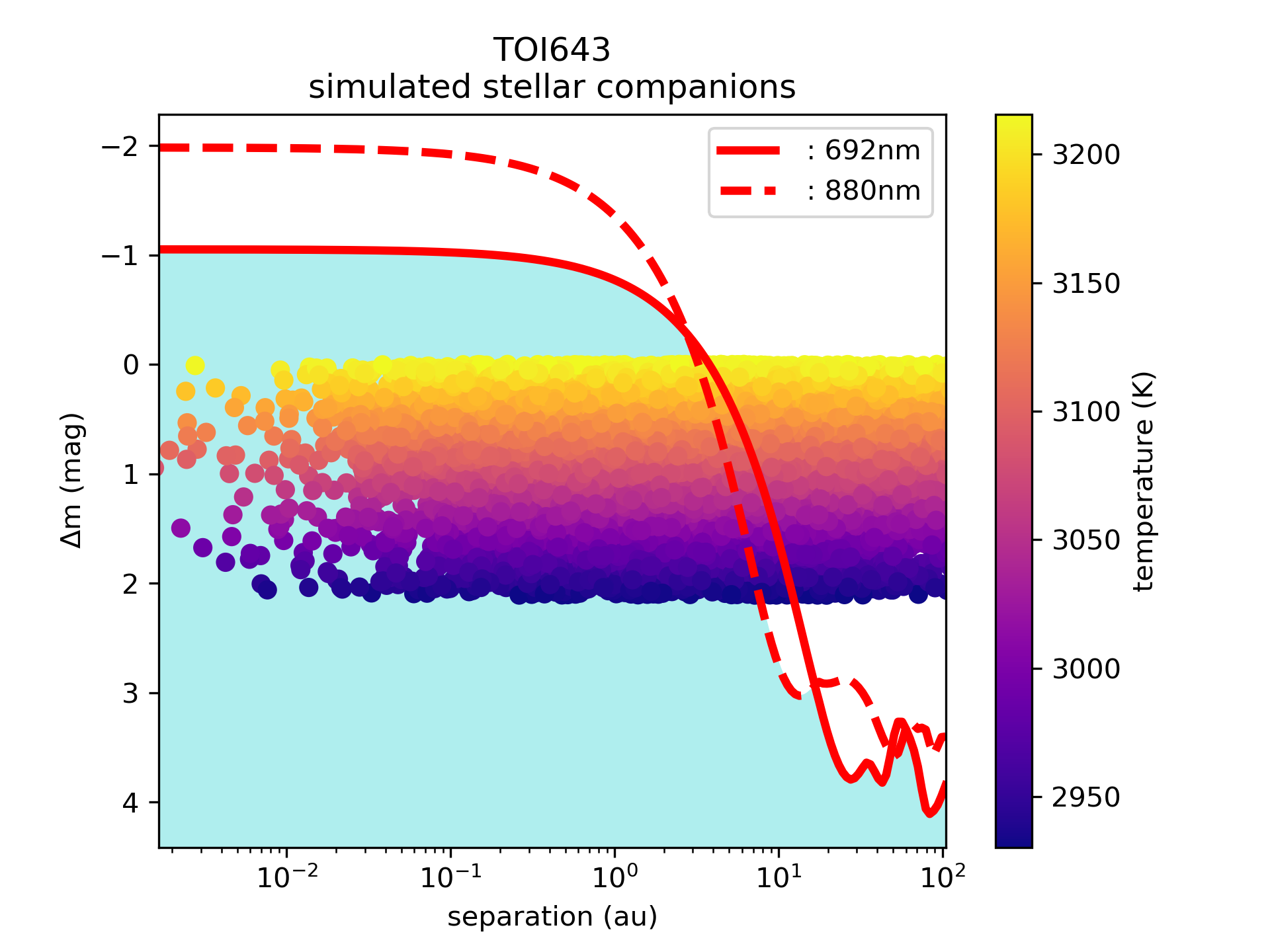}
	  \includegraphics[width=0.3\textwidth]{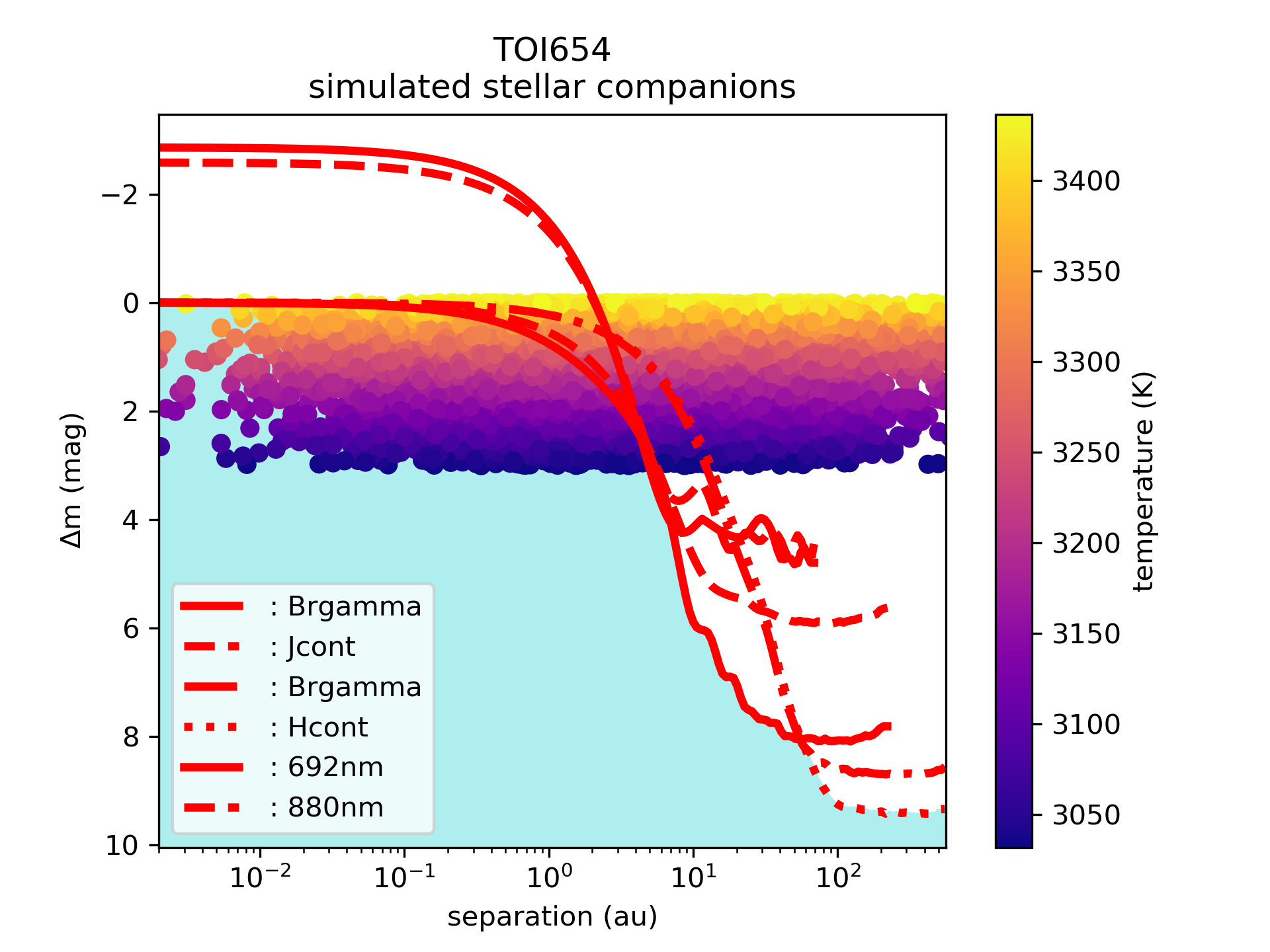}
	  \includegraphics[width=0.3\textwidth]{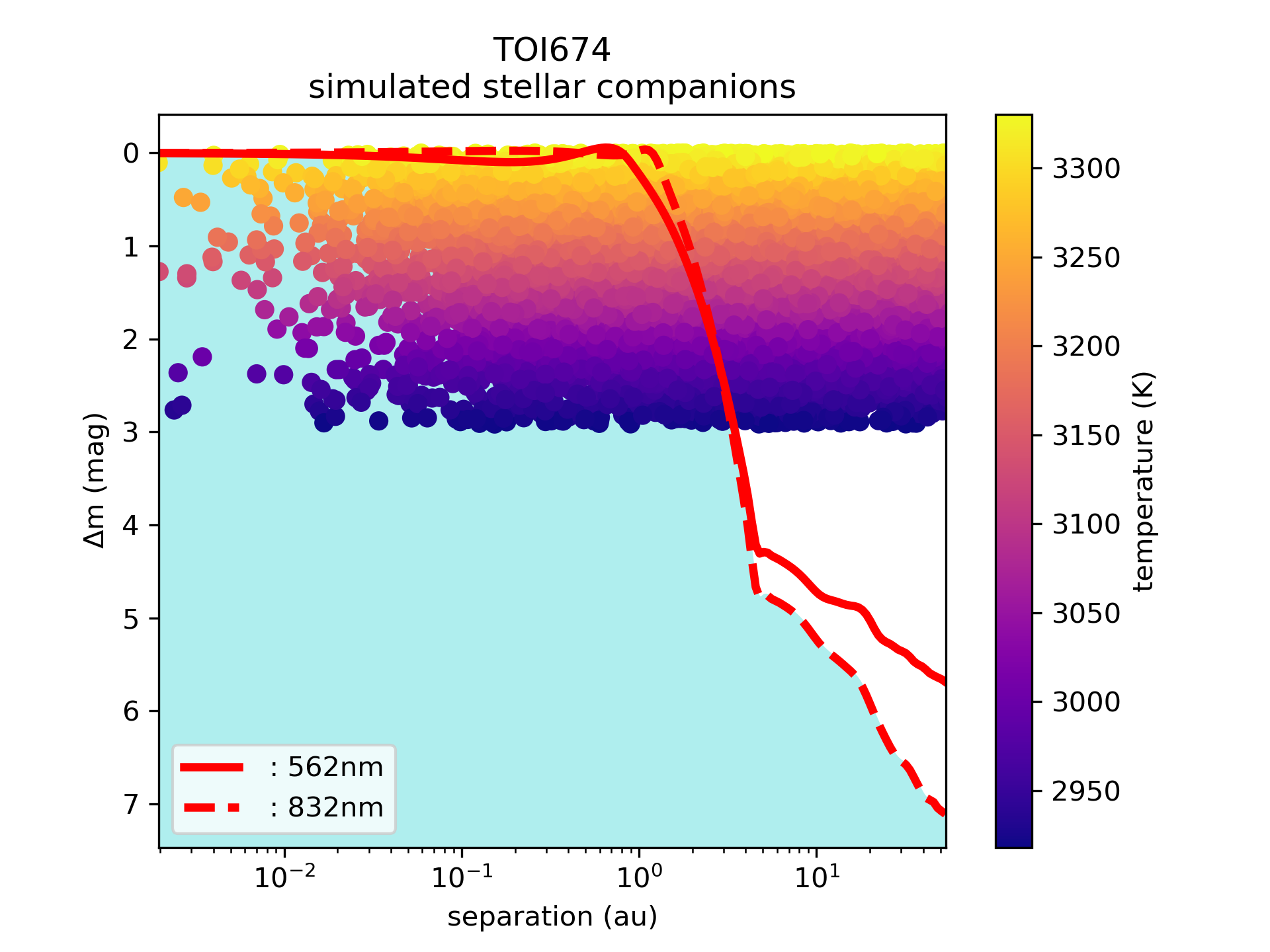}
	  \includegraphics[width=0.3\textwidth]{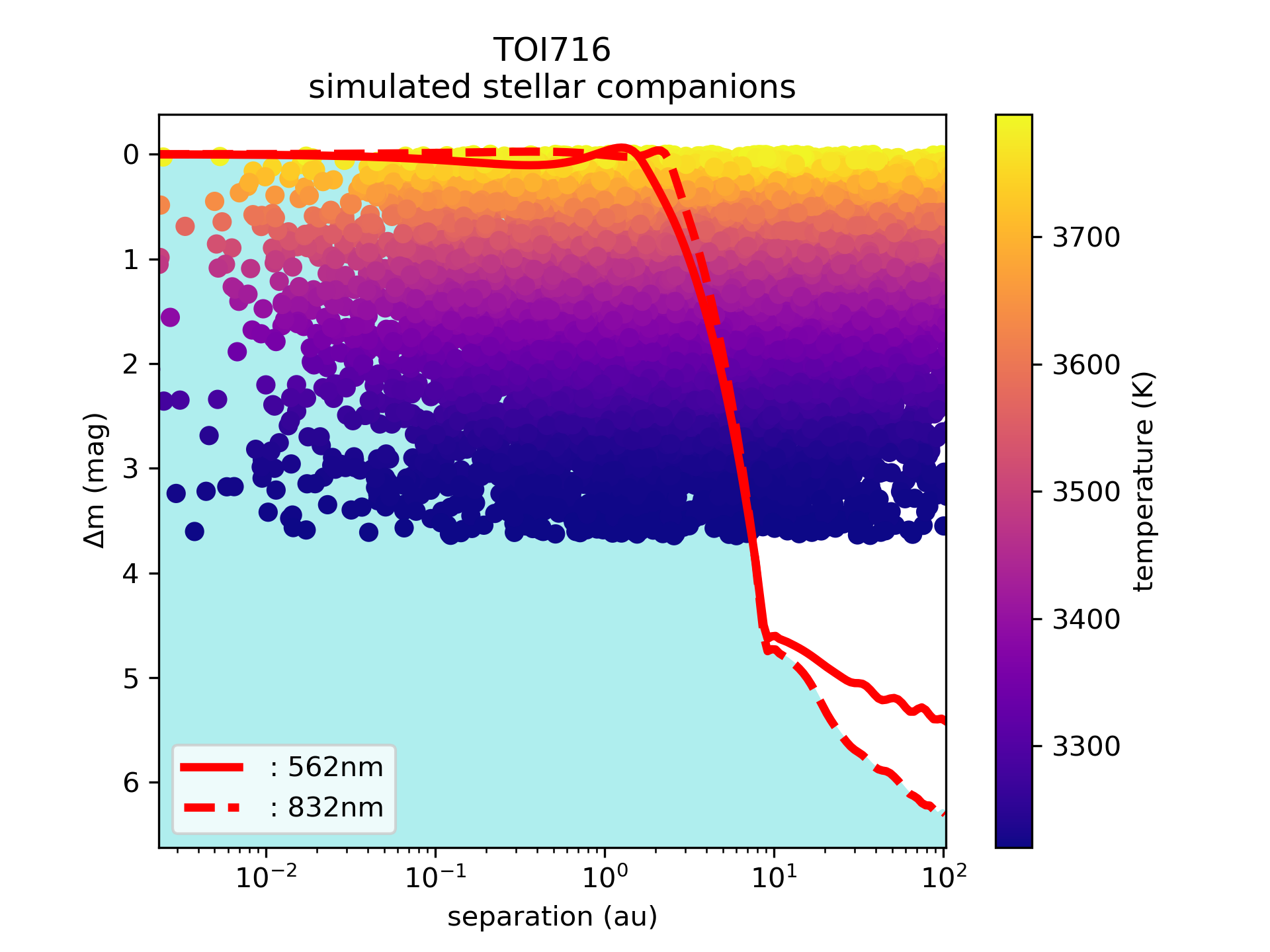}
	  \includegraphics[width=0.3\textwidth]{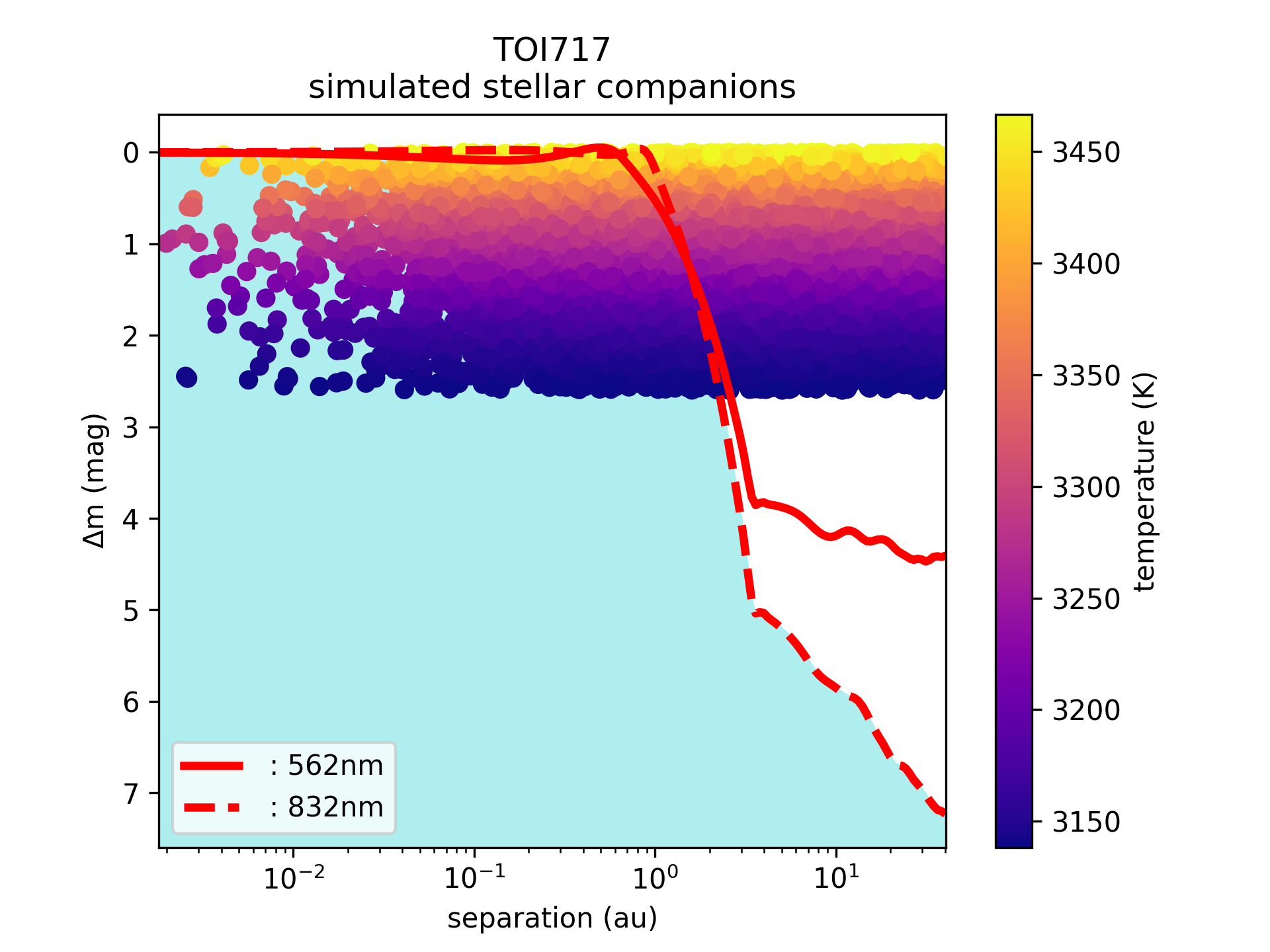}
	  \includegraphics[width=0.3\textwidth]{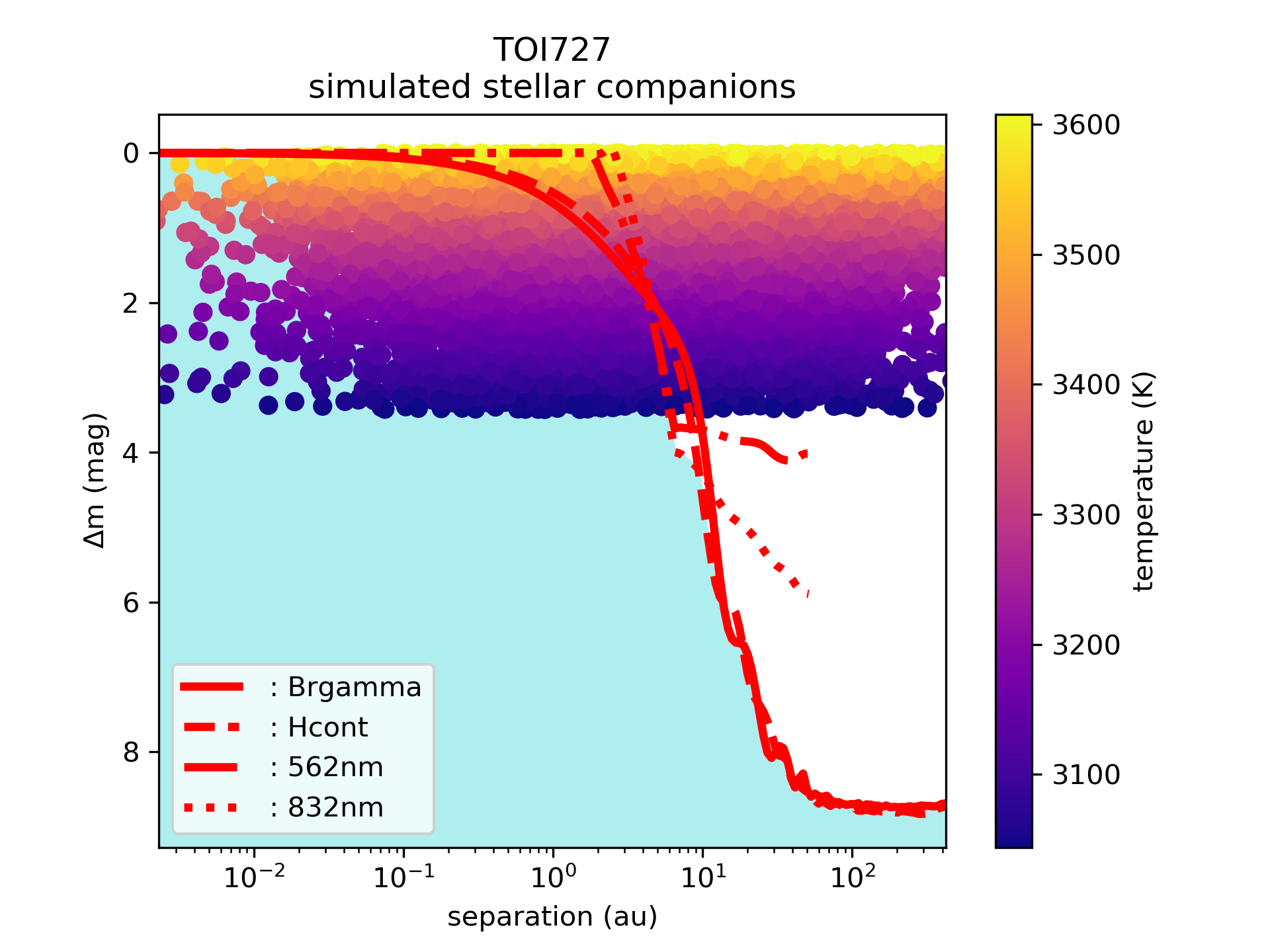}
	  \includegraphics[width=0.3\textwidth]{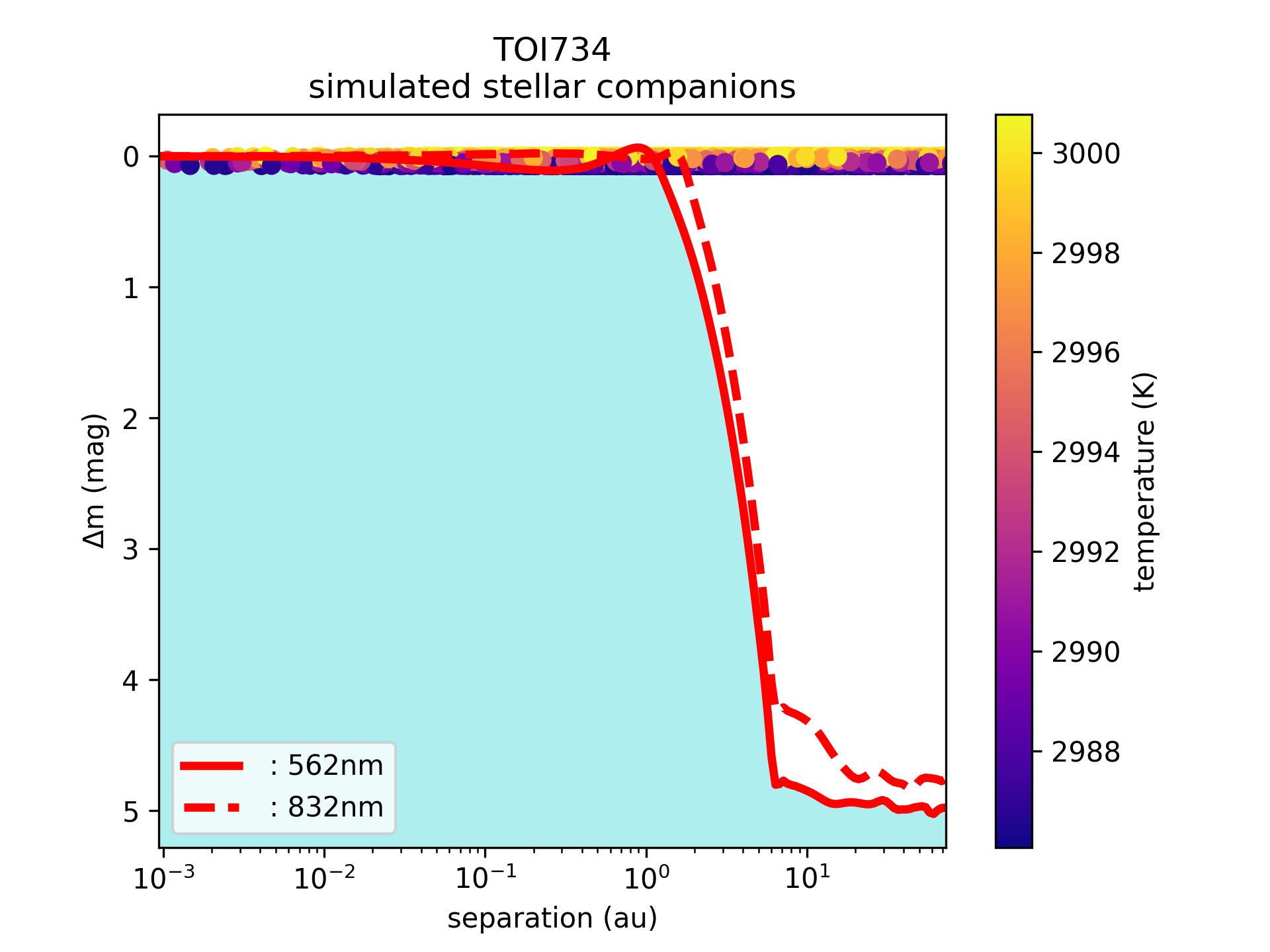}
	  \includegraphics[width=0.3\textwidth]{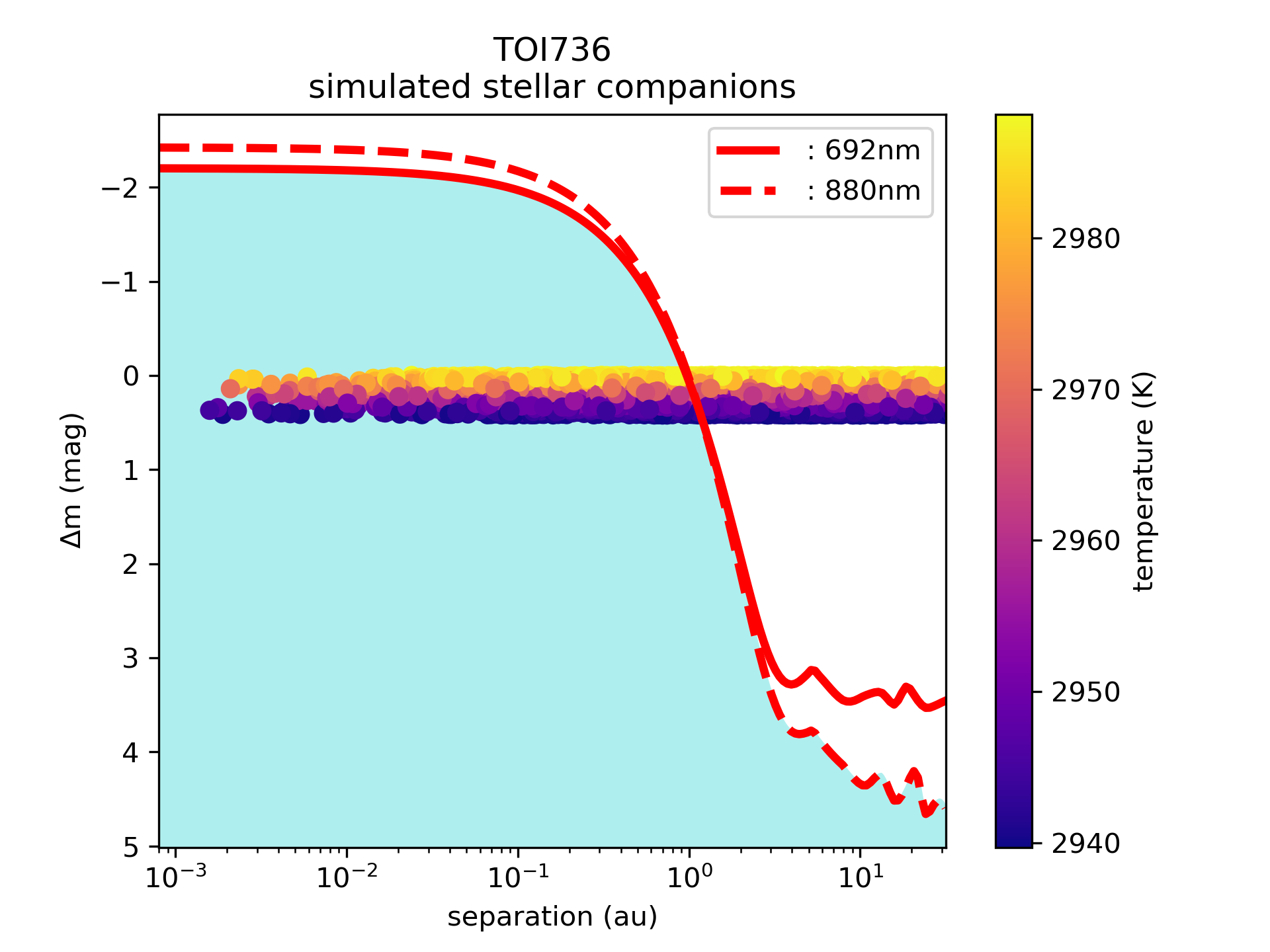}
	  \includegraphics[width=0.3\textwidth]{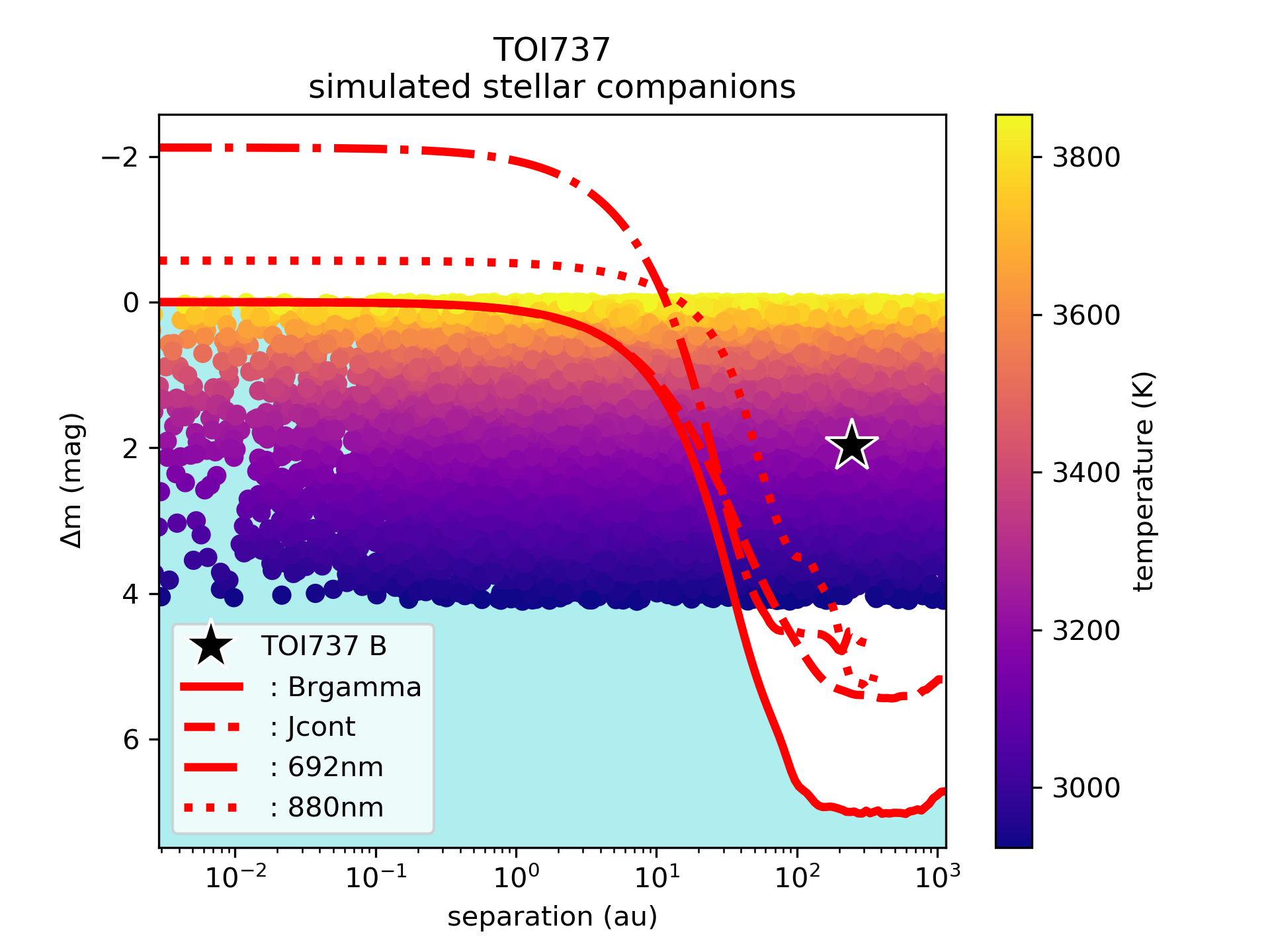}
	  \includegraphics[width=0.3\textwidth]{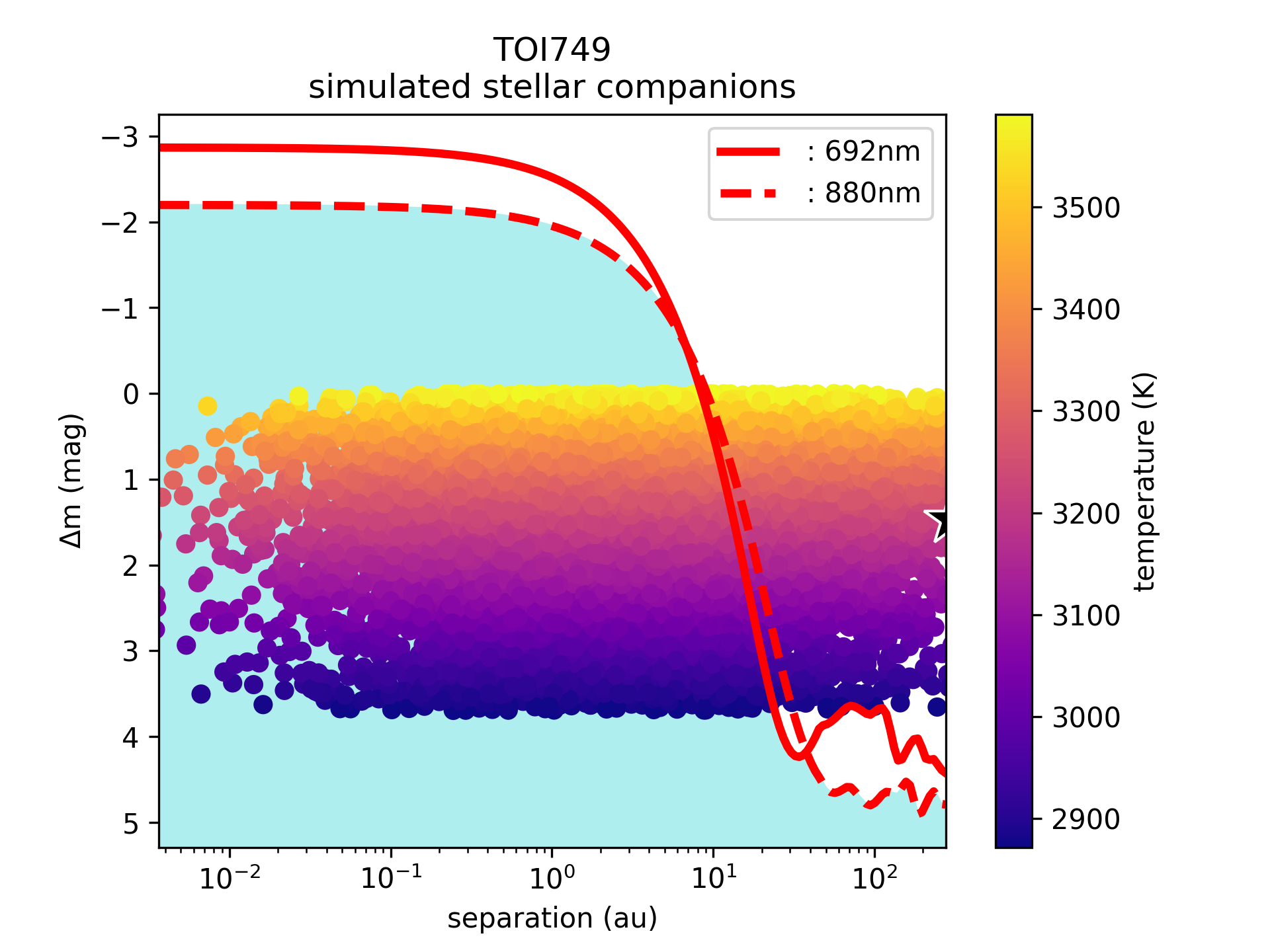}
	  \includegraphics[width=0.3\textwidth]{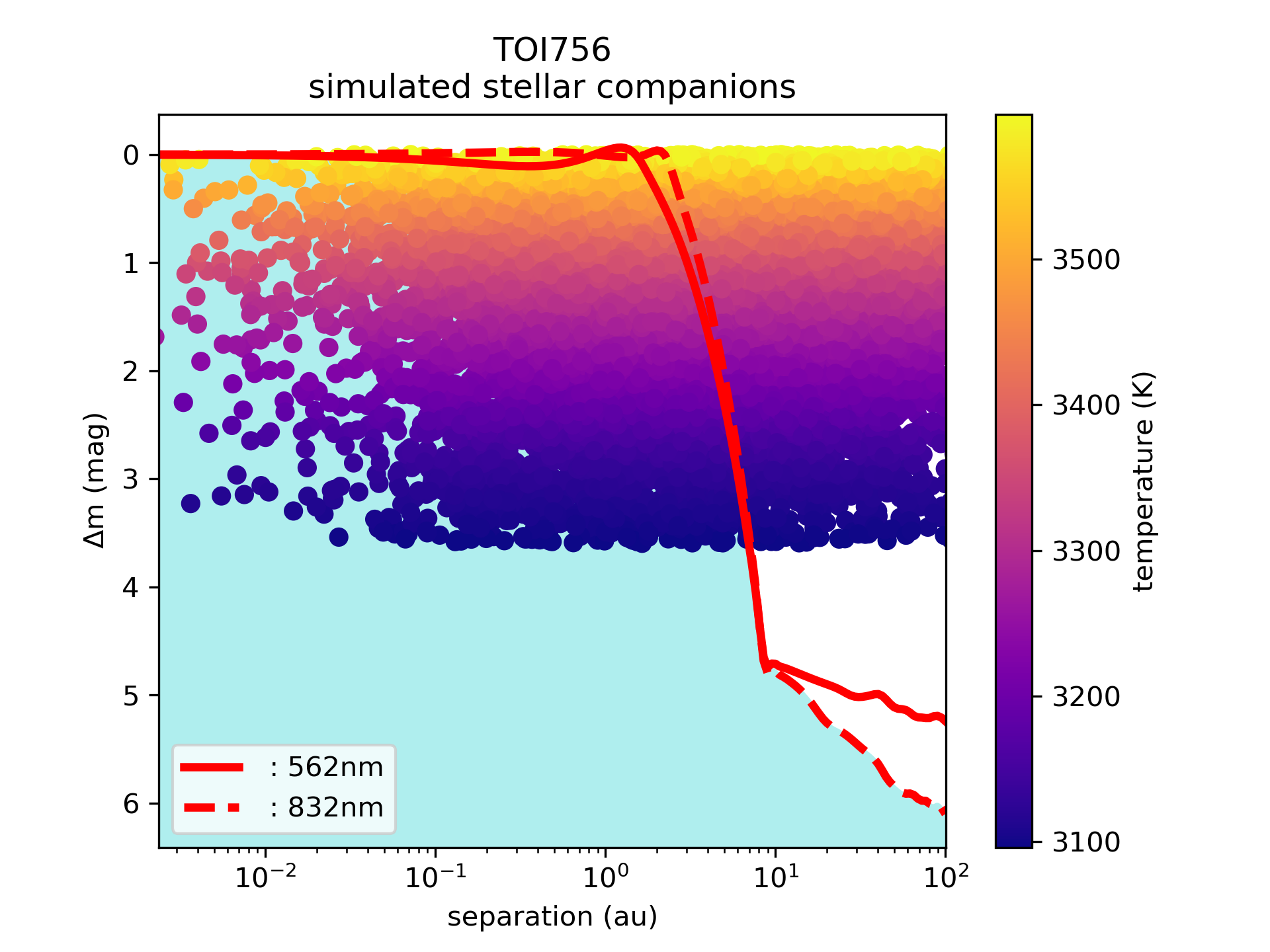}
	  \includegraphics[width=0.3\textwidth]{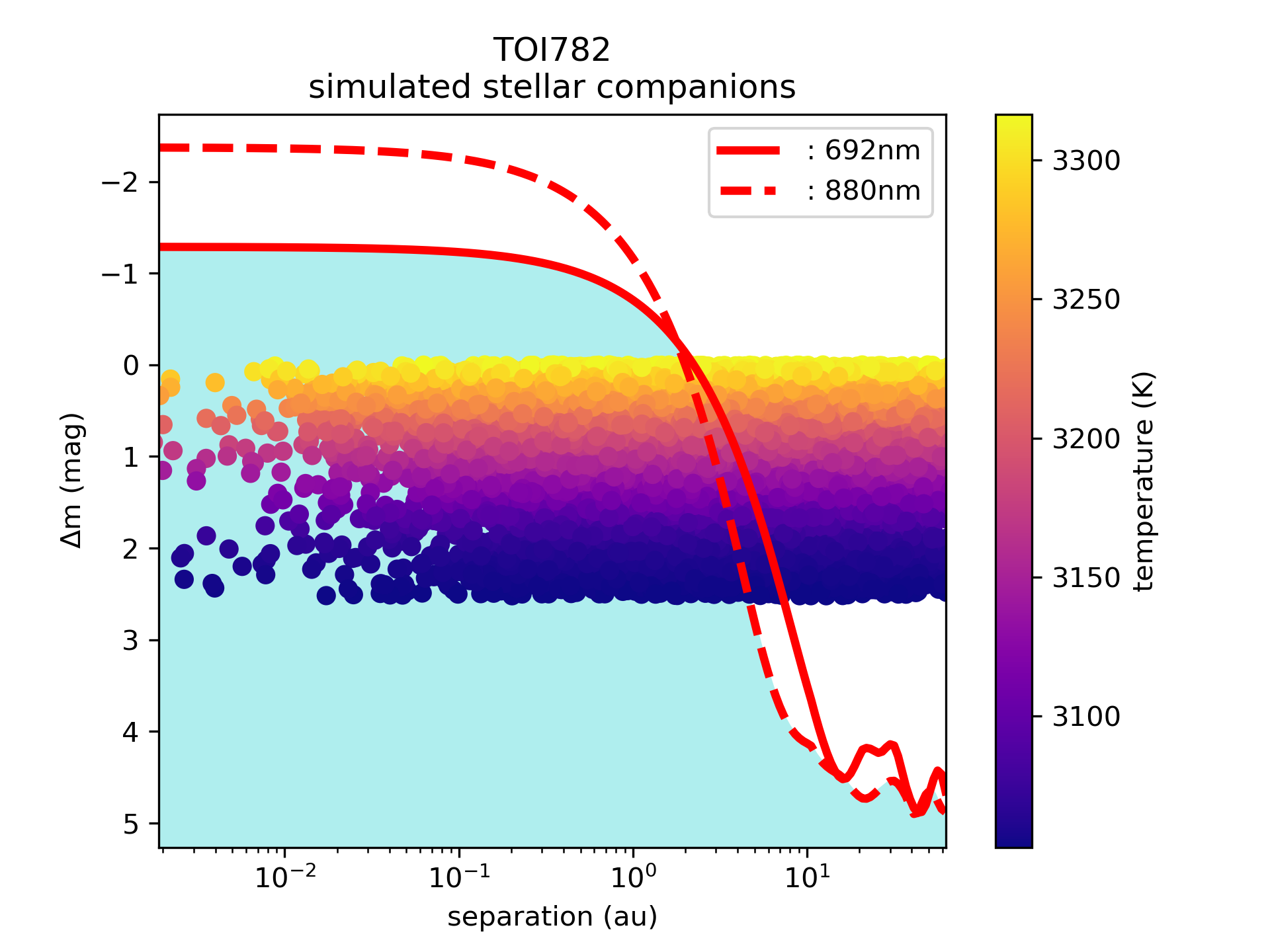}
	  \includegraphics[width=0.3\textwidth]{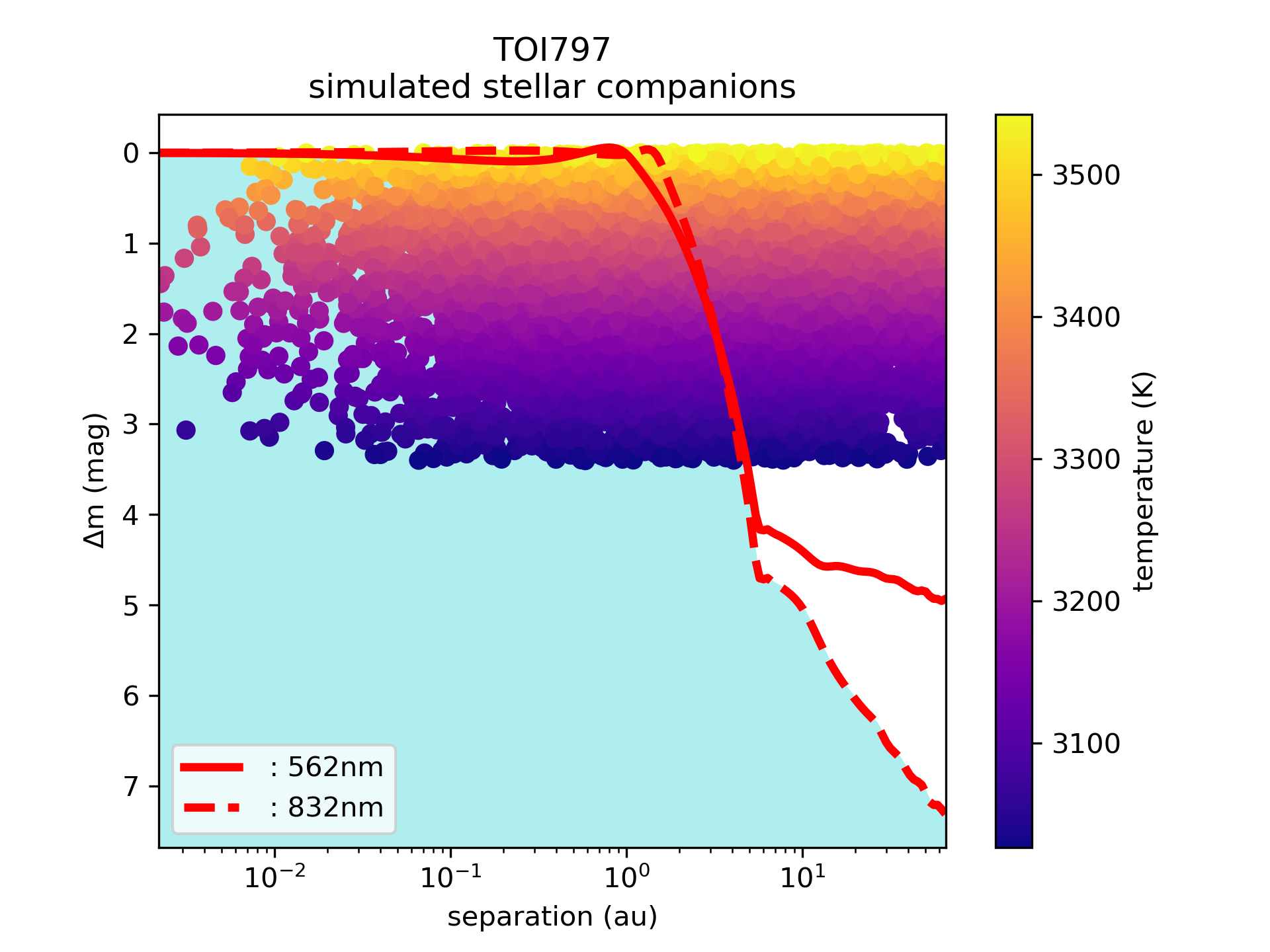}
	  \includegraphics[width=0.3\textwidth]{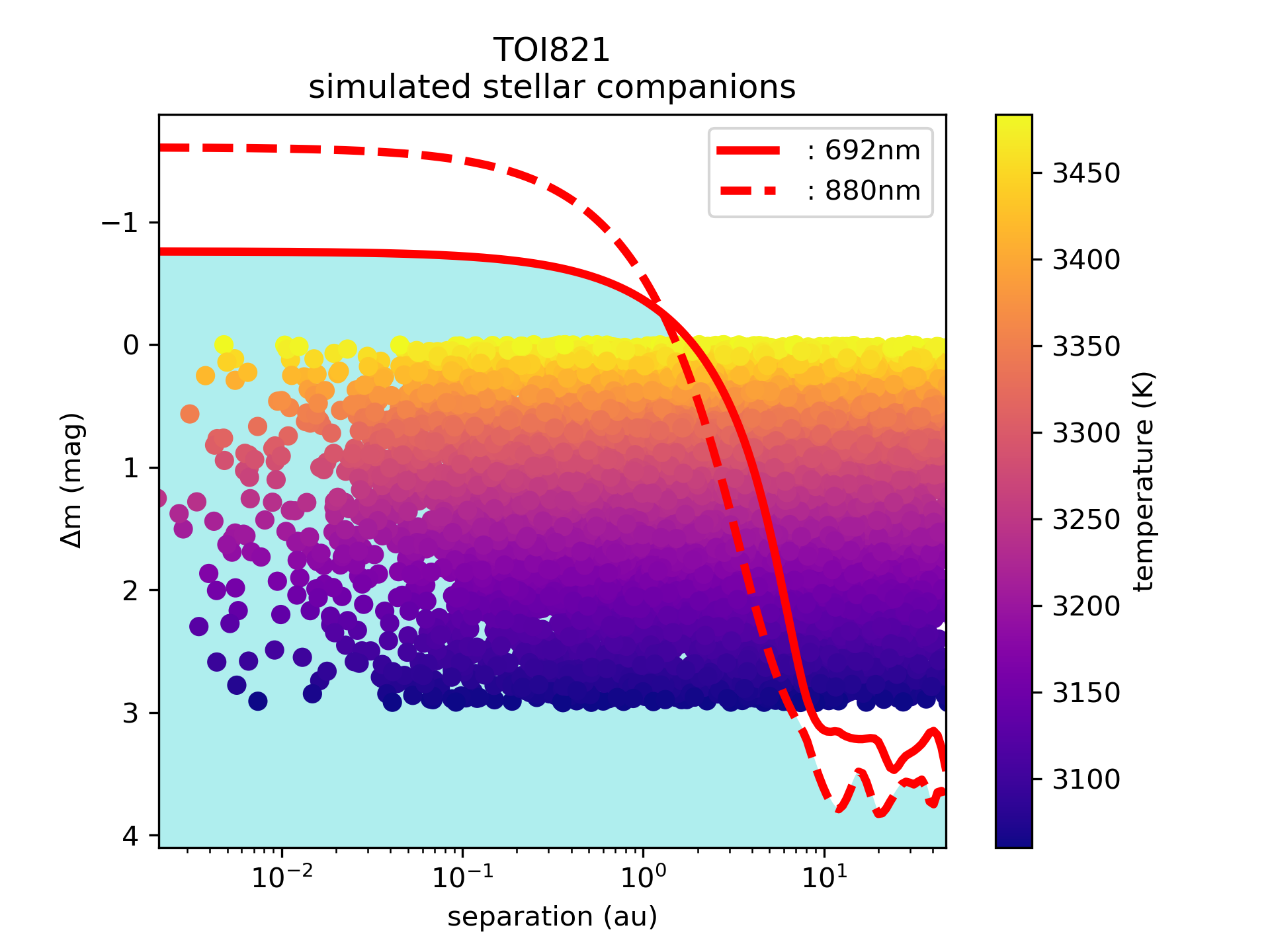}
  \end{center}
  \caption{Simulated Companions: TOI633 to TOI821}
  \label{fig:Sim_Comp_3}
\end{figure*}
\begin{figure*}[!htb]
  \begin{center}
	  \includegraphics[width=0.3\textwidth]{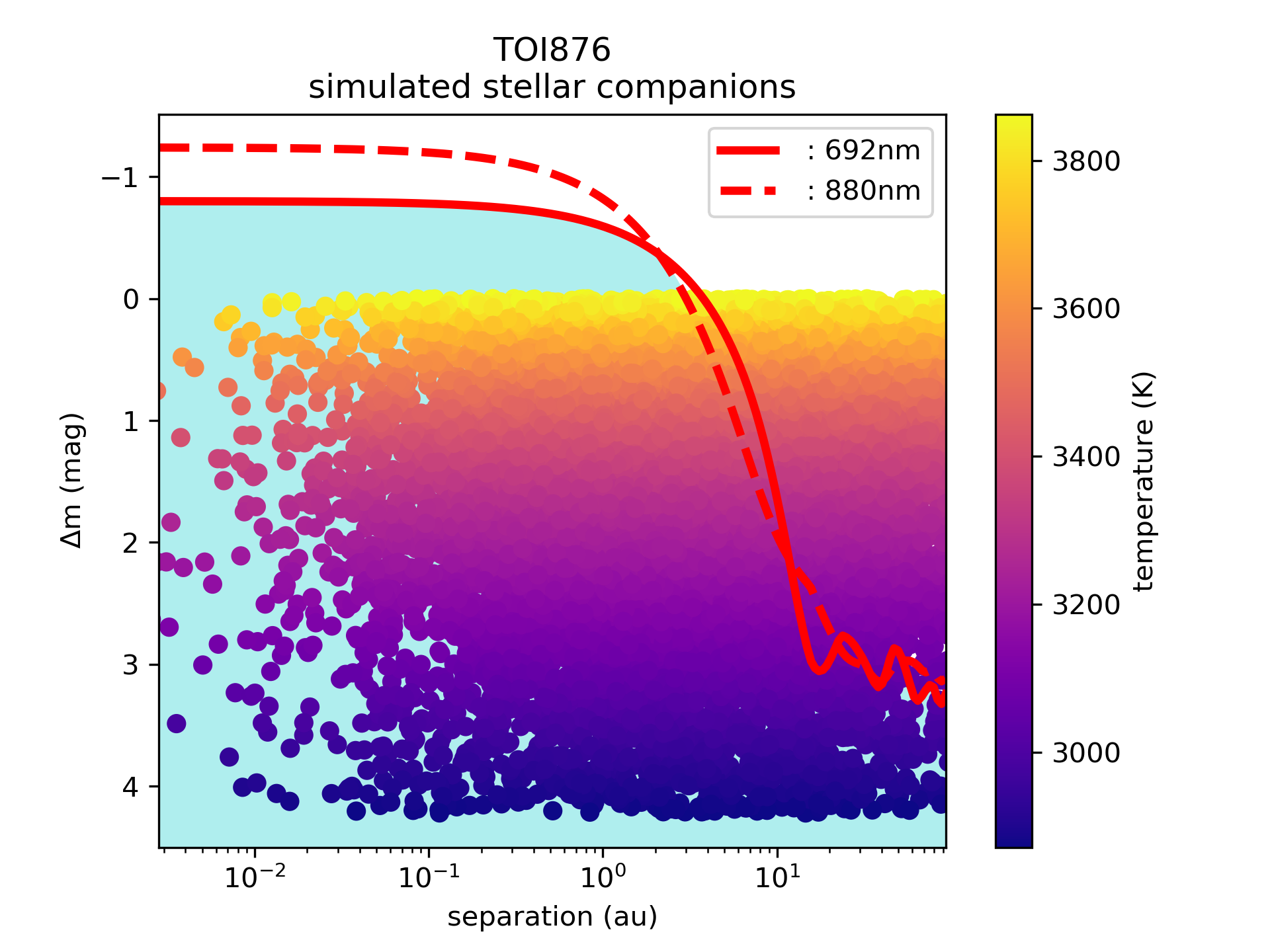}
	  \includegraphics[width=0.3\textwidth]{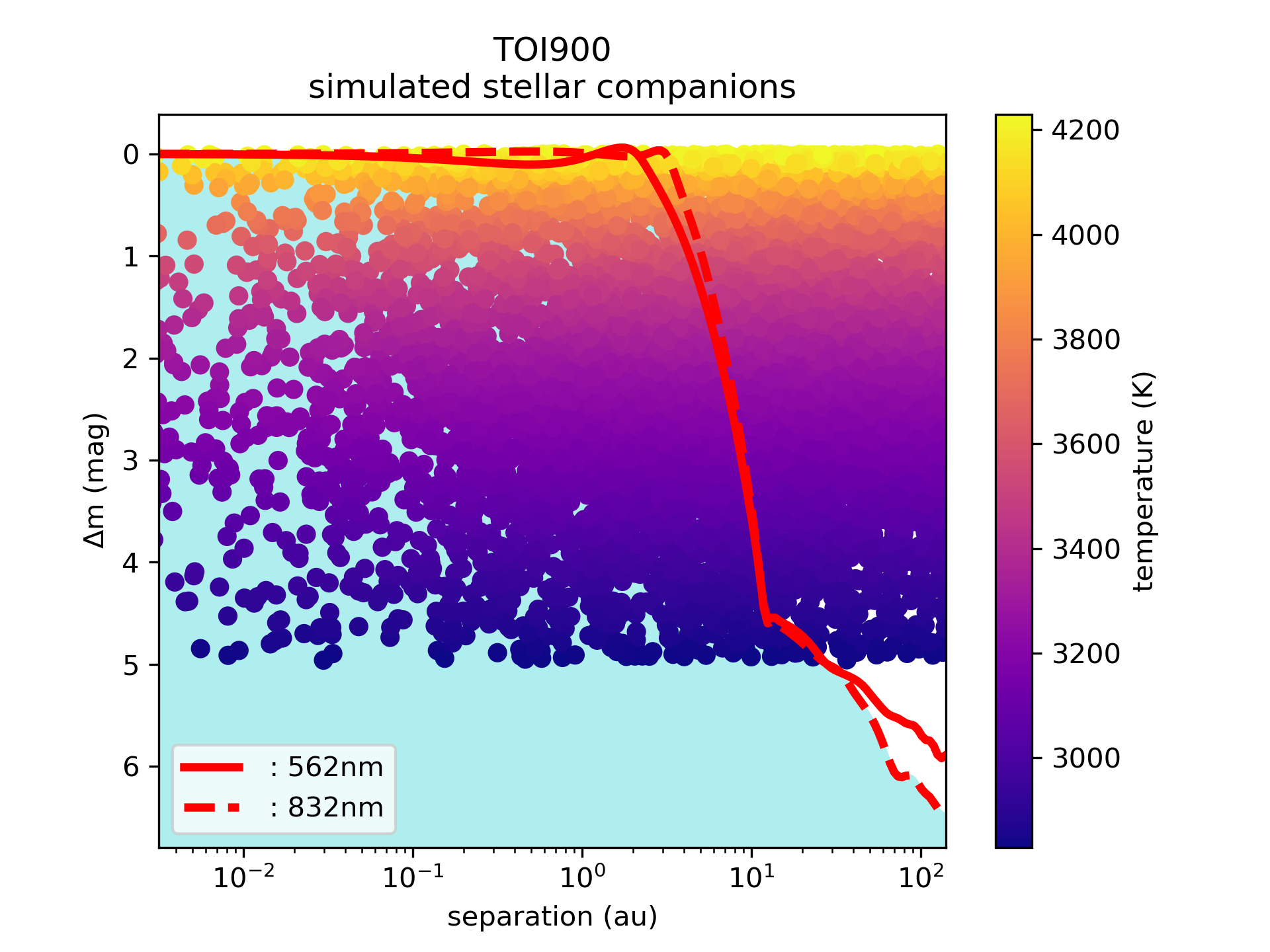}
	  \includegraphics[width=0.3\textwidth]{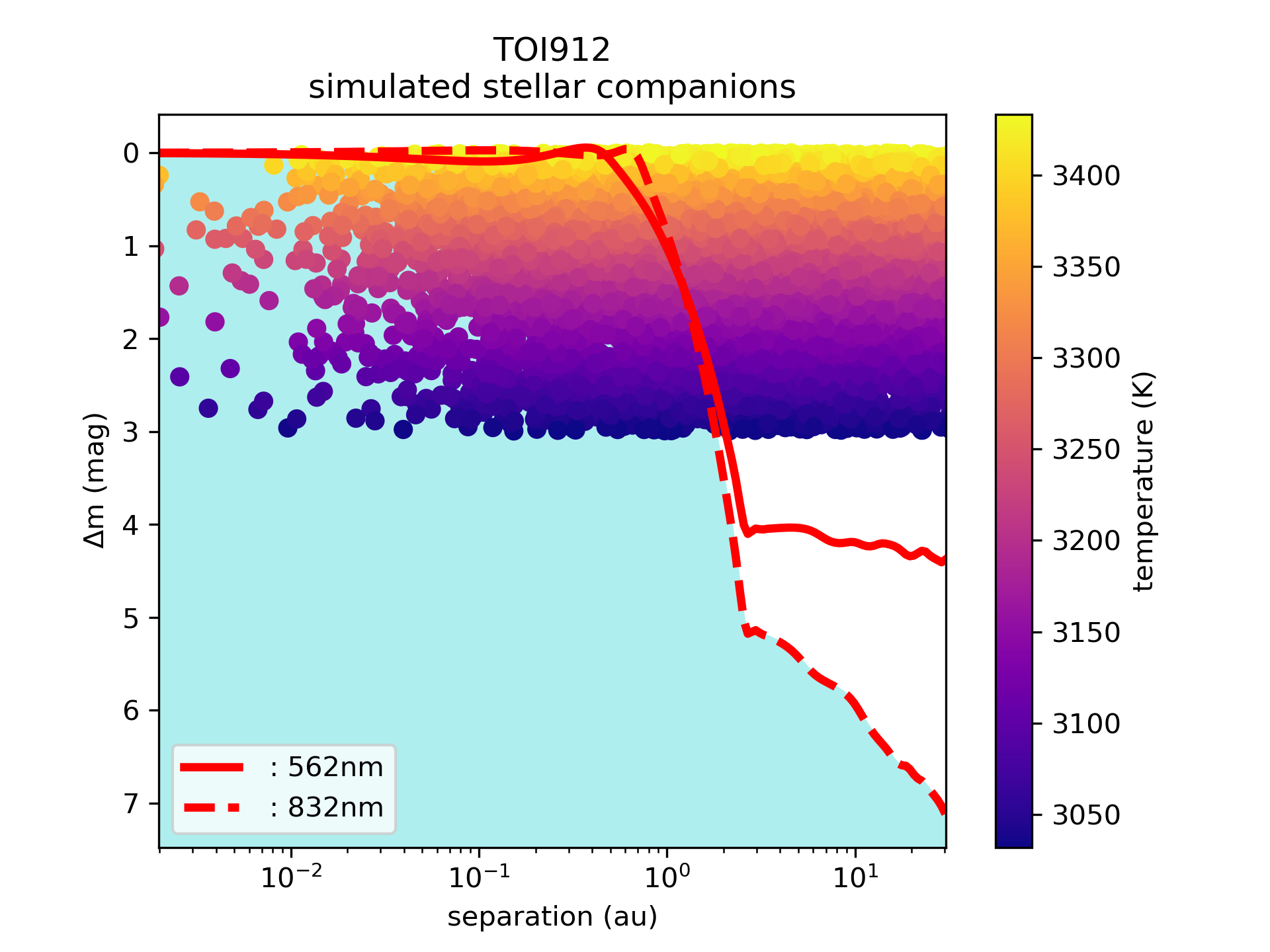}
	  \includegraphics[width=0.3\textwidth]{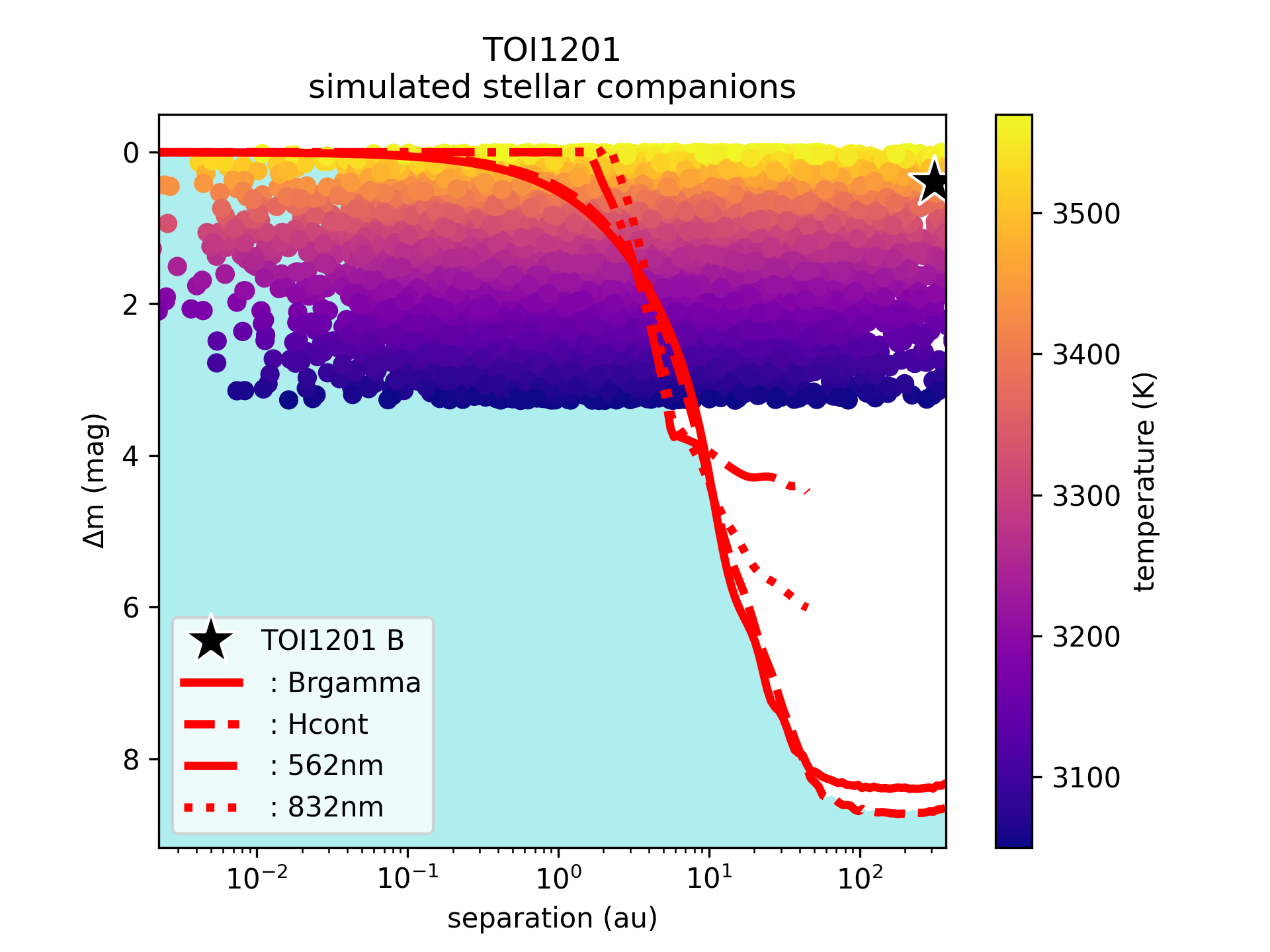}
	  \includegraphics[width=0.3\textwidth]{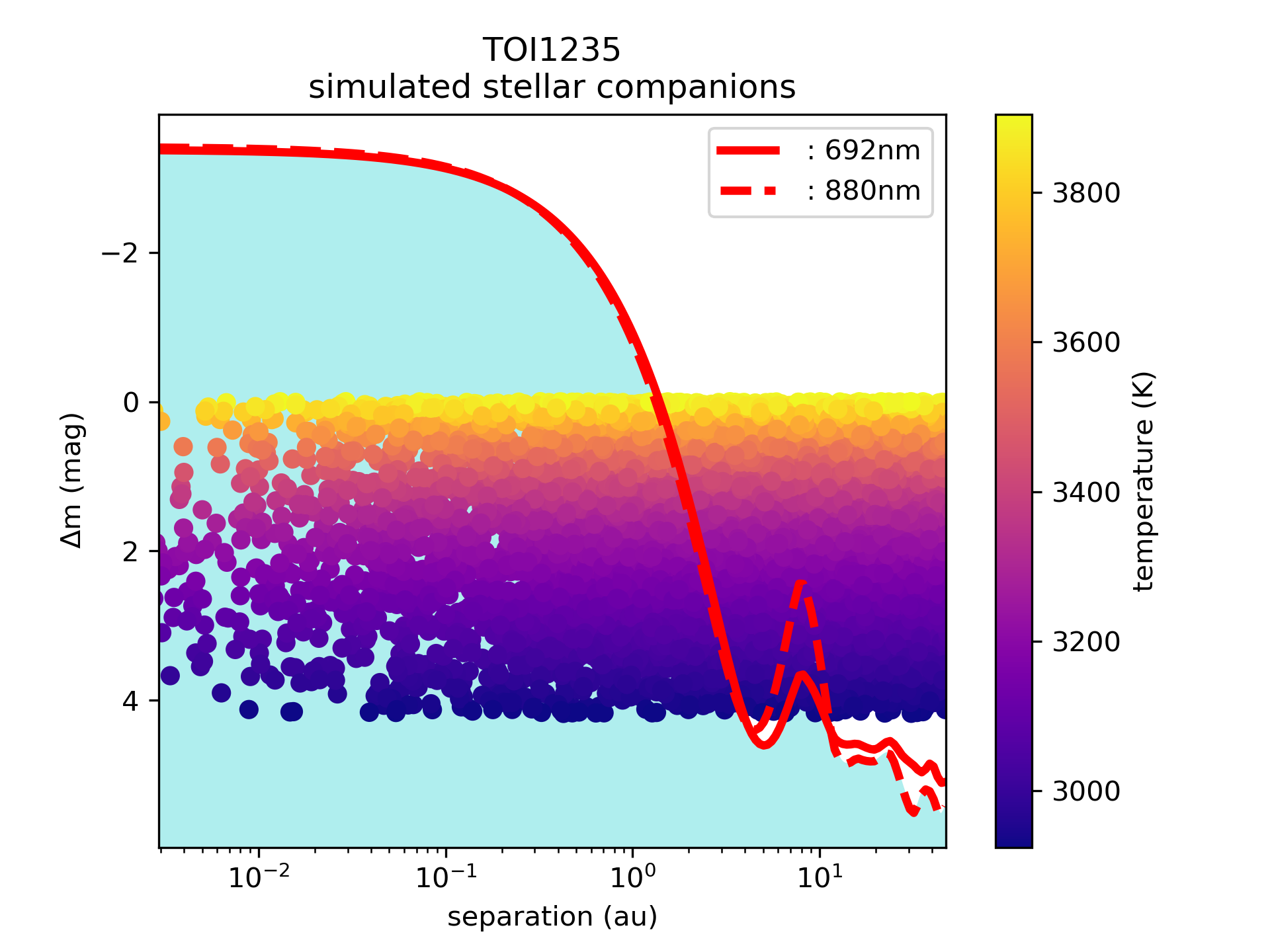}
	  \includegraphics[width=0.3\textwidth]{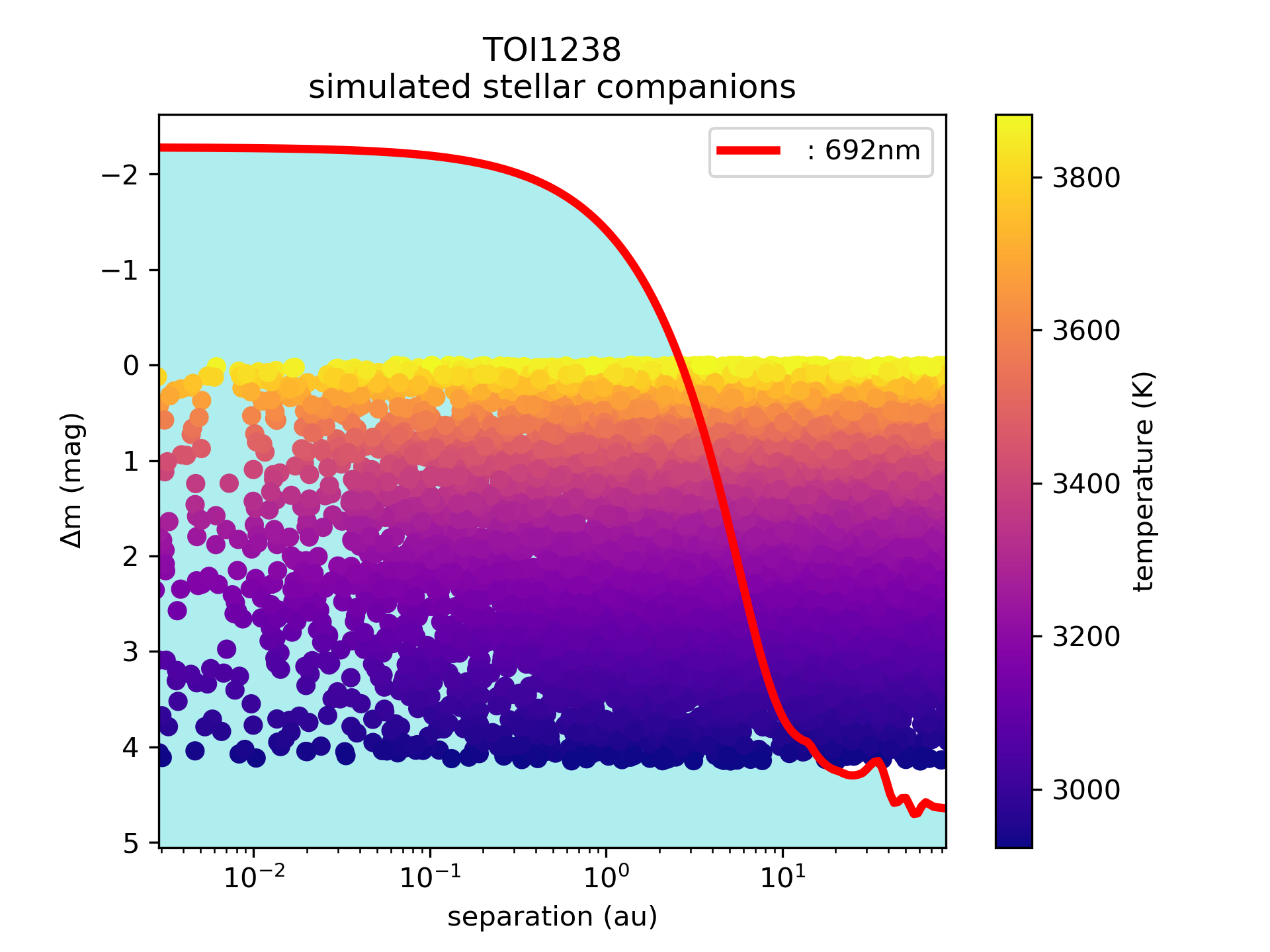}
	  \includegraphics[width=0.3\textwidth]{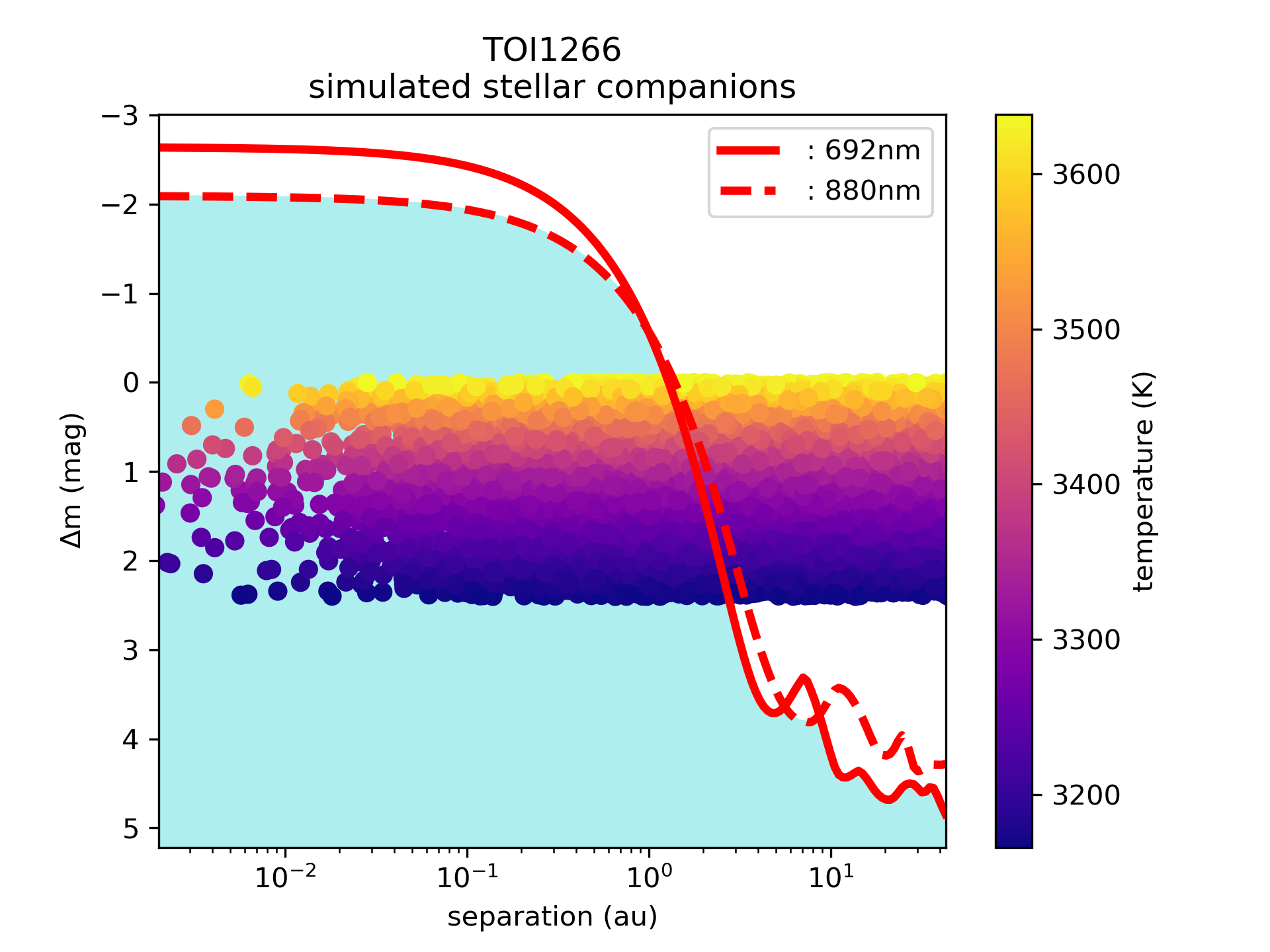}
	  \includegraphics[width=0.3\textwidth]{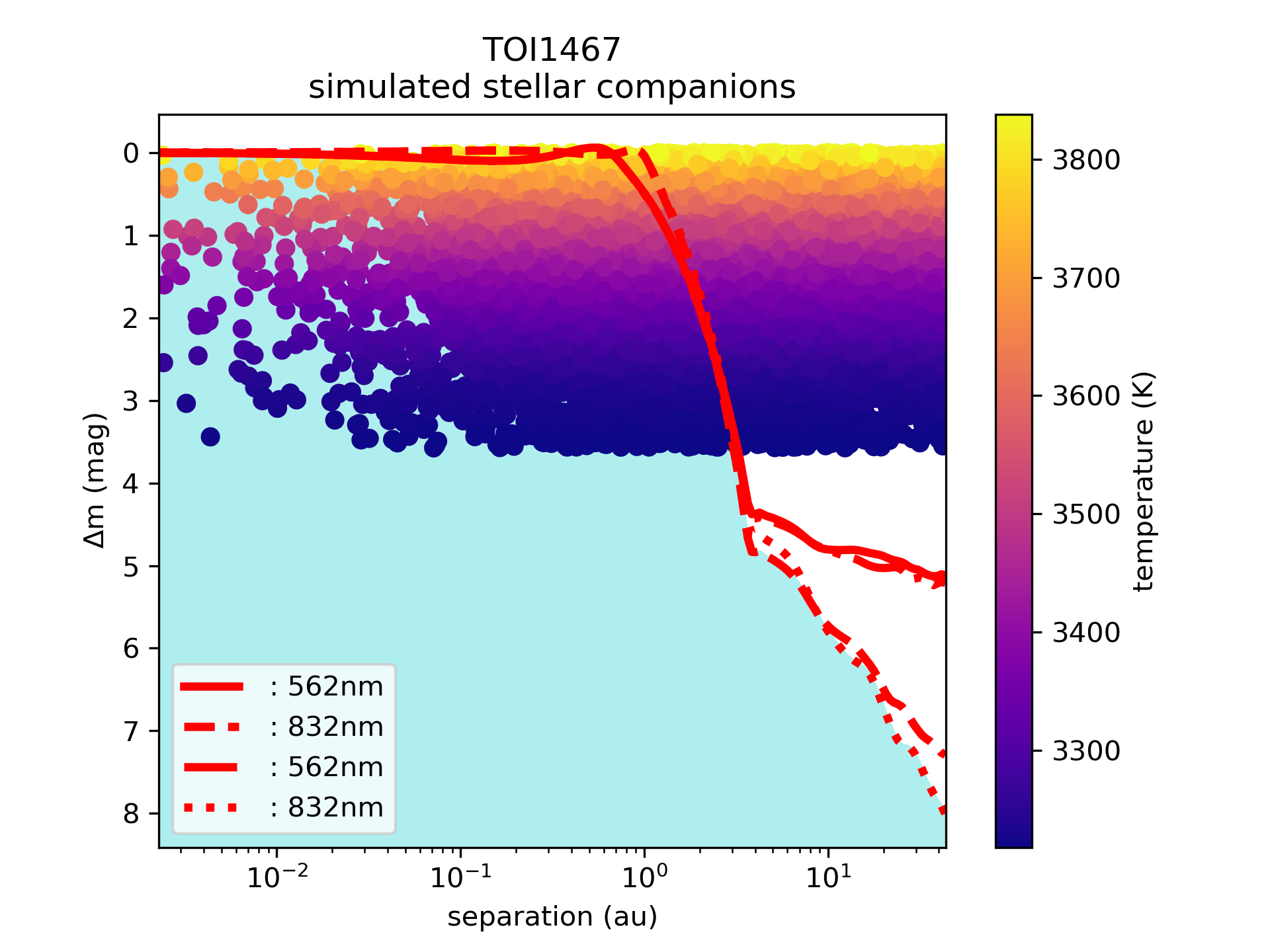}
	  \includegraphics[width=0.3\textwidth]{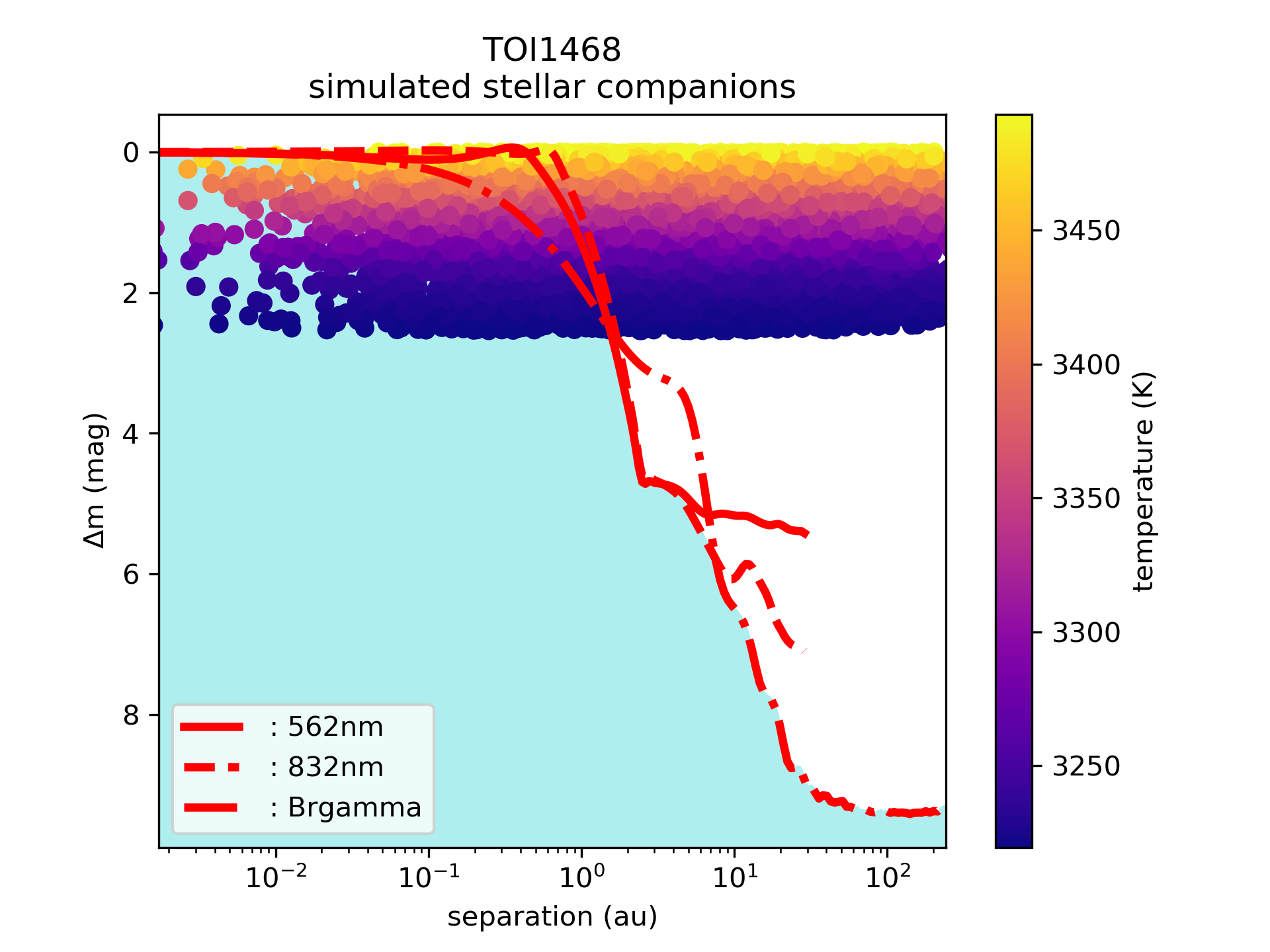}
	  \includegraphics[width=0.3\textwidth]{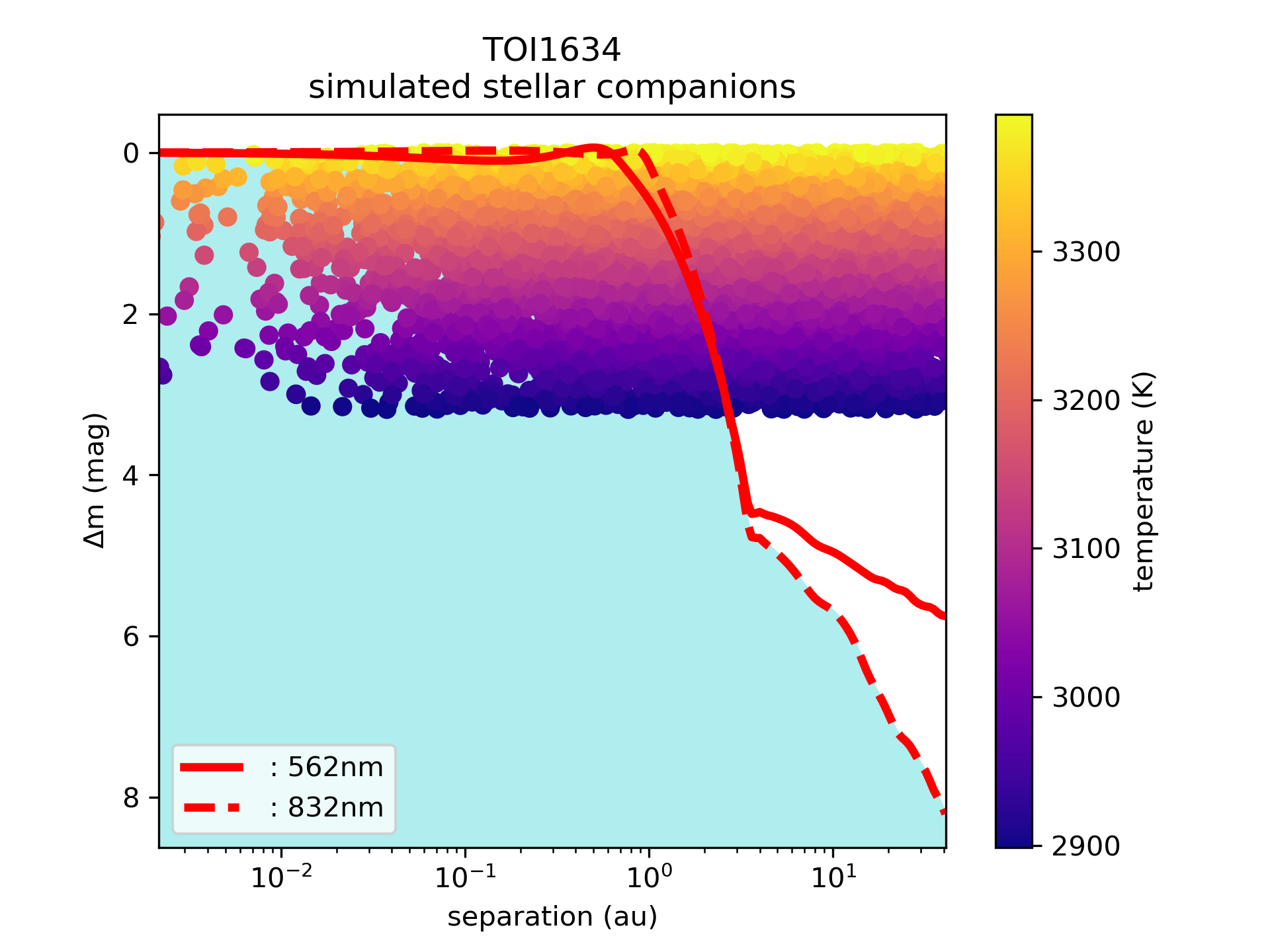}
	  \includegraphics[width=0.3\textwidth]{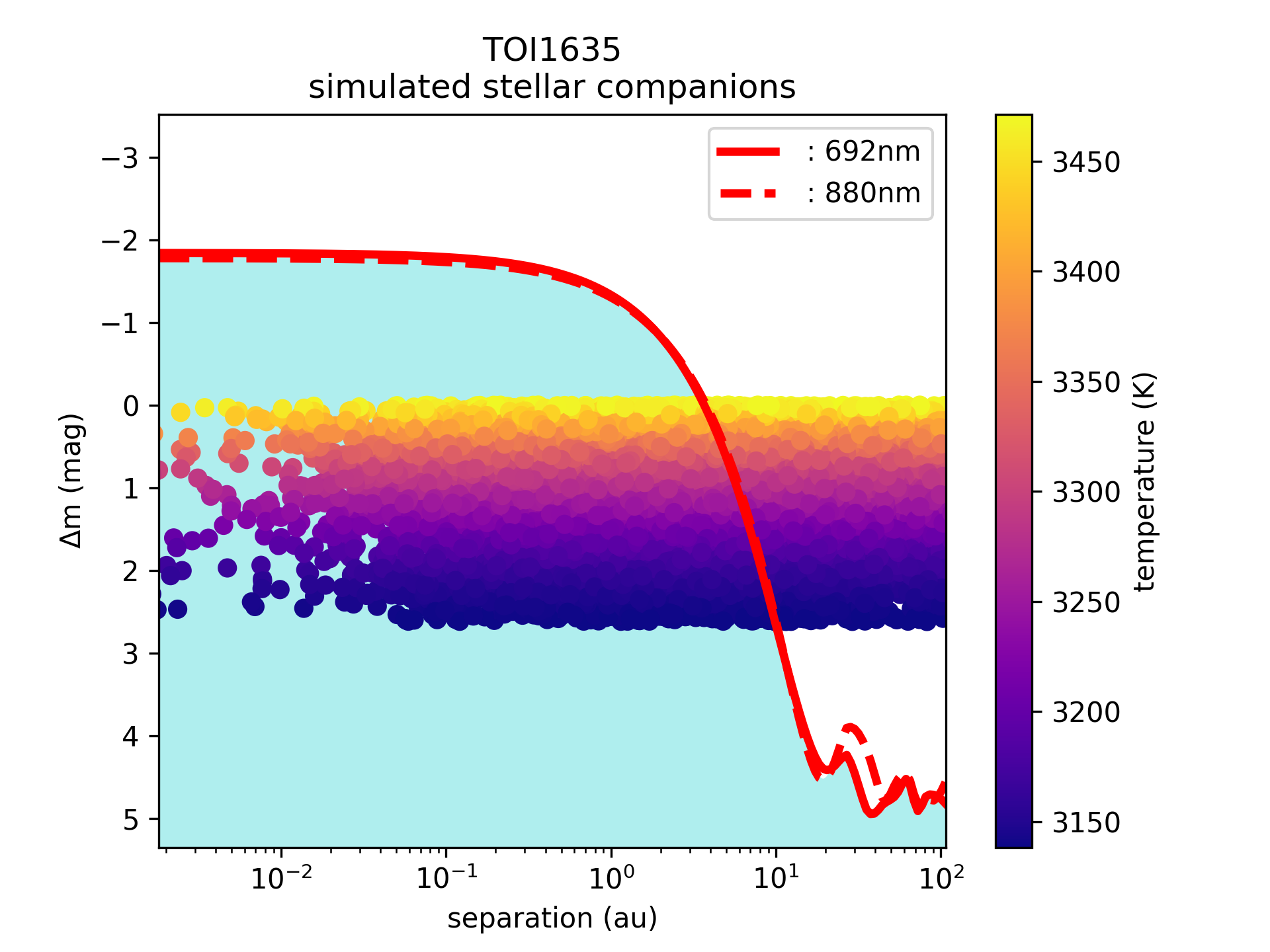}
	  \includegraphics[width=0.3\textwidth]{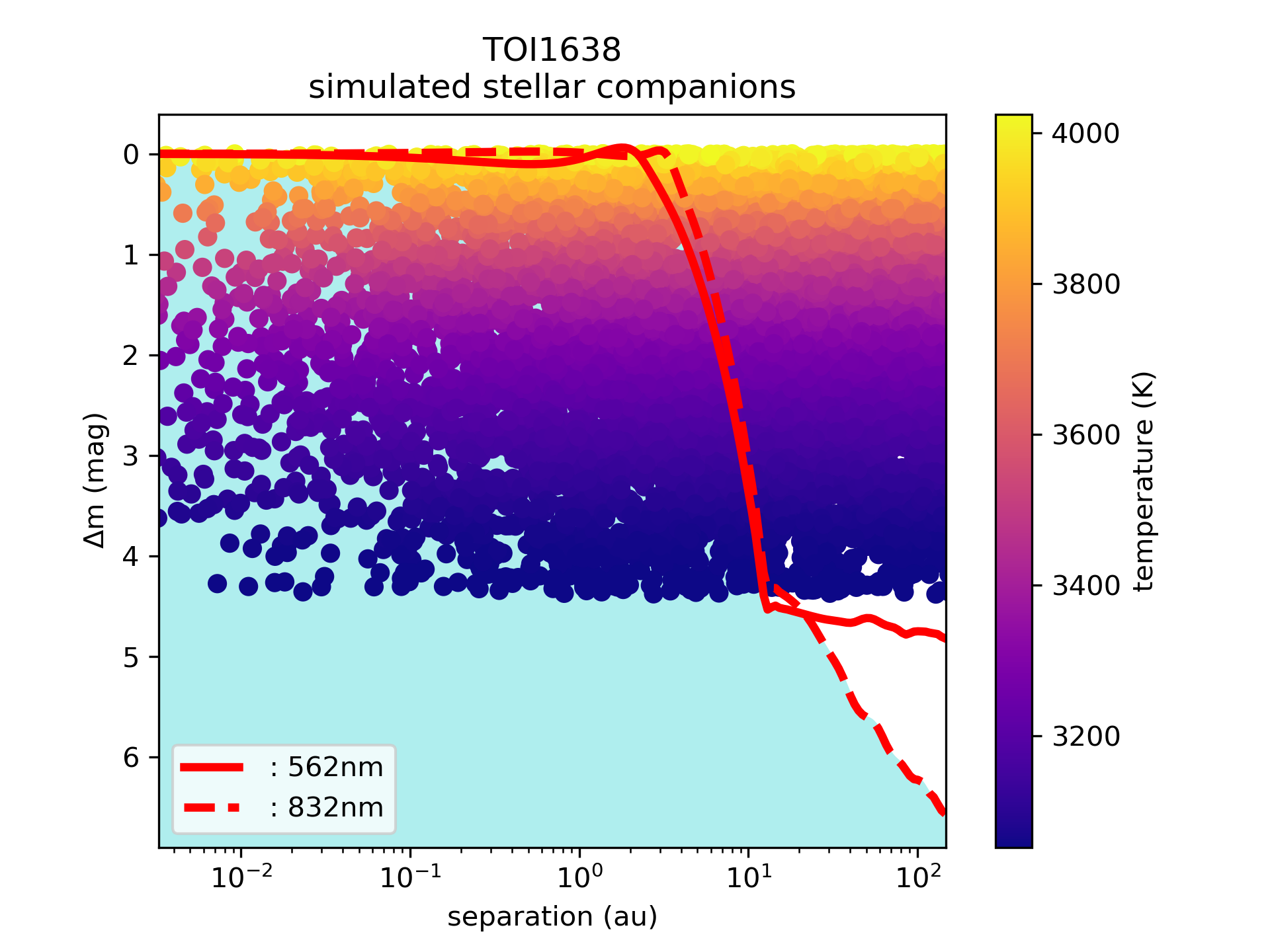}
	  \includegraphics[width=0.3\textwidth]{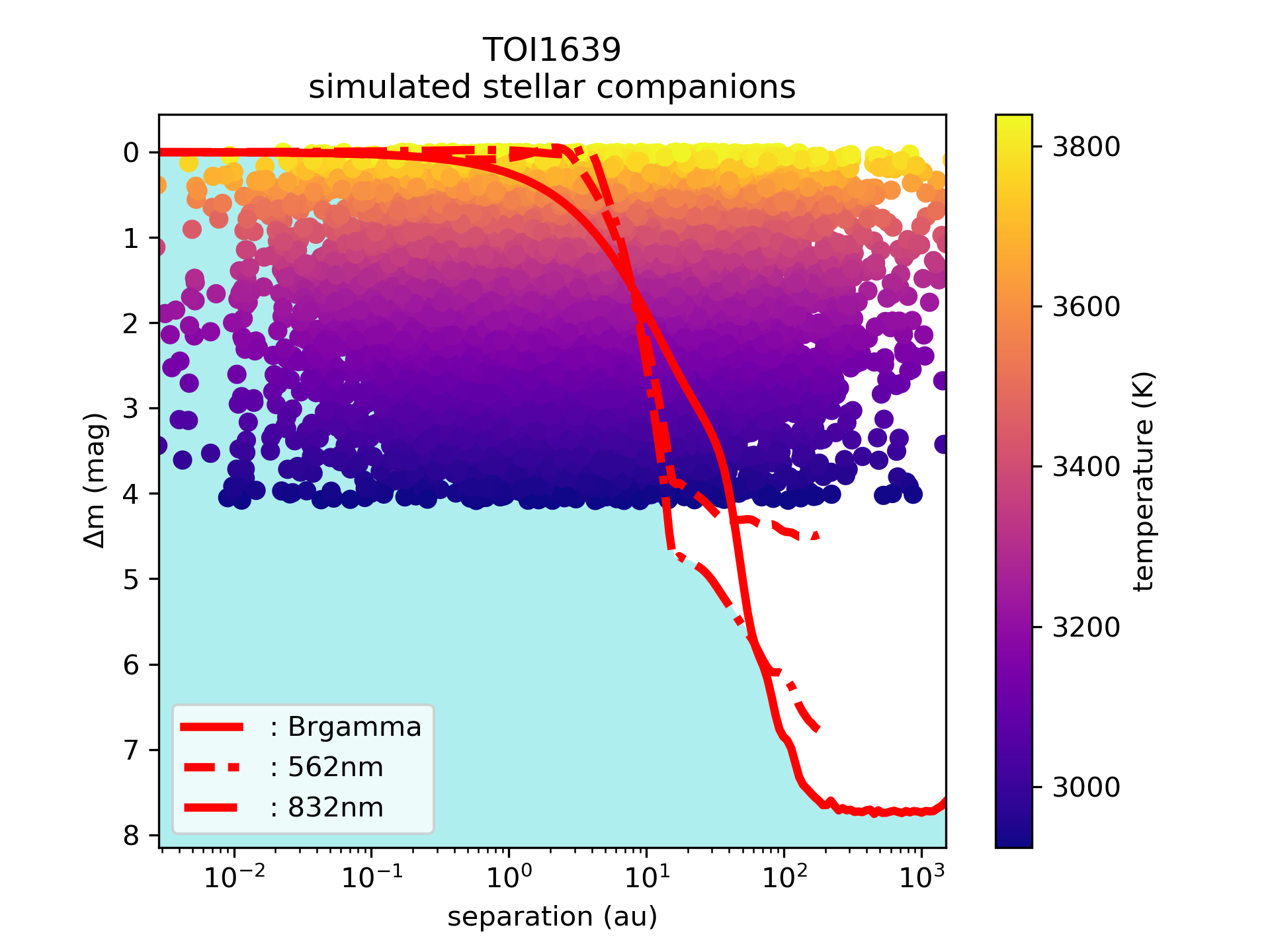}
  \end{center}
  \caption{Simulated Companions: TOI876 to TOI1639}
  \label{fig:Sim_Comp_4}
\end{figure*}



\end{document}